\documentclass[authoryear,preprint,12pt]{elsarticle}

\usepackage{amssymb}
\usepackage{amsmath}
\usepackage{fancyhdr}
\usepackage{graphicx}
\usepackage{epsfig}
\usepackage{natbib}
\usepackage{multicol}
\usepackage{pstricks,pst-grad}
\usepackage{aas_macros}
\usepackage{multirow}
\usepackage{amssymb}
\usepackage{graphics}
\usepackage{xspace}
\usepackage{lineno}
\usepackage{color}

\newcommand{\unitSI}[2][{}]{\ensuremath{\ifx\\#1\else#1\,\fi\mathrm{#2}}}
\newcommand{\unit}[2][{}]{\ensuremath{\ifx\\#1\else#1\,\fi{#2}}}

\newcommand{\pdiff}[2][{}]{\ensuremath{\partial_{#2}\ifx\\#1\else^{#1}\fi}}

\newcommand{\deriv}[2][{}]{\ensuremath{\frac{d #1} {d {#2} }}}

\newcommand{\Eq}[1]{Eq.~({#1})}
\newcommand{\Eqs}[1]{Eqs.~({#1})}
\newcommand{\Fig}[1]{Fig.~{#1}}
\newcommand{\Figs}[1]{Figs.~{#1}}
\newcommand{\Table}[1]{Table~{#1}}

\newlength{\figwidth}
\newcommand{\Rev}[1]{{\textcolor{black}{#1}}} 
\newcommand{\FAC}{\renewcommand{\FAC}{FAC\xspace}field-aligned current
(FAC)\xspace}

\journal{Planetary and Space Science}

\begin{document}

\begin{frontmatter}



\title{Influence of upstream solar wind on thermospheric flows at Jupiter}



\author[ad1,ad2]{J. N. Yates\corref{cor1}}
\ead{japheth@star.ucl.ac.uk}

\author[ad1,ad2]{N. Achilleos}

\author[ad1,ad2]{P. Guio}

\address[ad1]{Department of Physics and Astronomy, University College London, Gower Street, London, UK}
\address[ad2]{Centre for Planetary Sciences at UCL / Birkbeck, University College London, Gower Street, London, UK}
\cortext[cor1]{Corresponding author: Tel.: +44 (0)20 7679 4349; fax: +44 (0)20 7679 2328.}

\begin{abstract}

  The coupling of Jupiter's magnetosphere and ionosphere plays a vital role in creating its auroral 
  emissions. The strength of these emissions is dependent on the difference in speed of the rotational flows 
  within Jupiter's high-latitude thermosphere and the planet's magnetodisc. Using an azimuthally symmetric global 
  circulation model, we have simulated how upstream solar wind conditions affect the energy and direction of 
  atmospheric flows. In order to simulate the effect of a varying dynamic pressure in the upstream solar wind, we 
  calculated three magnetic field profiles representing compressed, averaged and expanded `middle' magnetospheres. 
  These profiles were then used to solve for the angular velocity of plasma in the magnetosphere. This angular 
  velocity determines the strength of currents flowing between the ionosphere and magnetosphere. We examine the 
  influence of variability in this current system upon the global winds and energy inputs within the Jovian 
  thermosphere. We find that the power dissipated by Joule heating and ion drag increases by ${\sim}\unitSI[190]{\%}$ 
  and ${\sim}\unitSI[185]{\%}$ from our compressed to expanded model respectively. We investigated the effect 
  of exterior boundary conditions on our models and found that by reducing the radial current at the outer edge 
  of the magnetodisc, we also limit the thermosphere's ability to transmit angular momentum to this region.

\end{abstract}


\end{frontmatter}

\linenumbers

\section{Introduction} 
\label{intro}

  Amongst the eight planets in the solar system, Jupiter, in addition to being the largest planet, also 
  has the largest magnetic moment and the largest magnetosphere. The magnetosphere interacts with both the solar 
  wind and the conducting layer or ionosphere, in the planet's upper atmosphere. These interactions can be quite 
  complex and we may use models with some simplifying assumptions (e.g. axial symmetry) to gain insight into the 
  dynamics of the magnetosphere, upper atmosphere and their physical interactions with the solar wind.\\

  Several models of Jupiter's magnetosphere and ionosphere have been developed in recent studies 
  \citep{nichols04,cowley05,cowley07,smith09}. These models range from detailed studies of the middle 
  magnetosphere only \citep{nichols04} to global studies of the entire magnetosphere \citep{cowley05, cowley07} and 
  investigations of the coupled magnetosphere, thermosphere and ionosphere systems \citep{smith09} (henceforth 
  SA09). \\
  
  One of the important observations for guiding models is the dominance of Jupiter's magnetosphere by the 
  rapid planetary rotation. Angular momentum is transferred from the planet to the disc-like \Rev{middle} magnetosphere via 
  ion-neutral collisions in the ionosphere. The magnetospheric plasma exhibits a wide range of angular velocities, 
  corresponding \Rev{to} a modest departure from rigid corotation with the planet at distances near Io 
  (\unitSI[6\mbox{--}10]{R_J}) out to regions beyond \unitSI[20]{R_J} which rotate at ${\sim}\unitSI[50]{\%}$ of the 
  planetary rate \citep{McNutt79, hill79, hill1983a, pontius97, vasyliunas1983}. This angular momentum and energy 
  transfer between the ionosphere and magnetosphere is conveyed by two principal current systems. The first of 
  these is related to the rotation of the middle magnetosphere. The second is related to the 
  interaction of the solar wind with the magnetosphere at the high-latitude magnetopause \citep{hill1983bURANUS, 
  isbell1984}. \\
  
  The principal source of plasma for the middle magnetosphere (${\sim}\unitSI[20]{R_J}$ to several tens of 
  \unit{R_J}) is the satellite Io \citep{bagenal1981} which ejects about \unitSI[500\mbox{--}1000]{kg\;s^{-1}} of 
  sulphur dioxide gas which is then ionised \citep{kivelBOOK2004}. Iogenic plasma initially near corotation will 
  lag further behind corotation as it diffuses radially outwards from the Io torus, due to the finitely conducting 
  ionosphere being unable to supply all of the necessary angular momentum via the coupling currents. The electric 
  field in the neutral atmosphere's rest frame depends on the difference in angular velocity between the polar 
  thermosphere and the magnetically conjugate plasma disc, and drives a flow of equatorially directed Pedersen 
  currents. Due to current continuity, \FAC in the steady-state must flow both upwards and 
  downwards along the magnetic field lines which connect the ionosphere and magnetospheric plasma disc. Downward 
  FACs flow from the outermost magnetosphere to the ionosphere. The upward directed FACs are carried by downward 
  precipitating electrons from the magnetosphere \citep{cowbun2001, hill2001, 
  khurana2001, southkiv2001}. These electrons excite emissions in the upper atmosphere and produce the main auroral 
  oval at ${\sim}\unitSI[15]{^{\circ}}$ co-latitude \citep{satoh1996, clarke1998, clarke2004, prange1998, 
  vasavada1999, pallier2001, grodent2003a}. Currents flow radially outward in the equatorial plane of the 
  magnetosphere and, via the \textbf{\emph{J\unit{\times}B}} force, accelerate the plasma towards corotation. The Pedersen, 
  radial and FACs thus represent a complete current `circuit' coupling the magnetosphere and 
  ionosphere. \\
  
  The thermospheric angular velocity at Jupiter partly controls the ionospheric Pedersen 
  currents and thus the dynamics of the magnetosphere. We do not, however, have many measurements of these 
  thermospheric velocities. Studies such as \citet{haungHill89} \Rev{and} \citet{pontius95} have \Rev{attempted} to 
  model these velocities by 
  coupling the magnetosphere, ionosphere and thermosphere, with the assumption that angular momentum was 
  transported through the thermosphere solely by vertical viscous transport. These studies yielded two main 
  conclusions: (i) the relationship between thermospheric and magnetospheric angular velocities was a linear one 
  and (ii) thermospheric dynamics could be parameterised using an `effective' ionospheric conductivity, which 
  represented the effects of the difference in angular velocity of the thermosphere and deep atmosphere (i.e. 
  planetary value). \\

  \citet{smith08saturn} \Rev{and} SA09 showed that, for both Saturn and Jupiter, meridional advection rather than 
  vertical viscous transfer is the main process by which angular momentum is distributed to the high latitude 
  thermosphere. They also showed that the neutral atmosphere super-corotates, ultimately due to the sub-corotation 
  of the middle and outer magnetosphere, at latitudes just equatorwards of the boundary between the field lines 
  of the middle magnetosphere and the Dungey-Vasyliunas (D-V) layer (region II in section~\ref{sec:angular}). 
  Super-corotation occurs in a region where ion drag forces (promote sub-corotation) are insignificant compared to 
  Coriolis forces (promote corotation). These regions, where ion drag tends to zero, lead to the upwelling of gas 
  which expands and cools adiabatically. This then causes \Rev{a pressure gradient} which drives poleward flows 
  at altitudes less than \unitSI[600]{km} (pressures higher than \unitSI[0.04]{\mu bar}). As ion drag is 
  insignificant 
  in this region, the Coriolis force has no obstruction and can accelerate the gas to super-corotate. `Hotspots' 
  were created in these models by converging meridional winds at the poles while lower latitude regions were cooled. 
  For more detailed conclusions the reader is referred to SA09. \\

  The modelling of SA09 combined the advanced middle magnetosphere model of \citet{nichols04} with the 
  axisymmetric model of the entire magnetosphere presented in \citet{cowley05}. These magnetospheric inputs were 
  then coupled to a global two-dimensional circulation model of the Jovian thermosphere. The auroral region in this 
  coupled model is represented by the one-dimensional auroral thermosphere and ionosphere model by 
  \citet{grodent01}. This auroral profile is linearly scaled at each time step according to the global pattern of 
  auroral conductance (see SA09). It is this coupled model of the magnetosphere, ionosphere and 
  thermosphere that we use in the present study of the effects of solar wind variability on Jupiter's thermospheric 
  flows. There are some necessary minor differences between the original SA09 model and the version used 
  herein. These will be discussed in sections \ref{sec:coupledmodel} and \ref{sec:solving}.\\
  
  The aim of this study is to see how magnetospheric compressions and expansions due to changes in solar wind 
  pressure affect steady-state thermospheric flows and temperatures, and the ensuing effect on predicted Jovian 
  auroral 
  activity. This aspect, as well as the inclusion of a realistic atmospheric model in our study, implies that this 
  is a natural extension of the previous studies that we have mentioned. Our basic approach is as follows. We start 
  with a `baseline' magnetodisc of equatorial radius, $\unit{R_{MM}}{=}\unitSI[65]{R_J}$, 
  where $\unit{R_J}{=}\unitSI[71492]{km}$ is taken as the radius of Jupiter. We then produce compressed and expanded 
  disc configurations (section~\ref{sec:magmodel}). Using these magnetospheric models as input to the atmospheric 
  model, we run for 50 Jovian rotations until steady-state is reached. \\

  The theoretical background for our study is given in section~\ref{sec:background}. In section~\ref{sec:results} 
  we show and discuss our results, in section~\ref{sec:pathogen} we show cases where our radial current boundary 
  condition is changed, and we summarise the findings in section~\ref{sec:conclusion}. At present we run our model 
  until steady-state is reached, but in the future we will aim to simulate the transient, time-dependent effects 
  caused by relatively rapid variations in solar wind pressure and thus magnetospheric size (e.g. 
  \citet{cowley07}). \Rev{These rapid variations in solar wind pressure cause significant changes in magnetospheric size, 
  and thus plasma angular velocity, on time scales of \unitSI[2{-}3]{hours}. On the other hand, the large inertia of the 
  neutral thermosphere implies that changes in plasma angular velocity would affect the thermosphere on longer time scales, 
  such as ${\sim}\unitSI[5{-}20]{hours}$. This condition allows for the approximation that plasma angular momentum is 
  conserved on the shorter time scales associated with solar wind pressure changes \citep{cowbun2003a}.} 
  Time-dependent effects are usually neglected in most studies of global energy 
  transport in the Jovian system. Such studies may thus have bearing on the `energy crisis' at Jupiter that has 
  evaded a definitive answer for four decades\Rev{,} the fact that the planet's exospheric temperatures cannot be 
  maintained by solar heating alone (e.g. \citet{miller2006}). As a first step, the present study focuses on the 
  steady state response of the Jovian thermosphere to different magnetospheric configurations (i.e. different solar 
  wind pressures).

\section{Theoretical Background}
\label{sec:background}

  In this section we present a summary of some basic theoretical principles that we use throughout this study. 
  We rely on work that has been conducted in previous studies by \citet{hill79, pontius95, pontius97, 
  nichols04, cowley05, cowley07}\Rev{,} and SA09. 

\subsection{Ionospheric Currents}
\label{sec:ioncur}

  The frictional drag due to ion-neutral collisions within the thermosphere causes a lag from corotation of the 
  thermosphere that can be represented by a `slippage parameter' $K$ \citep{haungHill89} 
  
    \begin{align}
	  \unit{\left(\Omega_T - \Omega_J\right)} &= \unit{-K}\unit{\left(\Omega_J - \Omega_M\right)}, 
	\end{align}
	
  \noindent\Rev{or equivalently}
  
    \begin{align}
	  \unit{\left(\Omega_T - \Omega_M\right)} &= \unit{\left( 1 - K\right)\left(\Omega_J - \Omega_M\right)}.
    \end{align}

  \noindent Here \unit{\Omega_M}, \unit{\Omega_J} are the angular velocities of the magnetosphere and Jupiter 
  itself (deep planetary angular velocity) respectively. \unit{\Omega_T} is the effective rotation \Rev{angular} velocity of 
  the thermosphere (see SA09). $K$ thus represents the \Rev{`slippage' of the neutrals from rigid corotation}. \\
  
  The coupling of the magnetosphere and ionosphere induces an electric field within the thermosphere's rest 
  frame which then causes ionospheric currents to flow. The ionospheric height-integrated Pedersen current density, 
  \unit{i_{P}}, and the total, azimuthally integrated form of this current, \unit{I_P(\theta_i)}, are 
  \Rev{(\citet{cowley07}, SA09)}
  
  \begin{align}
	\unit{ i_{P} } =& \unit{ \rho_{i} \Sigma_P \left( \Omega_T - \Omega_M\right) B_i} \label{eq:pedcur},
  \end{align}
  
  \noindent\Rev{and}
  
  \begin{align}
	\unit{ I_P(\theta_i) } =& \unit{ 2\pi \rho_{i}^{2} \Sigma_P \left( \Omega_T - \Omega_M\right) B_i} \label{eq:AIpedcur},
   \end{align}

  \noindent where \unit{\Sigma_P} is the height-integrated Pedersen conductance, \unit{B_i} is the assumed radial 
  ionospheric magnetic field, \unit{\theta_i} is the ionospheric co-latitude\Rev{,} and 
  \unit{\rho_i} is the perpendicular distance to the planet's magnetic / rotation axis.\\

  Current continuity requires that there also exists in the magnetodisc a radial current density, \unit{i_{\rho}}, 
  which can also be azimuthally integrated, represented as \unit{I_{\rho}} \Rev{(\citet{nichols04}, SA09)}. We write 

    \begin{align}
	\unit{\rho_e i_{\rho}} =& \unit{2\rho_i i_P}\label{eq:radcur},\\
	\unit{ I_{\rho} } =& \unit{ 8\pi \Sigma_P F_e \left( \Omega_T - \Omega_M\right) }, \label{eq:AIradcur}
    \end{align}
  
  \noindent where $\unit{\rho_i}{=}\unit{ R_i \sin\theta_i}$ (\unit{R_i} is the ionospheric radius), $\unit{ B_i 
  }{=}\unit{ 2 B_J }$ (\unit{B_J} is the equatorial magnetic field strength at the planet's surface) and the 
  function $\unit{F_e(\rho_e)}{=}$ $\unit{F_i(\theta_i)}$ ${=}\unit{B_J \rho_i^2}$ on a magnetic flux shell which 
  intersects the ionosphere 
  at co-latitude \unit{\theta_i}. \unit{\rho_e} is the equatorial distance from the planet centre to the field 
  lines lying in this shell. We adopt $\unit{B_J}{=}\unitSI[426 400]{nT}$ as Jupiter's dipole equatorial field 
  \citep{connerney1998}, and $\unit{R_i}{=}\unitSI[67 350]{km}$ as the radius of the polar Pedersen layer (radius of 
  the 
  ionosphere) \citep{cowley07}. Note that the auroral ionosphere is at high latitudes, where the planet's radius is 
  ${\sim}\unitSI[66 854]{km}$ at the \unitSI[1]{bar} surface. \unit{F_e} and \unit{F_i} are the equatorial and 
  ionospheric flux functions respectively (discussed further in section \ref{sec:magmodel}). The mapping between 
  \unit{\theta_i} and \unit{\rho_e} is represented by the equality $\unit{F_e(\rho_e)}{=}\unit{F_i(\theta_i)}$.\\

  Another result of current continuity with regard to the variation of the Pedersen current with latitude is the 
  creation of FACs which flow from the ionosphere to the magnetosphere. The density of these 
  currents (at the ionospheric footpoint of the relevant field line) is  

    \begin{align}
	\unit{j_{||i}(\theta_i)} =& \unit{-\frac{1}{2\pi R_i^2 \sin\theta_i}} \unit{\frac{d I_P}{d \theta_i}}, 
	\label{eq:jpar}
    \end{align}

  \noindent where \unit{j_{||i}(\theta_i)} is the \FAC density and the sign corresponds to the 
  northern hemisphere where the magnetic field points outward from the planet \citep{cowley07}.

\subsection{Effective neutral rotation velocity}
\label{sec:omega_thermo}
  
  SA09 define \unit{\Omega_T} as a weighted average of the effective rotation \Rev{angular} velocity 
  throughout the thermosphere-ionosphere. In this section we clarify what is meant by this, for the sake of 
  completeness.\\
  
  \citet{smith08saturn} showed that the equatorward current density in the ionosphere consists of two 
  contributions: (1) Pedersen current associated with the azimuthal thermospheric velocity \unit{u_{\phi}} and (2) 
  Hall current associated with the meridional thermospheric velocity \unit{u_{\theta}}. \unit{u_{\phi}} and 
  \unit{u_{\theta}} will vary, in general, with altitude $z$ in the thermosphere. We now define a local effective 
  angular velocity \unit{\omega_T} \Rev{(\citet{smith08saturn}, SA09)} as follows:
  
  \begin{align}
	\unit{ \rho_i \omega_T } &= \unit{ \rho_i \Omega_J + u_{\phi} + \frac{\sigma_H}{\sigma_P} u_{\theta} }, 
	\label{eq:omT_loc}
  \end{align}
  
  \noindent where \unit{\sigma_{P}} and \unit{\sigma_{H}} are the local \Rev{Pedersen and Hall conductivities} respectively. 
  Integrating over the height of the thermosphere-ionosphere to get the total equatorward current, we find that 
  \Rev{\unit{\Omega_T} can be defined as }

  \begin{align}
	    \unit{ \Sigma_P \Omega_T } &= \unit{ \int \sigma_P \, \omega_T dz }, \label{eq:omT_def}
  \end{align}
  
  \noindent where \unit{\Sigma_P} is the height-integrated Pedersen conductivity
  
  \begin{align}
	    \unit{ \Sigma_P } &= \unit{ \int \sigma_P dz }, \label{eq:sigp_def}
  \end{align}
  
  \noindent where $z$ is altitude. In these expressions \unit{\Omega_T} is a weighted average of the effective 
  neutral angular velocity \unit{\omega_T} throughout the thermosphere-ionosphere, which also contains 
  contributions from meridional winds.

\subsection{Energy Transfer}
\label{sec:nrgtrans}

  In this section we introduce equations which describe the energy transfer from planetary rotation to: (i) 
  magnetospheric rotation, and (ii) heating of the neutral atmosphere. According to \citet{hill2001}, the total 
  power per unit area of the ionosphere extracted from planetary rotation, \Rev{$P$} is given by 

    \begin{align}
      \unit{P} =& \unit{ \Omega_J \tau }, \label{eq:totpower}\\
      \unit{ \tau } =& \unit{ \rho_i i_{P} B_i }, \label{eq:torque}
    \end{align}

  \noindent where \unit{\tau} is the torque per unit area of the ionosphere exerted by the 
  \textbf{\emph{J\unit{\times}B}} force. The smaller component of this total used to accelerate the 
  magnetospheric plasma is

    \begin{align}
       \unit{P_M} = \unit{ \Omega_{M} \tau }. \label{eq:magpower}
    \end{align}

  \noindent The remainder of this power is dissipated in the upper atmosphere as heat and mechanical work

    \begin{align}
      \unit{P_A} = \unit{ (\Omega_{J} - \Omega_{M} ) \tau }. \label{eq:atmpower}
    \end{align}

  \noindent The power \unit{P_A} consists of two components, as shown by \citet{smith05}. One of these is Joule 
  heating, \unit{P_J}, and the other is ion drag power, \unit{P_D}, which is dependent on the sub-corotation of the 
  neutral atmosphere and is then viscously dissipated as heat\Rev{. These are given by}

    \begin{align}
      \unit{P_J} =& \unit{ (\Omega_{T} - \Omega_{M} ) \tau } \label{eq:jouleheating},
    \end{align} 
  
  \noindent\Rev{and}
  
  	\begin{align}
      \unit{P_D} =& \unit{ (\Omega_{J} - \Omega_{T} ) \tau }. \label{eq:iondrag}
    \end{align}
  
  \noindent These expressions can then be integrated over the appropriate region of the 
  ionosphere to obtain total (global hemispheric) powers.

\subsection{Magnetosphere Model}
\label{sec:magmodel}
  
  The magnetosphere model component used in this study is essentially the same as that used by SA09, 
  based on the \citet{cowley05} axisymmetric model for the entire magnetosphere and the more advanced middle 
  magnetosphere model proposed by \citet{nichols04}. The difference between the SA09 model and the one 
  used in this study is that we also use the formalism from \citet{cowley07} to calculate equatorial magnetic 
  profiles for compressed and expanded configurations of the magnetosphere. Our coupled model requires as input an 
  equatorial profile of magnetic field strength, along with the corresponding flux function (the flux function is 
  the magnetic flux per radian of azimuth integrated from the given location to infinity). For the axisymmetric, 
  poloidal field models which we employ, surfaces of constant flux function define a shell of field 
  lines with a common equatorial radial distance \unit{\rho_e} and ionospheric co-latitude \unit{\theta_i}. This 
  allows us to magnetically map the ionosphere to the equatorial plane using $\unit{F_i (\theta_i)}{=}\unit{F_e 
  (\rho_e)}$ \citep{nichols04}. The ionospheric form of the flux function is given by
  
  \begin{equation}
      \unit{F_i} = \unit{B_J \rho_{i}^{2}} = \unit{B_J R_{i}^{2} \sin^2\theta_i}.\label{ionflux}
  \end{equation}
  
  \noindent The equatorial magnetic field in the middle magnetosphere, \unit{B_{ze}}, and corresponding flux 
  function, \unit{F_{e}}, in this region are given by the equations below \citep{nichols04}

    \begin{align}
      \unit{B_{ze}(\rho_{e})} =&  \nonumber \\
	      & \unit{- B_{o} \left( \frac{R_{J}}{\rho_{e}}\right)^{3} \exp \left[ - \left( 
	      \frac{\rho_{e}}{\rho_{eo}}\right)^{5/2}\right] - A \left( \frac{R_{J}}{\rho_{e}} \right)^{m}}, 
	      \label{eqmag}\\
      \unit{F_{e}(\rho_{e})} =& \,\, \unit{F_{\infty}}  \nonumber \\
		 &  + \unit{\frac{B_{o}R_{J}^{3}}{2.5 \rho_{eo}}\Gamma \left[ -\frac{2}{5}, \left( 
		 \frac{\rho_{e}}{\rho_{eo}}\right)^{5/2} \right] + \frac{AR_{J}^{2}}{m-2} \left( 
		 \frac{R_{J}}{\rho_{e}}\right)^{m-2}}, \label{eqflux}
    \end{align}
  
  \noindent where $\unit{B_{o}}{=}\unitSI[3.335]{\times}\unitSI[10^5]{nT}$, 
  $\unit{\rho_{eo}}{=}\unitSI[14.501]{R_J}$, 
  $\unit{A}{=}\unitSI[5.4]{\times}\unitSI[10^4]{nT}$, $\unit{m}{=}\unitSI[2.71]{}$, 
  $\unit{F_{\infty}}{\approx}\unitSI[2.841]{}$ ${\times}\unitSI[10^4]{nT\,R^{2}_{J}}$, 
  and $\unit{\Gamma(a,z)}{=}\unit{\int^{\infty}_{z} t^{a-1} e^{-t} dt}$ is the incomplete gamma function. These 
  parameters represent an analytical fit to spacecraft magnetometer data \citep{connerney1981,khurkiv1993}. \Rev{ The 
  magnetic field model has a grid resolution of \unitSI[0.01]{R_J} which, when magnetically mapped to the ionosphere, 
  produces footprints of the field lines separated by angles equal to or smaller than the thermospheric model's 
  latitudinal grid spacing. This is a sufficient condition to sample realistic \FAC profiles and thermospheric flow 
  patterns within the ionospheric part of the model.} \\

  Using \Eqs{\ref{eqmag}-\ref{eqflux}} as a starting point we are able to calculate model magnetic fields and flux 
  functions corresponding to states of differing magnetospheric size. These models are valid within the range of 
  ${\sim}\unitSI[5]{R_J}$ to near the magnetopause, however in this study we employ a middle magnetosphere with 
  maximum radial distance of \unitSI[85]{R_J}. \citet{cowley07} assume that Jupiter's magnetosphere consists of two 
  components; the middle and the outer regions. They take the equatorial magnetic field strength in the outer 
  magnetosphere (beyond \unitSI[65]{R_J} for their `baseline' case) to be constant between ${\sim}\unitSI[5]{}$ and 
  ${\sim}\unitSI[15]{nT}$. Using \Eqs{\ref{eqmag}-\ref{eqflux}}, valid only within the middle magnetosphere, we 
  apply their method of compressing and expanding this region's magnetic field configuration. We then use our 
  middle magnetosphere field model to obtain solutions for plasma angular velocity \unit{\Omega_M} in this region 
  (section~\ref{sec:angular}). For the outer magnetosphere we shall use constant, assumed values of 
  \unit{\Omega_M}.\\

  Using the principles of magnetic flux conservation described \Rev{by} \citet{cowley07}, we were able to calculate 
  equatorial field profiles for Jupiter's magnetosphere for different values of solar wind dynamic pressure. To 
  compress (resp. expand) a magnetodisc (middle magnetosphere) from an initial radius \unit{R_{MMO}}, a uniform 
  southward (resp. northward) perturbation field, \unit{\Delta B_z}, is applied to our initial magnetospheric model 
  (described by equations \Eqs{\ref{eqmag}{-}\ref{eqflux}}). The formalism in \citet{cowley07} enables us to 
  calculate \Rev{\unit{\Delta B_z}} as a function of magnetodisc radius \unit{R_{MM}}. At a given \unit{R_{MM}} (`final' disc 
  radius), the flux conservation condition is
  
  \begin{align}
	\unit{ - \pi R_{MM}^{2}} \unit{\Delta B_z} = \unit{2\pi \left( F_{O}\left(R_{MM}\right) - 
	F_{O}\left(R_{MMO}\right) \right)}, \label{deltamag}
  \end{align}

  \noindent where \unit{F_{O}} is the initial profile of the flux function (given by \Eq{\ref{eqflux}}). 
  Rearranging to solve for \unit{\Delta B_z} 
  
  \begin{align}
	\unit{\Delta B_z} = \unit{ \frac{-2\Delta F}{R_{MM}^{2}} }, \label{eq:deltaMF}
  \end{align}
  
  \noindent where $\unit{\Delta B_z}{<}0$ for a southward field perturbation, and
  
  \begin{align}
	\unit{\Delta F} = \unit{F_O\left(R_{MM}\right) - F_{O}\left(R_{MMO}\right)}. \label{deltaflux}
  \end{align}
  
\begin{figure}[h!]
    \centering
    \includegraphics[width=0.99\figwidth]{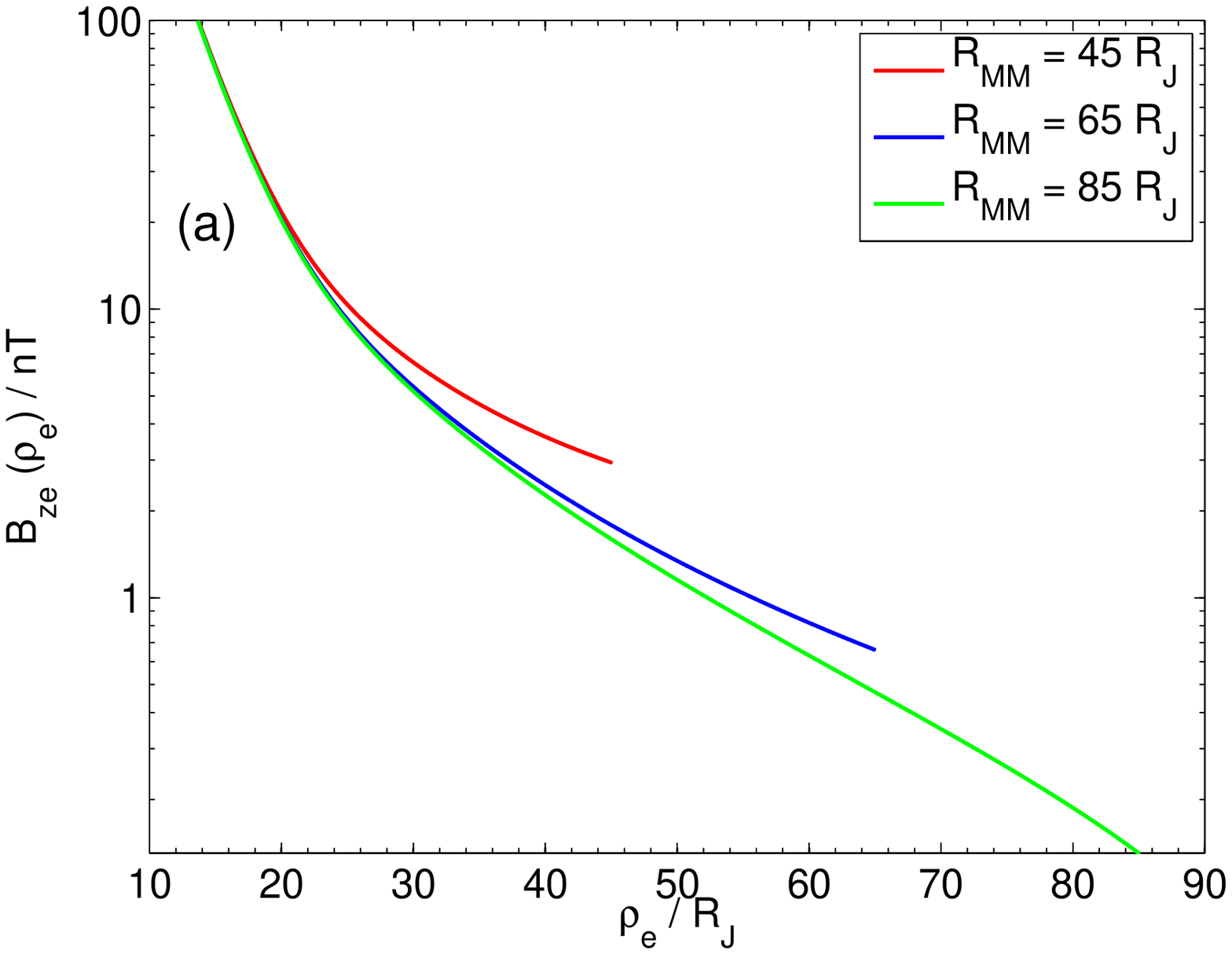}\\
    \includegraphics[width=0.99\figwidth]{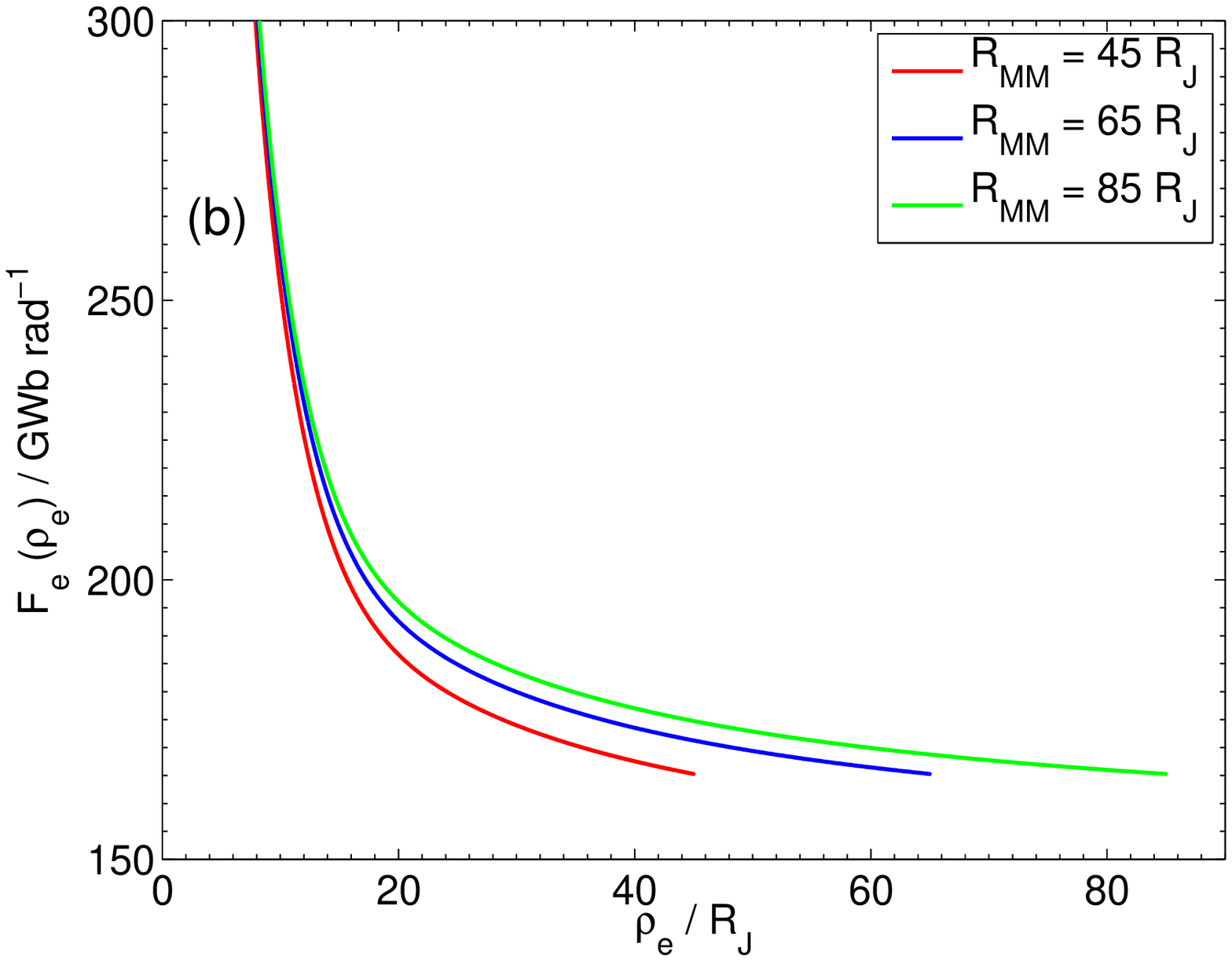}
    \caption{ \label{fig:magpro} 
    		\Rev{(a) Variation} of \Rev{the} magnetic field strength (log scale) with equatorial radial distance within the 
    		magnetodisc for the three configurations used. Case A is represented by the red solid 
    		line, whilst cases B and C are represented by the blue and green solid lines respectively. \Rev{(b)} The 
    		corresponding flux functions for the three magnetospheric cases are plotted against equatorial radial 
    		distance using the same colour code. }
      
  \end{figure}

  \noindent Using \Eqs{\ref{eqmag}-\ref{deltaflux}} we calculated equatorial magnetic field and flux function 
  profiles for three different magnetospheric configurations\Rev{, namely} a compressed system, case A with 
  $\unit{R_{MM}}{=}\unitSI[45]{R_J}$, a baseline system, case B with $\unit{R_{MM}}{=}\unitSI[65]{R_J}$ and case C, 
  an 
  expanded system with $\unit{R_{MM}}{=}\unitSI[85]{R_J}$. We choose $\unit{R_{MM0}}{=}\unitSI[65]{R_J}$ (as used by 
  \citet{cowley07}). These configurations are listed in \Table{\ref{tb:cases}} and the respective profiles are 
  shown in \Fig{\ref{fig:magpro}}.\\
  
  
  \Fig{\ref{fig:magpro}a} shows how the magnetic field strength varies with equatorial distance in the magnetodisc 
  for the three cases. The red and green lines show compressed (case A) and expanded (case C) magnetic field profiles 
  respectively. Case A with a disc radius of $\unit{R_{MM}}{=}\unitSI[45]{R_J}$ corresponds to a relatively high 
  solar wind pressure and a strong equatorial magnetic field. Case C, representing a relatively low solar wind 
  pressure has a magnetodisc radius of $\unit{R_{MM}}{=}\unitSI[85]{R_J}$ and a comparatively weak magnetic field. 
  \Fig{\ref{fig:magpro}b} shows how the corresponding flux functions vary with equatorial distance. By definition 
  the value of the flux function at $\unit{\rho_e}{=}\unit{R_{MM}}$ has the same value for all cases.

\subsection{Ionosphere Model} 
\label{sec:ionosphere}

  For simplicity, we use an auroral ionosphere model from the literature to derive a global conductivity model. 
  This conductivity model consists of both vertical and horizontal variations which we shall briefly summarise in 
  sections~\ref{sec:v_cond} and \ref{sec:h_cond}. For a more detailed description of the conductivity model \Rev{employed,} 
  the reader is referred to the following studies\Rev{, SA09, \citet{nichols04} and \citet{grodent01}}.

\subsubsection{Vertical dependence of conductivity} \label{sec:v_cond}

  The one-dimensional auroral ionosphere model by \citet{grodent01} (hereafter GG) developed for Jupiter, is used 
  to establish the altitude dependence of ionospheric conductivity. The auroral model uses a two-stream electron 
  transport code to calculate auroral electron and ion densities. There are two versions of this model\Rev{, i.e. }`diffuse' 
  and `discrete' but for our studies, both versions produce similar results. \Rev{Consequently}, the diffuse version is used, 
  as it covers a greater region of the main auroral oval and polar cap. \\

  The GG model outputs Pedersen and Hall conductivity profiles for a specific thermal structure. However our 
  thermosphere model has a variable thermal structure which is a function of latitude. In order to maintain 
  realistic height-integrated conductivities in the model, \Rev{at each pressure level we calculate} the conductivity 
  per unit mass as follows (SA09) 

  \begin{align}
	\unit{s_i} &= \unit{\frac{\sigma_i}{\rho}}, \label{eq:condpmass}
  \end{align}

  \noindent where $\unit{i}{=}\unit{P\mbox{ or }H}$ representing Pedersen or Hall, \unit{\sigma} is the 
  conductivity and \unit{\rho} is the neutral mass density. Adjacent pressure levels enclose constant masses of 
  thermospheric gas (hydrostatic equilibrium assumption). Therefore, the height-integrated Pedersen 
  (\unit{\Sigma_P}) and Hall (\unit{\Sigma_H}) conductivities depend solely on the profiles of \unit{s_i} with 
  respect to pressure and not thermal structure. This leads to a Pedersen conducting layer, which we define as the 
  region with conductivity greater than \unitSI[10]{\%} of the Pedersen conductivity at the auroral ionisation peak\Rev{,} 
  located at pressures of ${\sim}\unitSI[0.8{-}0.04]{\mu bar}$ or at altitudes of 
  ${\sim}\unitSI[350{-}600]{km}$ above the \unitSI[1]{bar} level.

\subsubsection{Horizontal conductivity model} \label{sec:h_cond}

  The height-integrated conductivity in the inner and middle magnetospheres is dependent on 
  the \FAC density according to the following equations \citep{nichols04}

  \begin{align}
      \unit{\Sigma_P (j_{|| i})} &= \unit{ \Sigma_{PO} + \Sigma_{Pj}(j_{|| i})}, \label{eq:pedcond_jpar}
  \end{align}
  
 \noindent\Rev{where}
  
  \begin{align}
      \unit{\Sigma_{Pj}(j_{|| i})} &= \unit{0.16 j_{|| i}} + \nonumber \\
	    & \unit{\left\{2.45\left[ \frac{(j_{|| i}/0.075)^2}{1 + (j_{|| i}/0.075)^2} \right]  
	      \times \frac{1}{[1 + exp(-(j_{|| i} - 0.22)/0.12)]}\right\} }, \label{eq:ped_jpar_full}
  \end{align}

  \noindent where $\unit{\Sigma_{PO}}{=}\unitSI[0.0275]{mho}$ \citep{nichols04} is the background conductivity 
  due to solar photoionisation\Rev{,} and \unit{\Sigma_{Pj} (j_{|| i})} in \Rev{\unitSI{mho}} is an auroral 
  enhancement due to the \FAC density \unit{j_{|| i}} \Rev{in \unitSI{\mu A\;m^{-2}}}. \Rev{ The dependence of 
  \unit{\Sigma_H} on \unit{j_{|| i}} is calculated from \Eq{\ref{eq:ped_jpar_full}} using standard formulae 
  (e.g. \citet[p.201]{kivbook1995}). The total conductivity in the ionisation region is dominated by \unit{\Sigma_P} due to 
  the small values of \unit{\Sigma_H}.}\\

  In the outer magnetosphere and polar cap regions\Rev{,} conductivity enhancement is likely to be present since UV and IR 
  auroral emissions are detected in these regions. \citet{cowley05} set $\unit{\Sigma_P}{=}\unitSI[0.2]{mho}$ 
  (effective Pedersen conductivity) in these regions in accordance with the theory of \citet{isbell1984}. To allow 
  for comparison, we employ the same conductivity value in these regions.

\subsection{Coupled Model} 
\label{sec:coupledmodel}

  We couple our magnetosphere model with a global numerical model of the thermosphere and a global conductivity 
  model of the ionosphere as described in SA09 and section~\ref{sec:ionosphere}. The resolution of the model grid 
  is \unitSI[0.2]{^{\circ}} in latitude, and \unitSI[0.4]{} pressure scale heights in the vertical direction. That 
  is, we use pressure as a vertical coordinate, with the lower boundary at \unitSI[2]{\mu bar} (\unitSI[300]{km} 
  above the \unitSI[1]{bar} level) and the upper boundary 
  at \unitSI[0.02]{nbar}. Altitudes are updated in the model assuming hydrostatic equilibrium. For simplicity all 
  models are axisymmetric with respect to the planet's axis of rotation. This assumption does not 
  greatly influence the basic physics underlying the conclusions of this study (SA09). 
  It is important to emphasise that the assumption of axisymmetry implies zero azimuthal gradients in the model 
  variables. This allows us to represent model outputs in two-dimensions (latitude and altitude) while still using 
  the three-dimensional Navier-Stokes equations. \\

  Section 6 in SA09 describes the method of coupling the magnetosphere, thermosphere and ionosphere 
  models. We employ essentially the same \, method, with a few minor changes. The same value of the Jovian 
  radius, \unit{R_J}${=}$\unitSI[71 492]{km} is used for both our flux function calculations and atmospheric 
  modelling. The coupled model in SA09 ran for 200 Jovian rotations to reach steady-state. Comparisons 
  of height, temperature and azimuthal velocity in the inertial frame data for case B were made for run-times of 
  200 and 50 rotations. Calculations show that between both run-times there was a maximum relative difference of 
  ${\sim}\unitSI[0.4]{\%}$, ${\sim}\unitSI[0.8]{\%}$ and ${\sim}\unitSI[1.2]{\%}$ for height, temperature and 
  azimuthal 
  velocity respectively. This difference causes no significant change in any other parameters obtained from the 
  model and running the model for 50 rotations would save considerable CPU time. Thus for the purposes of this 
  study, running the model for 50 rotations was considered sufficient to reach steady state.

\subsection{Solving the coupled equations of thermospheric and magnetospheric momentum} 
\label{sec:solving}

  Studies such as \citet{hill79} and \citet{pontius97} have shown that for the middle magnetosphere to be 
  in a steady state, the radial gradient of the outward angular momentum flux of iogenic plasma must be equal in 
  magnitude to the torque per unit radial distance on that plasma. The plasma model that describes the middle 
  magnetosphere is based on four equations

  \begin{align}
      \unit{ \frac{1}{\rho_{e}} \deriv[]{\rho_{e}} \left( \rho_{e}^{2}\Omega_M\right) } &= \unit{ 
      \frac{8\pi\Sigma_{P}F_e|B_{ze}|}{\dot{M}} \left( \Omega_T - \Omega_M\right)}, \label{eqhillpon}\\
      \unit{j_{||i}} &= \unit{ \frac{4B_J}{\rho_e|B_{ze}|} \deriv[]{\rho_e} \left[ \Sigma_P F_e \left( \Omega_T - 
      \Omega_M\right)\right] }, \label{eqjpar}\\
      \unit{\Sigma_P} &= \unit{\Sigma_P (j_{||i})}, \label{eqsigjpar}\\
      \unit{\Omega_T} &= \unit{\Omega_T(\Omega_M , \Sigma_P)}, \label{eqomt2omt}
  \end{align}

  \noindent where $\unit{\dot{M}}{=}\unitSI[1000]{kg\;s^{-1}}$ is the assumed mass outflow rate from the Io torus and 
  \unit{j_{||i}} is the upward \FAC density in the ionosphere.\\

  These equations describe the inter-dependence of magnetospheric angular momentum per unit mass 
  (\unit{\rho_{e}^{2}\Omega_M}), \FAC density (\unit{j_{||i}}) and Pedersen conductance 
  (\unit{\Sigma_P}). \Eq{\ref{eqomt2omt}} represents the output from the thermospheric model component, which is 
  forced by magnetospheric inputs of \unit{\Omega_M(\rho_e)}. \Eq{\ref{eqhillpon}} is the Hill-Pontius equation 
  \citep{hill79, pontius97} with a modification by SA09 to include effects of neutral thermosphere flow, 
  represented by \unit{\Omega_T}. This equation balances torques caused by the outward diffusion of the disc plasma 
  and the \textbf{\emph{J\unit{\times}B}} force associated with the  magnetosphere-ionosphere coupling currents. 
  \Eq{\ref{eqjpar}} is used to calculate the \FAC in the  ionosphere. An increase in field-aligned 
  current should have an effect on angular velocities \unit{\Omega_M (\rho_e)}, through enhancement of the 
  ionospheric conductivities. We account for this in \Eq{\ref{eqsigjpar}}, representing \Eqs{\ref{eq:pedcond_jpar} 
  and \ref{eq:ped_jpar_full}}, which describes how enhancements in \unit{j_{||i}} also affect the Pedersen 
  conductance \unit{\Sigma_P} (higher flux of precipitating auroral electrons increases the production rate of 
  ionospheric plasma). \\
  
  Our method for solving these equations is the same as that in SA09 and \citet{nichols04} (who originated this 
  model). This is essentially a 
  shooting method which varies the value of \unit{\Omega_M} at the outer edge of the disc until the solution,  
  integrated inwards from this location, smoothly joins an appropriate `inner disc' analytical solution at 
  \unitSI[12]{R_J}. We set the azimuthally integrated radial current at the outer edge of the disc to a value of 
  \unitSI[100]{MA} as our outer boundary condition (following \citet{nichols04}), whilst we have near-rigid 
  corotation of plasma as an inner 
  boundary condition. We need, however, to ensure that the height-integrated Pedersen conductivities at the 
  poleward ionospheric boundary of the magnetodisc field line region, \unit{\Sigma_P}(disc), and at the equatorward 
  boundary of the outer magnetosphere region, \unit{\Sigma_P}(outer), join smoothly together to avoid 
  discontinuities at this interface. This is particularly important for large compressions such as that of case A. 
  A Gaussian function was used to extrapolate \unit{\Sigma_P} from the magnetodisc into the outer magnetosphere 
  region. We ensured the Gaussian function would terminate with a polar value equal to the chosen background 
  \unit{\Sigma_P} in the outer magnetosphere, and that this transition would occur with a small latitudinal scale 
  (\unitSI[0.2]{^{\circ}}). The amplitude and centre of the Gaussian function were calculated using the gradient of 
  \unit{d\Sigma_P / d\theta} at the poleward edge of the disc region. We further discuss the resulting profiles of 
  \unit{\Sigma_P} in section~\ref{sec:conductivities}.

\section{Results and Discussion} 
\label{sec:results}
  In this section we present the results obtained from our modelling. We firstly discuss results concerning angular 
  velocities, conductivities and currents. Then we proceed to discuss the thermospheric flows and energies.

\subsection{Angular velocities, conductivities and currents} 
\label{sec:angcur}

\subsubsection{Angular velocities} 
\label{sec:angular}

  Combining models from \citet{nichols04} and \citet{cowley05} of the middle and outer magnetospheres, one 
  can essentially divide the entire magnetosphere into four regions, labelled I-IV. The Dungey-type interaction of 
  the polar open field lines with the solar wind takes place in 
  region I. The closed field lines of the outer magnetosphere are involved in Dungey and Vasyliunas cycles 
  (associated with mass loss from the disc) in region II. Regions III \Rev{(shaded region in figures)} and IV represent 
  the middle magnetosphere (magnetodisc) and 
  the corotating inner magnetosphere respectively (see \Fig{\ref{fig:omegaplots}}). As stated in section 
  \ref{sec:omega_thermo}, \unit{\Omega_T} is a weighted average of the effective angular velocity throughout the 
  thermosphere-ionosphere, computed over all altitudes at each co-latitude \unit{\theta_i}. \unit{\Omega_M} in 
  region I has a constant value of ${\sim}\unitSI[0.1]{\Omega_J}$ \citep{isbell1984}. Region II also has a fixed 
  value of \unit{\Omega_M} 
  that depends on magnetospheric size, in accordance with observations \citep{cowley07}. The profiles of 
  \unit{\Omega_M} in regions I, II and III are joined smoothly across their boundaries with the use of hyperbolic 
  tangent functions. The plasma angular velocity profiles for regions III and IV are calculated using 
  \Eqs{\ref{eqhillpon} - \ref{eqomt2omt}} by the model. \\


  \Fig{\ref{fig:omegaplots}} shows how the thermospheric (solid lines) and magnetospheric (dashed lines) 
  angular velocities vary in Jupiter's high latitude region for our three cases. We also show the region 
  boundaries used in our model and the magnetically mapped location of Io in the ionosphere. Case B, our 
  `baseline' is shown in blue. At low latitudes, rigid corotation with Jupiter's deep atmosphere is 
  maintained. At the higher latitudes 
  ($>\unitSI[60]{^{\circ}}$) the magnetosphere (represented by $\unit{\Omega_M}$) sub-corotates to a greater 
  degree than the thermosphere (expressed by $\unit{\Omega_T}$). The shape of these $\unit{\Omega_M}$ 
  and $\unit{\Omega_T}$ profiles are similar to those obtained in the studies of SA09. \unit{\Omega_M} and 
  \unit{\Omega_T}  profiles for case C, our expanded case, are represented by green lines. These profiles 
  resemble those of case B but they possess slightly smaller angular velocities in region II. For case A, 
  $\unit{\Omega_M}$ and $\unit{\Omega_T}$ are shown by the red lines. Both $\unit{\Omega_M}$ and 
  $\unit{\Omega_T}$ indicate sub-corotation to a lesser extent than the respective profiles from cases B and 
  C, in agreement with the study of \citet{cowley07} who modelled \unit{\Omega_M}, assuming simplified 
  profiles for \unit{\Omega_T} (where $\unit{K}{=}\unitSI[0.5]{}$). We thus show that the thermosphere and 
  magnetosphere for compressed configurations corotate to a greater degree than in the case of expanded 
  configurations. Our plotted profiles quantify this result for both \unit{\Omega_M} and \unit{\Omega_T}.
   
\subsubsection{Conductivities and Currents} 
\label{sec:conductivities}
\begin{figure}
      \centering
      \includegraphics[width= 0.99\figwidth]{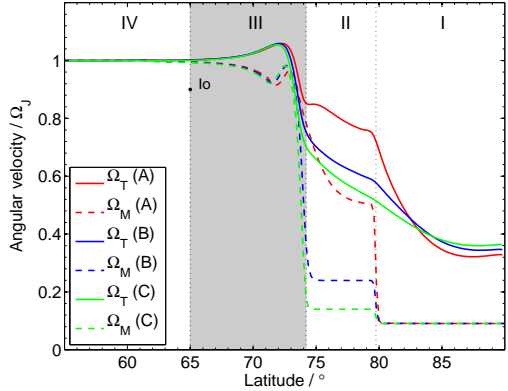}
      \caption{ Thermospheric and magnetospheric angular velocities for cases A-C are plotted in the high 		
      latitude region and are represented by solid and dashed lines respectively. Red lines represent case
      A, blue case B and green case C. The \Rev{black dot} labelled `Io' indicates the magnetically 
      mapped position of the moon Io's orbit \Rev{in} the ionosphere. The magnetospheric regions \Rev{(region III is shaded)} 
      considered in this study are labelled and separated by the dotted black lines.}

      \label{fig:omegaplots}
  \end{figure}

  Previous studies of the effect of solar wind-induced compressions and expansions of Jupiter's 
  magnetosphere have shown that magnetospheric compressions reduce ionospheric and parallel currents (in the steady 
  state). Expansions on the other hand, have the opposite effect due to the increased transport of angular momentum 
  to the magnetosphere \citep{southkiv2001, cowbun2003b, cowley07}. Our profiles in \Fig{\ref{fig:omegaplots}} 
  confirm and quantify the expected \Rev{angular velocity profiles of both the thermosphere and magnetospheric plasma in 
  the steady state, when the rate of addition of angular momentum to the plasma (at a given radial distance), due to the 
  magnetosphere-ionosphere currents, exactly balances the rate of removal due to the radial plasma outflow.} We 
  consider the solutions for \unit{\Omega_M} and \unit{\Omega_T} in more detail in \Rev{section~\ref{sec:angular}}. The 
  weaker average magnetic field for the expanded cases, combined with the finite ionospheric conductivity, leads to lower 
  \unit{\Omega_M} values, despite increased rates of angular momentum transport in the system. In this section we 
  present our quantitative findings regarding ionospheric conductivities and currents for the different 
  magnetospheric configurations of our coupled system.\\

\begin{figure}
    \centering
    \includegraphics[width=0.99\figwidth]{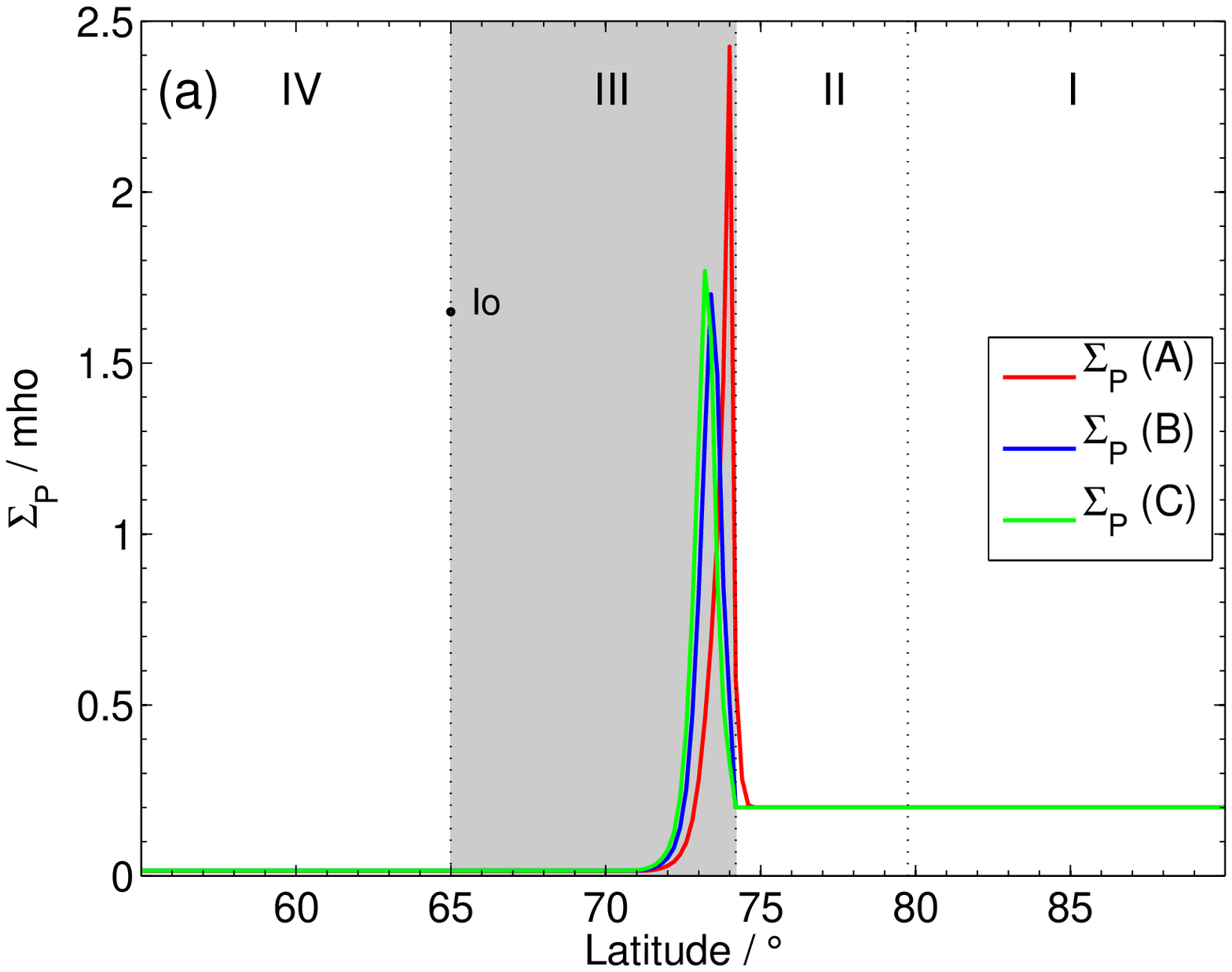}\\
    \includegraphics[width=0.99\figwidth]{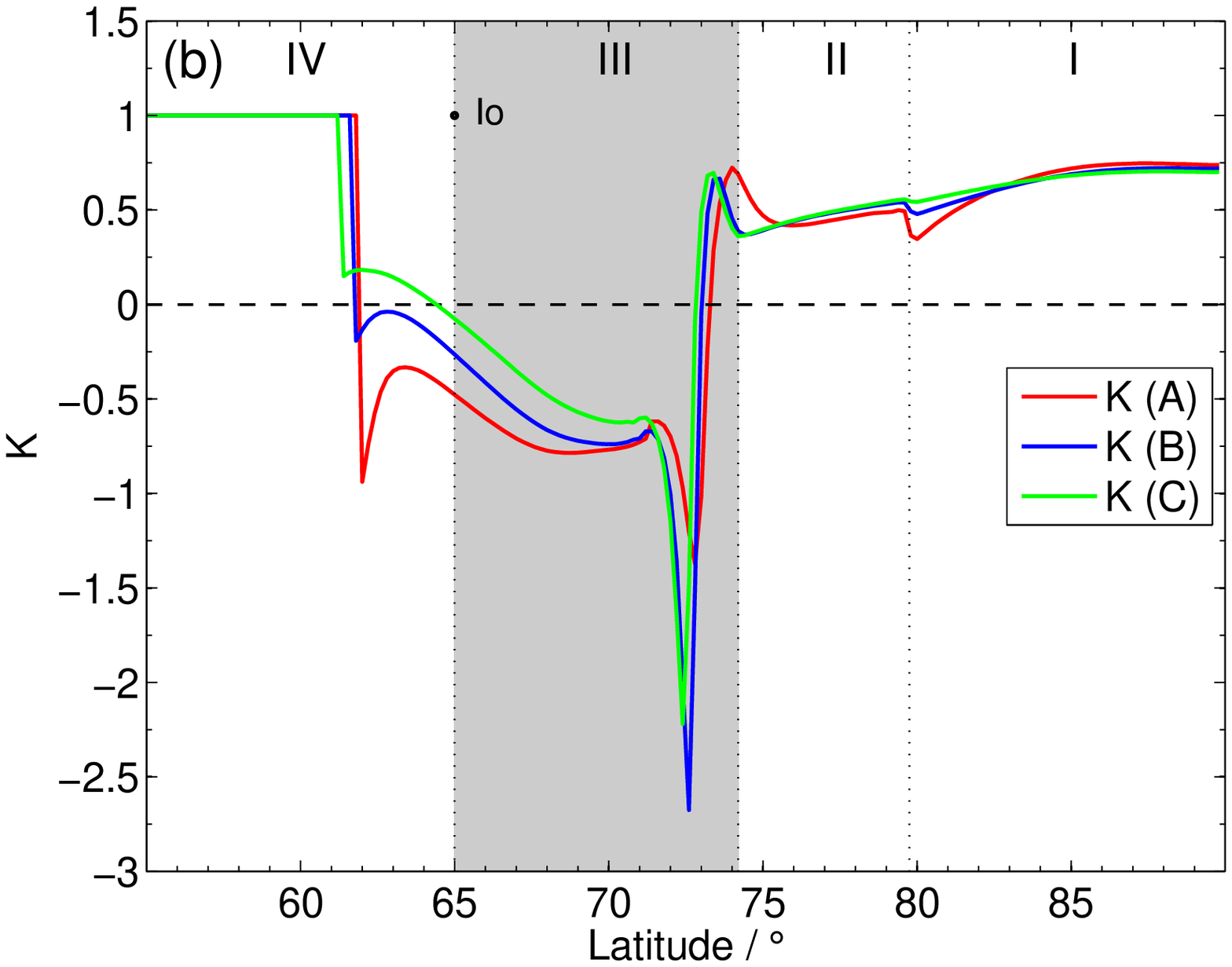}
    \caption{ \Rev{(a)} Height-integrated Pedersen conductivities for cases A-C plotted \Rev{versus} latitude. Case\Rev{s} A-C 
    are represented by red, blue and green solid lines respectively. The magnetically mapped location of Io \Rev{in} the 
    ionosphere is labelled and marked by the black \Rev{dot}. \Rev{Magnetospheric} regions \Rev{(region III is shaded)} are 
    labelled and separated by dotted black lines. \Rev{(b) `Slippage'} parameter $K$ plotted \Rev{versus} latitude for 
    cases A-C. The colour code for cases A-C remains the same as (a).}	
      \label{fig:pedcond}
  \end{figure}

  
  The variation of height-integrated true Pedersen conductivity \unit{\Sigma_P} for our three magnetospheric cases 
  is shown in \Fig{\ref{fig:pedcond}a}, where cases A-C are represented by red, blue and green lines respectively. 
  The magnetically mapped location of Io \Rev{in} the ionosphere is shown by the black dot and the 
  magnetospheric regions used in this study are marked and separated by black 
  dotted lines. All three cases have peaks just equatorward of the region III / II boundary --- characteristic 
  features of the \unit{\Omega_M} solutions (\Eqs{\ref{eqhillpon}{-}\ref{eqomt2omt}}) --- and then fall to the 
  assumed conductivity value in regions II and I. Cases B and C have similar profiles and peak values \Rev{close} to those 
  calculated in SA09\Rev{,} whilst case A has a peak that is significantly higher than both of these cases. The profile 
  for case A resembles that from \citet{nichols04} for the near-rigid corotation approximation where $ \left( 1 - 
  \Omega_M / \Omega_J\right) << 1$, which are conditions met by case A in regions IV and III. Another feature that 
  distinguishes case A is that the peak conductivity is shifted poleward slightly compared to cases B and C\Rev{. This} 
  is partly due to the model method which connects the Pedersen conductivity in region III with the fixed value in 
  region II for case A (see section~\ref{sec:solving}). The poleward shift is also due to the higher 
  \unit{\Sigma_P} required in case A in order to achieve the prescribed value of radial current at the outer edge 
  of the magnetodisc (poleward boundary of region III) (see section~\ref{sec:solving}).\\

  \Fig{\ref{fig:pedcond}b} shows how the slippage parameter \unit{K} varies with latitude for our three 
  magnetospheric cases. The profiles for \unit{K} indicate the ratio between thermospheric and 
  magnetospheric angular velocities with respect to Jupiter's planetary rotation velocity 
  ($\unit{K}{=}\unit{\left( \Omega_J - \Omega_T\right) / \left( \Omega_J - \Omega_M\right)}$). Positive values for 
  \unit{K} represent situations when both the thermosphere and magnetosphere are 
  sub-corotating or super-corotating with respect to the planet, as seen in region IV, II and I. Negative 
  \unit{K} values represent situations where the thermosphere and magnetosphere are 
  undergoing opposing motions i.e one is super-corotating whilst the other is sub-corotating. This is seen
  just equatorward of Io's magnetic footprint on the ionosphere and for the latitudinal majority of region III. 
  This distinction is important because the last half degree of latitude in region III maps to the largest 
  part of the equatorial magnetosphere.\\

\begin{figure}
    \centering
    \includegraphics[width=0.99\figwidth]{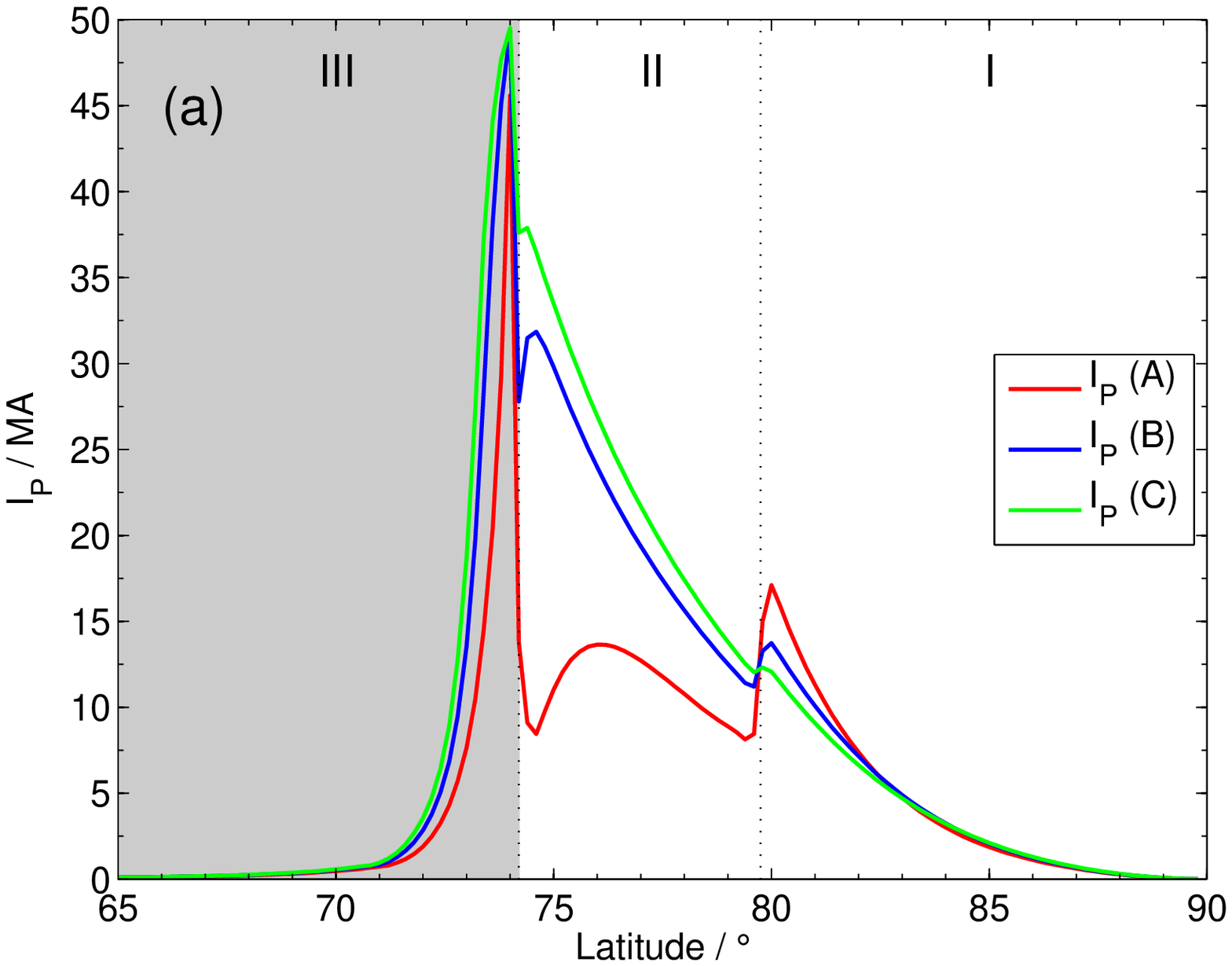}\\
    \includegraphics[width=0.99\figwidth]{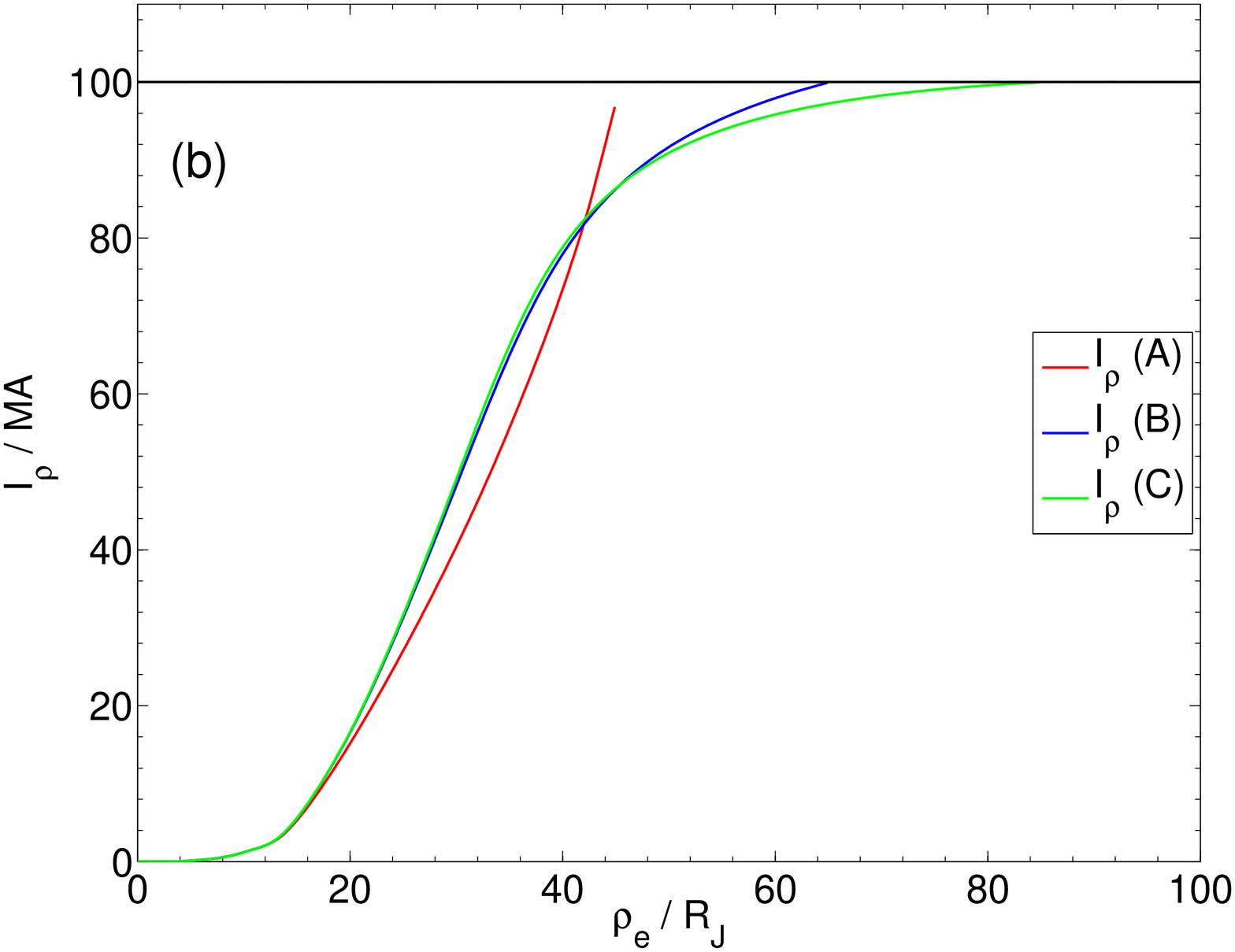}
    \caption{ \Rev{(a)} Azimuthally-integrated Pedersen current \Rev{shown} as a function of latitude for cases A-C. 
    Case A is represented by the solid red line, case B by the blue line and case C by the green line. The 
    magnetospheric regions \Rev{(region III is shaded)} are also marked and separated by the dotted black lines. 
    \Rev{(b) Azimuthally}-integrated 
    radial current plotted against equatorial radial distance from Jupiter for cases A-C. The 
    colour code is the same as in (a).}
		
      \label{fig:currents}
  \end{figure}

  \Fig{\ref{fig:currents}a} shows the corresponding variation of azimuthally\Rev{-}integrated Pedersen current with 
  latitude. The colour code is the same as that in \Fig{\ref{fig:omegaplots}}. Profiles for cases B and C follow a 
  similar trend to \Rev{the} steady-state Pedersen current profiles in \citet{cowley07}\Rev{,} whilst \Rev{the profile for} 
  case A is different within regions III and II, due to conditions comparatively nearer to corotation. \\

  \Fig{\ref{fig:currents}b} shows the azimuthally integrated radial currents through the magnetospheric equator for 
  all three cases in regions IV and III and how they vary with radial distance. The radial currents for cases B and 
  C show a `s-curve' structure which is consistent with previous studies such as SA09. Case A however, shows a more 
  linear relation between the equatorial radial distance and azimuthally integrated radial current which is not 
  seen in the more expanded case of SA09 but is consistent with the near-rigid corotation approximation conditions 
  presented in \citet{nichols04}. As previously noted, this near-rigid corotation condition applies to case A 
  throughout regions IV and III. We also note that, as mentioned in section~\ref{sec:solving}, our outer boundary 
  condition is that the radial current value at the region III / II boundary is \unitSI[100]{MA}. The case A curve in 
  \Fig{\ref{fig:currents}b} does not quite reach this value due to the joining of the Pedersen conductivity across 
  regions III and II (see section~\ref{sec:solving}). \Rev{A hyperbolic tangent function is used to smoothly join the 
  Perdesen conductivity across regions II and III, using information from a few points either side of this boundary. 
  This leads to a smoothing of the disc solution near its outer edge, leading to a slightly different value of the 
  azimuthally integrated radial current at this location.} This curve does demonstrate however, that 
  \unit{I_P} and \unit{I_{\rho}} in case A have to increase very rapidly in the outer magnetodisc in order to satisfy 
  the boundary condition. Since there is no a priori reason why \unit{I_{\rho \infty}} should be independent of 
  magnetosphere size, we will also investigate, later, the effect of varying the boundary condition upon the 
  resulting profiles of current and angular velocity (section~\ref{sec:pathogen}).\\

\begin{figure}
      \centering
      \includegraphics[width= 0.99\figwidth]{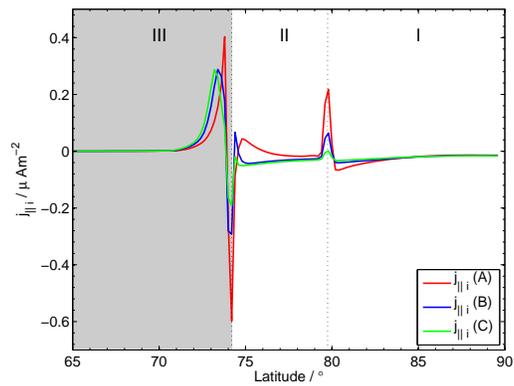}
      \caption{ \FAC densities in the high latitude region for cases A-C. Red solid lines 
      represent FACs for case A whilst blue and green solid lines represent FACs for cases B and C respectively. 
      The magnetospheric regions \Rev{(region III is shaded)} are labelled and separated by black dotted lines. }
      \label{fig:jpar}
  \end{figure}

  \FAC densities are plotted against latitude in \Fig{\ref{fig:jpar}}. For all three cases, 
  FAC densities have three positive peaks, the first two lying on either side of the region III / II boundary and 
  the third lying on the region II / I boundary. Positive peaks correspond to upward directed FACs that produce 
  aurorae. At the boundary between region III and II, the negative peaks indicate strong downward-directed FACs 
  whose magnitude is dependent on the equatorial radius of region III (\unit{R_{MM}}). The main auroral oval is 
  represented by the peak at ${\sim}\unitSI[73]{^{\circ}}$ latitude. Our model suggests that there would also be 
  weaker more distributed aurorae poleward of the main oval, represented by the second and third peaks at 
  ${\sim}\unitSI[75]{^{\circ}}$ and ${\sim}\unitSI[80]{^{\circ}}$ respectively. A relatively dark region would arise 
  from the trough at ${\sim}\unitSI[74]{^{\circ}}$ latitude, creating `dark rings'. The latter feature is also 
  obtained in previous studies by \citet{cowley05,cowley07} but at present, we lack the observations required to 
  constrain the value of \unit{j_{||i}} downward. The strong downward FACs at ${\sim}\unitSI[74]{^{\circ}}$ are due 
  to the significant changes in Pedersen current on crossing the boundary between regions II and III, which in turn 
  is due to the changes in magnetospheric and thermospheric angular velocities. The Pedersen conductivity in the 
  model also changes significantly across this boundary, which also contributes to a large magnitude for 
  \unit{j_{||i}}. The strongest downward FACs in our calculations are even less constrained by observations, but 
  they also occur in the modelling of \citet{tao09} who also used a coupled magnetosphere-\Rev{thermosphere} approach. \\

  Our calculations shown in \Figs{\ref{fig:omegaplots}-\ref{fig:currents}} all support the expected trends 
  described in \citet{southkiv2001}. The angular velocity profiles (\Fig{\ref{fig:omegaplots}}) for 
  both the thermosphere and magnetodisc show that there is a greater degree of sub-corotation for more expanded 
  magnetospheres, corresponding to lower solar wind dynamic pressures. This is due to the thermosphere being able
  to transfer momentum to a compressed magnetosphere (stronger field) with greater efficiency than a larger, 
  expanded one. The Pedersen conductivities (\Fig{\ref{fig:pedcond}a}), \FAC densities 
  (\Fig{\ref{fig:jpar}}) and azimuthally-integrated Pedersen and radial currents (\Figs{\ref{fig:currents}a-b}) all 
  show an increase \Rev{in region III (shaded)} for expanded magnetospheres. \Rev{ In this region, the integrated auroral 
  \FAC for case A is ${\sim}\unitSI[50{-}60]{\%}$ of that cases B and C suggesting that auroral emission would be greater 
  for an expanded magnetosphere than a compressed one.} Our currents naturally have similar values to those obtained in 
  SA09. They also show similar trends and profiles to studies such as \Rev{those of \citet{cowley05,cowley07} and 
  \citet{tao09}}. Our study is an extension of these works in the sense that we use an atmospheric circulation model 
  coupled to three distinct magnetospheric configurations.

\subsection{Thermospheric flows and energies}  
\label{sec:flownrg}


  \Rev{\Fig{\ref{fig:momentum}} shows momentum balances for our compressed and expanded configurations in both the 
  low and high altitude regions.} \Fig{\ref{fig:thermosphere}} shows the thermospheric flows, temperature distributions 
  and power dissipated per unit area for all three model configurations. Results for each case are displayed in the 
  columns of the figure. 

\subsubsection{Thermospheric flows}
\label{sec:flows}

\begin{figure}
    \centering
    \includegraphics[width=0.99\figwidth]{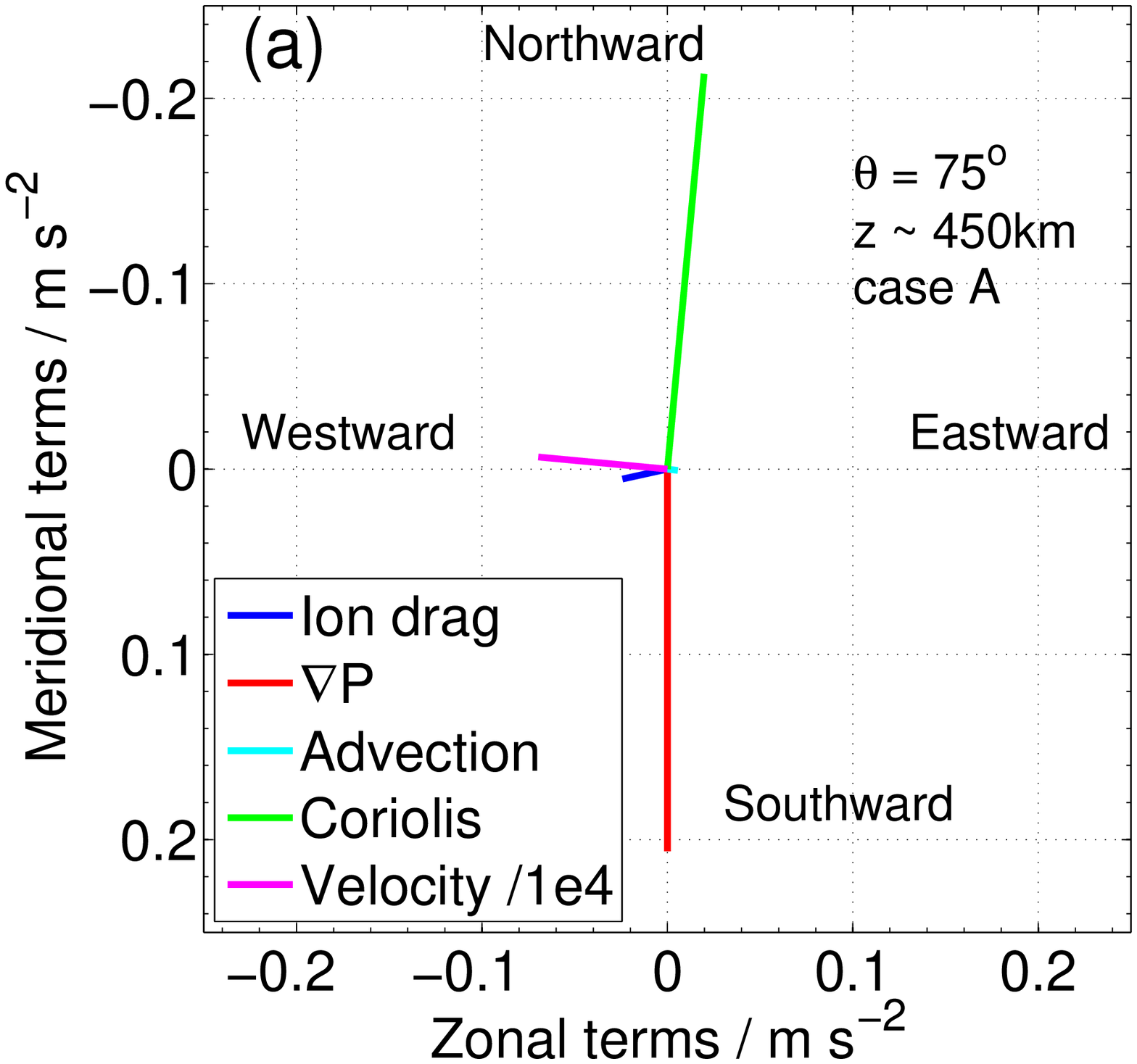}
    \includegraphics[width=0.99\figwidth]{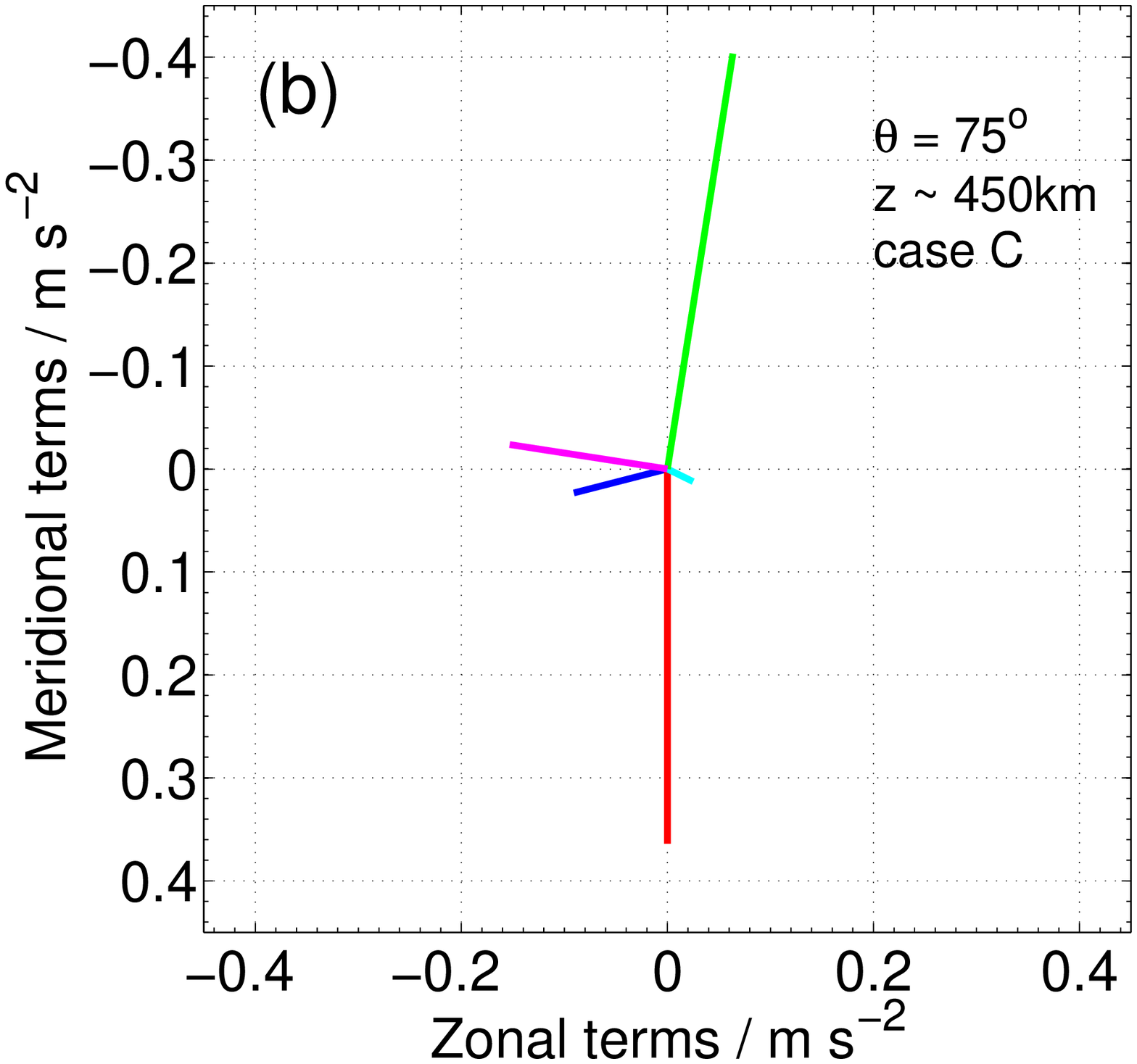}\\
    
    \includegraphics[width=0.99\figwidth]{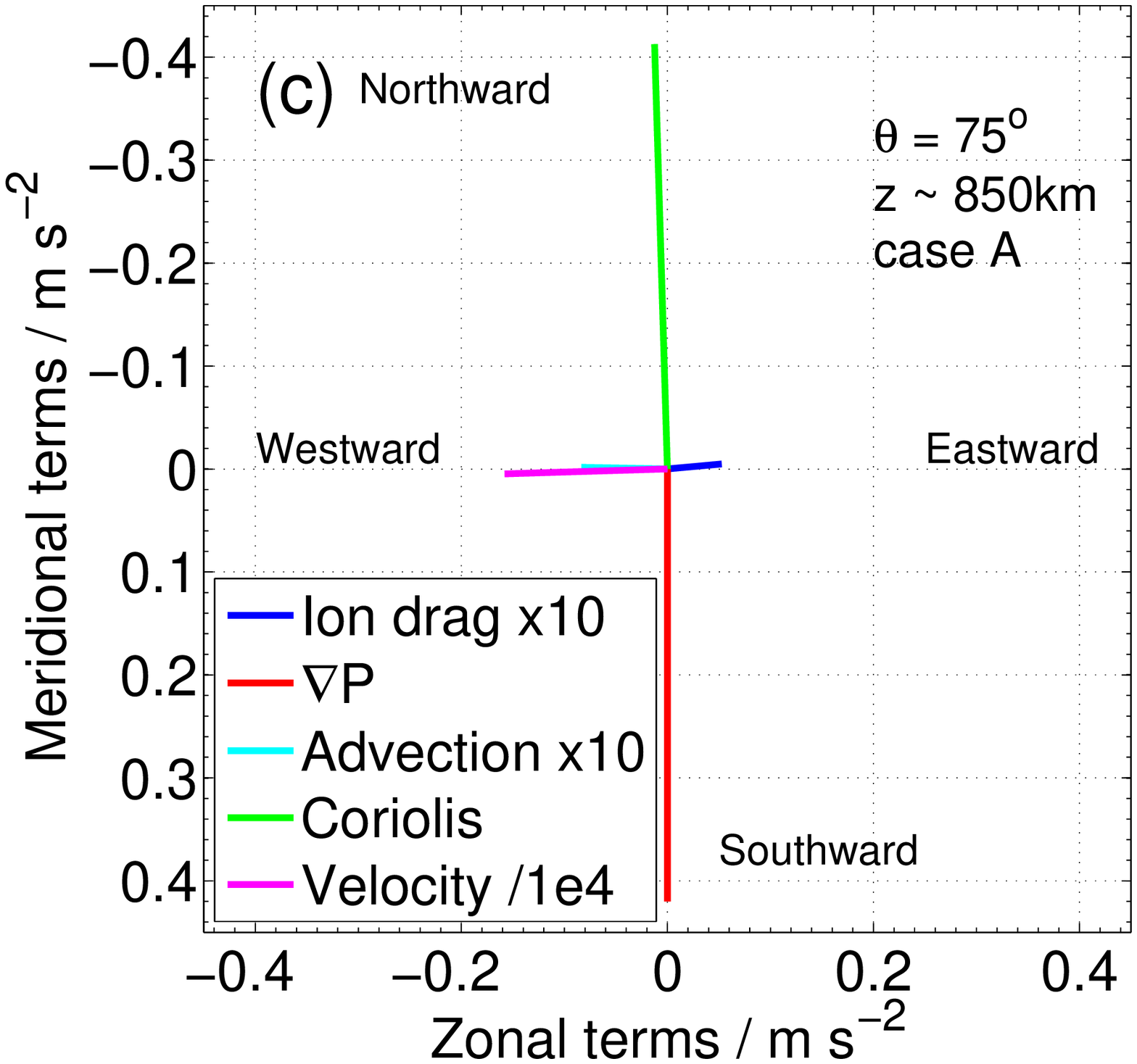}
    \includegraphics[width=0.99\figwidth]{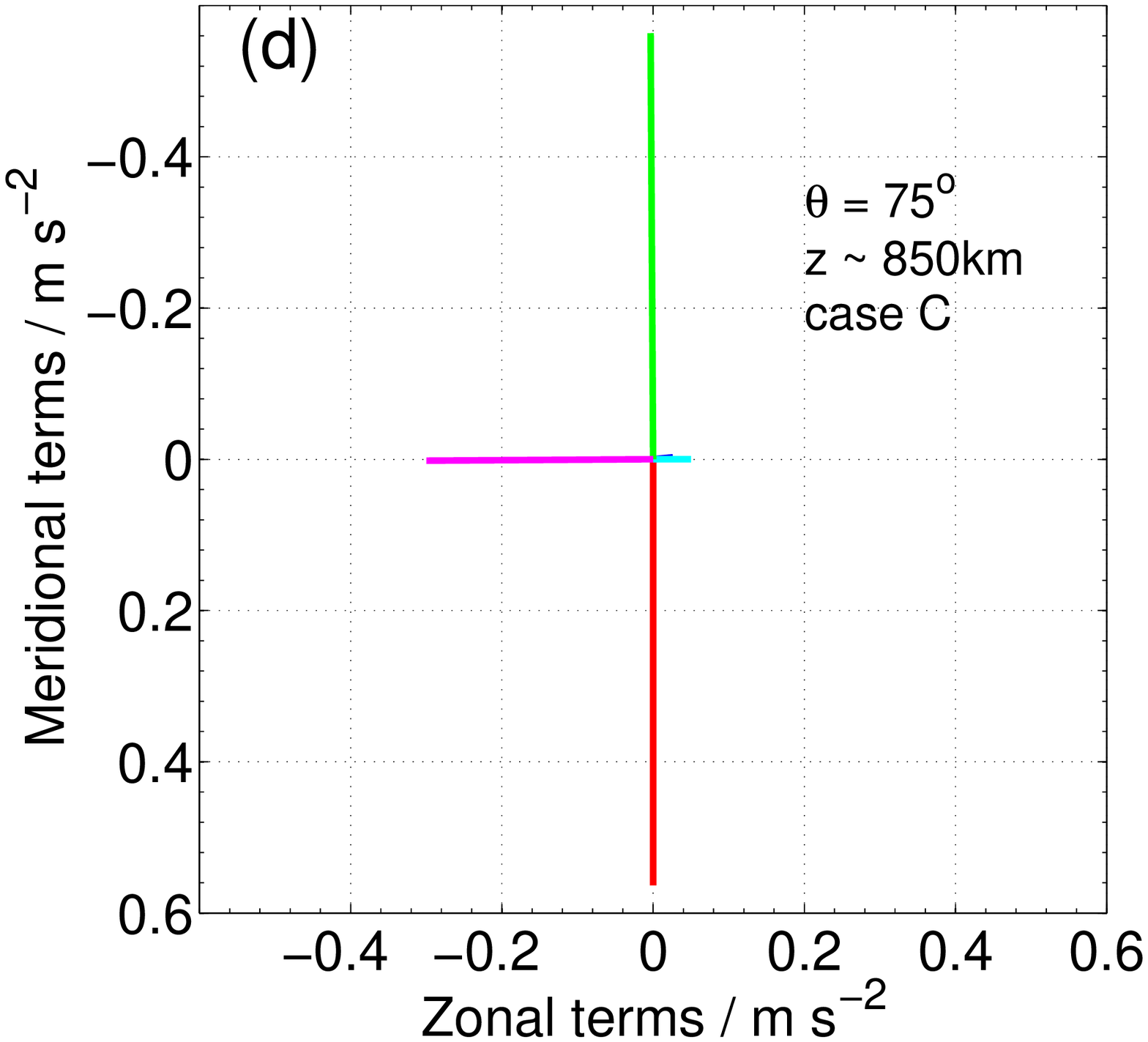}\\
    
    \caption{ Force balance diagrams for cases A (left column) and C (right column) at ionospheric co-latitude of 
    		\unitSI[75]{^{\circ}}. (a)-(b) show meridional and zonal force balance in the low altitude region whilst 
    		(c)-(d) show meridional and zonal force balance in the high altitude region. Ion drag forces are 
    		represented by blue lines, fictitious (Coriolis) forces by green lines, pressure gradient by red \Rev{and 
    		advection by the cyan line}. The velocity vector is also plotted and is 
    		represented by the magenta lines. Note that the magnitude of velocity components have been divided 
    		by a factor of $\unitSI[1{\times}10]{^4}$ to fit the plotted scale and that in (c)-(d) the components of 
    		ion drag and advection have been multiplied by a factor of \unitSI{10} to increase visibility. }
		
      \label{fig:momentum}
  \end{figure}

\begin{figure*}
    \centering 
    \includegraphics[width=0.66\figwidth]{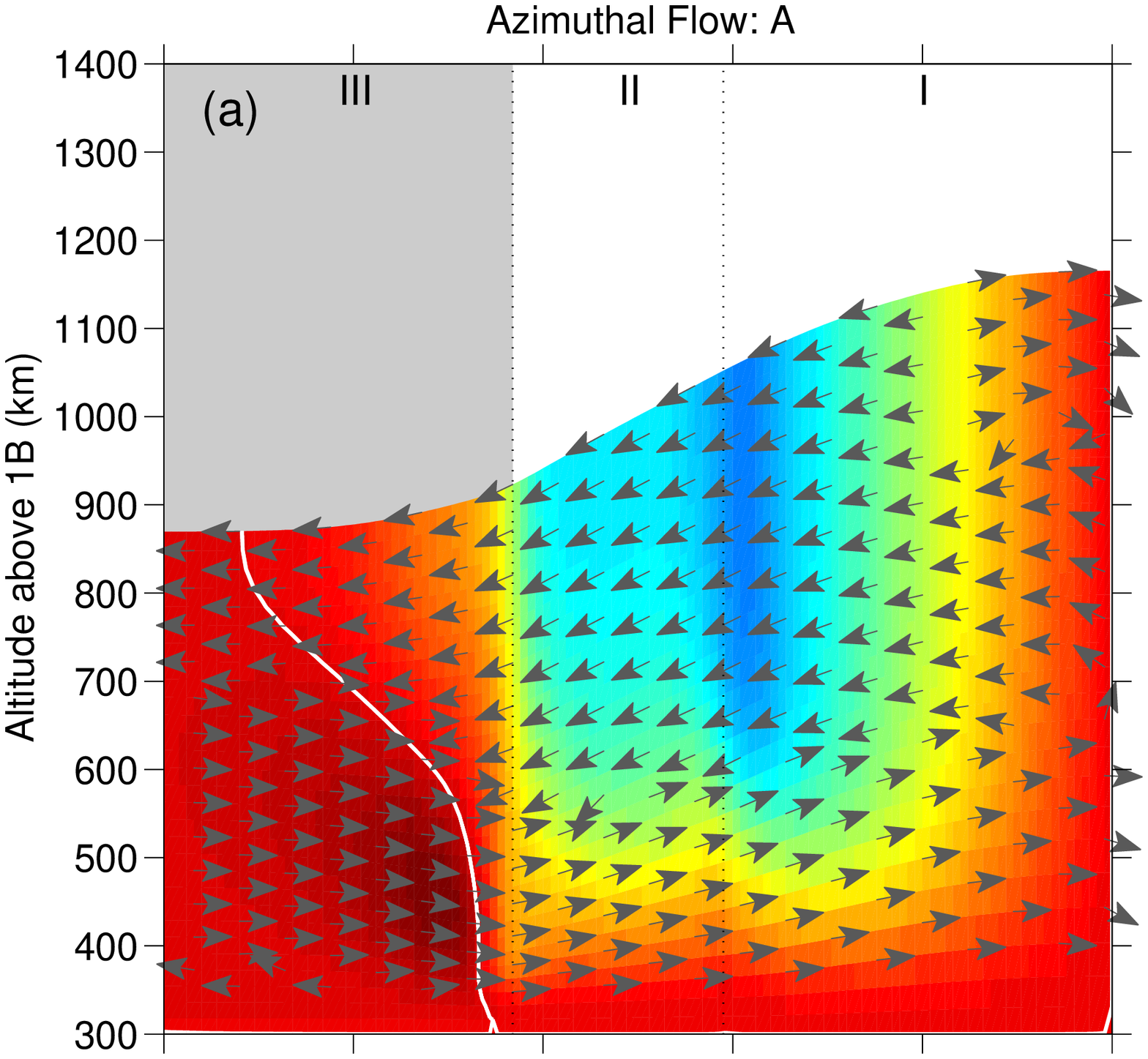}
    \includegraphics[width=0.582\figwidth]{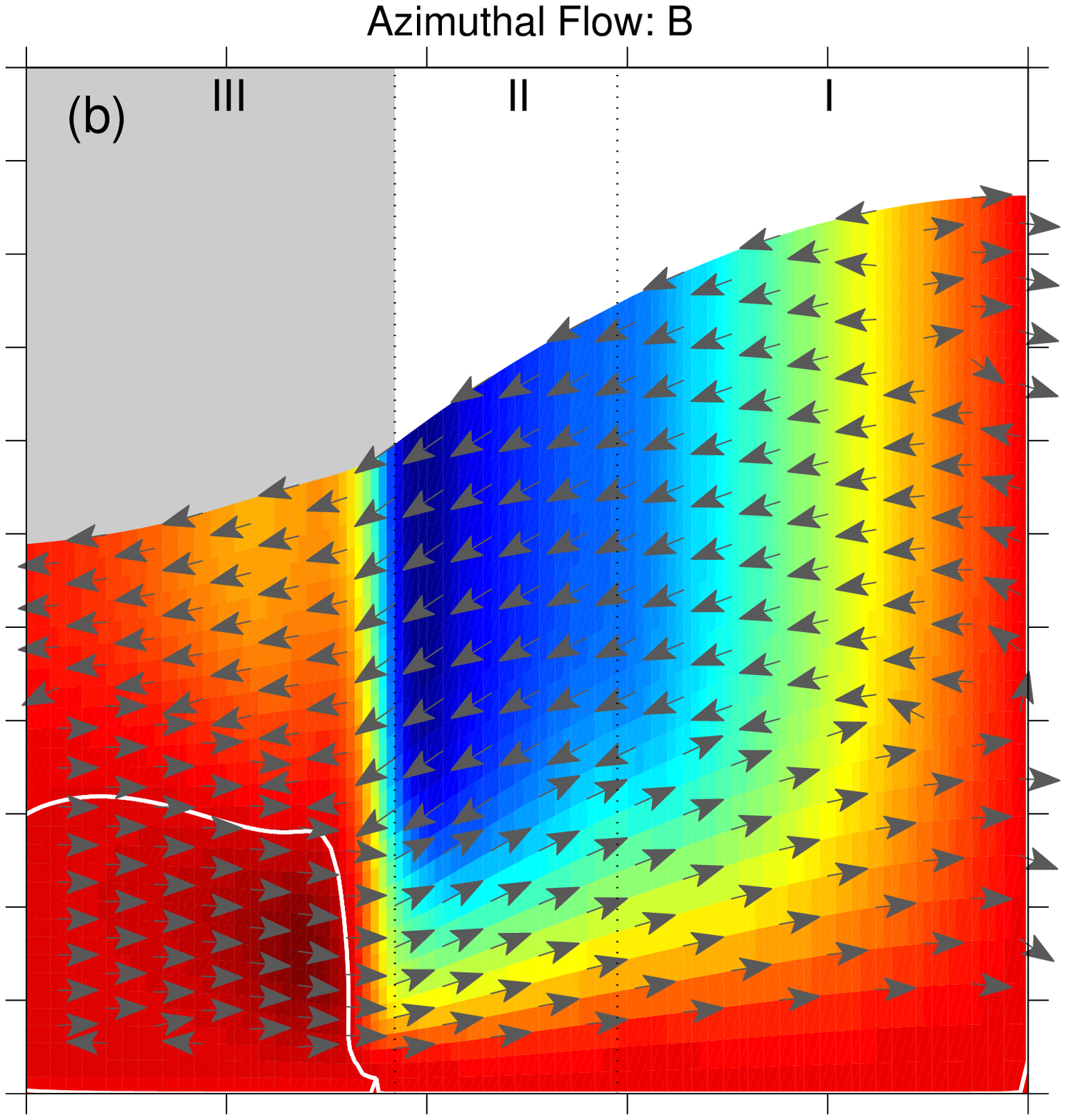}
    \includegraphics[width=0.61\figwidth]{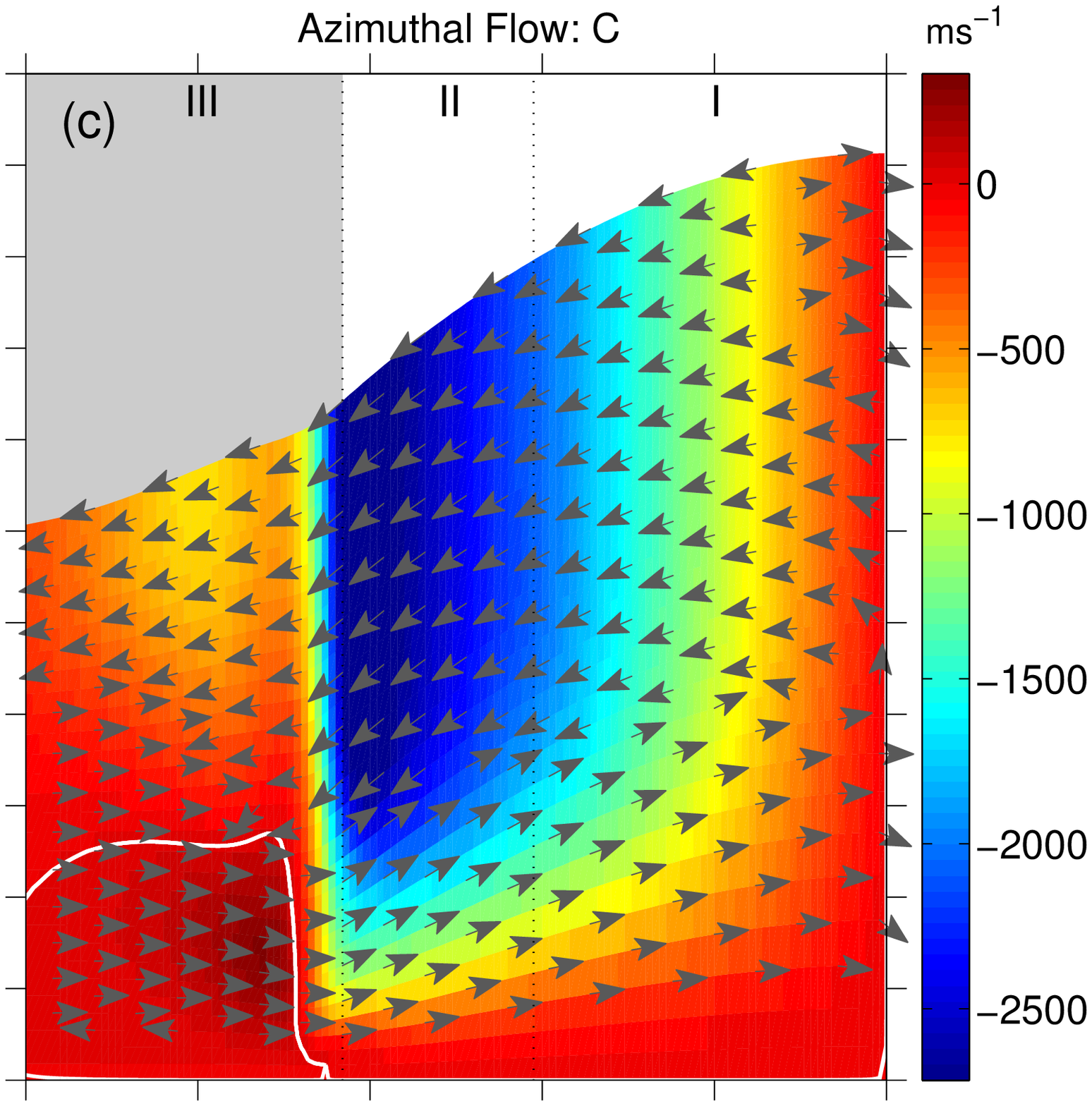}\\

    \includegraphics[width=0.66\figwidth]{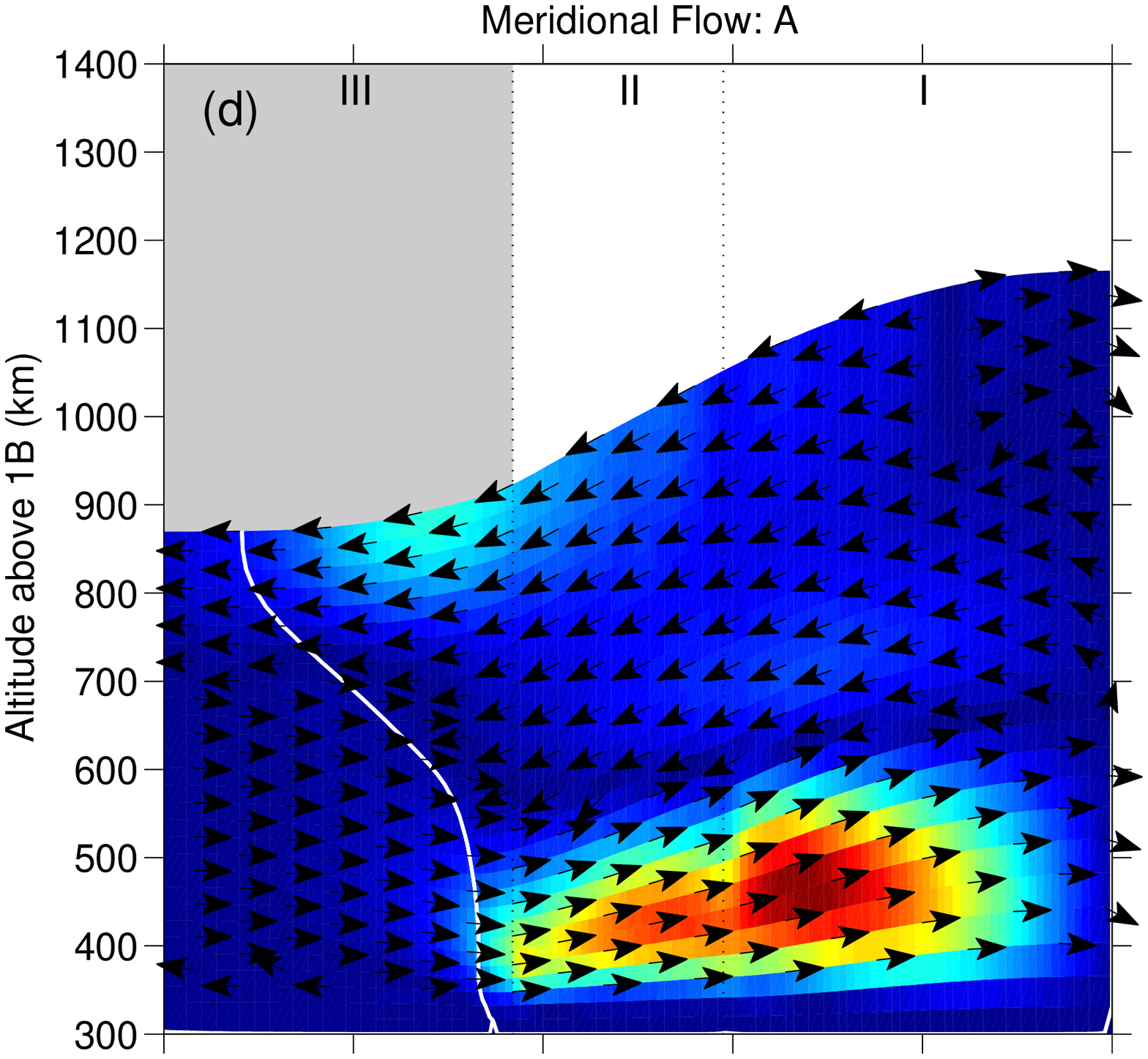}
    \includegraphics[width=0.582\figwidth]{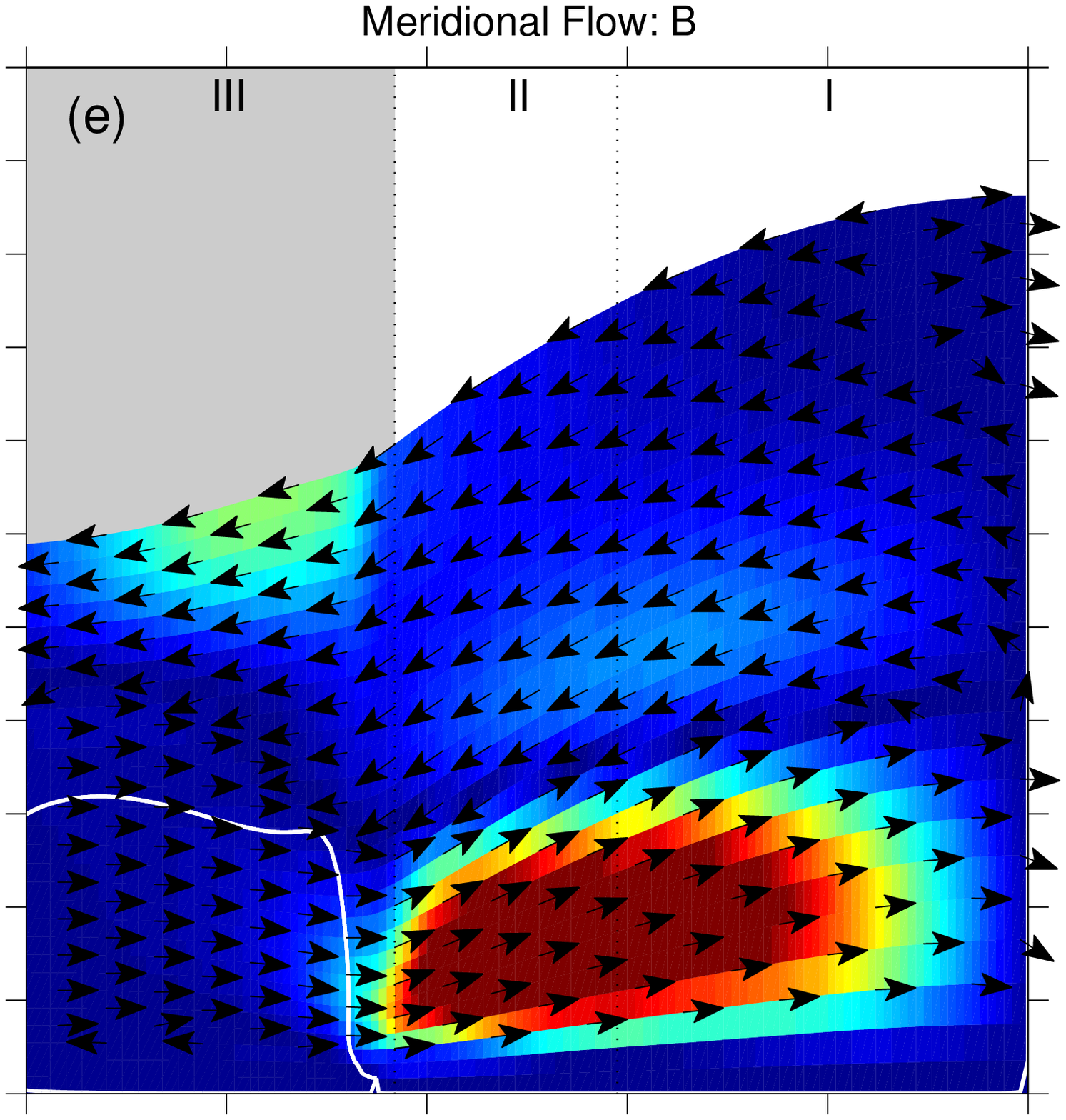}
    \includegraphics[width=0.582\figwidth]{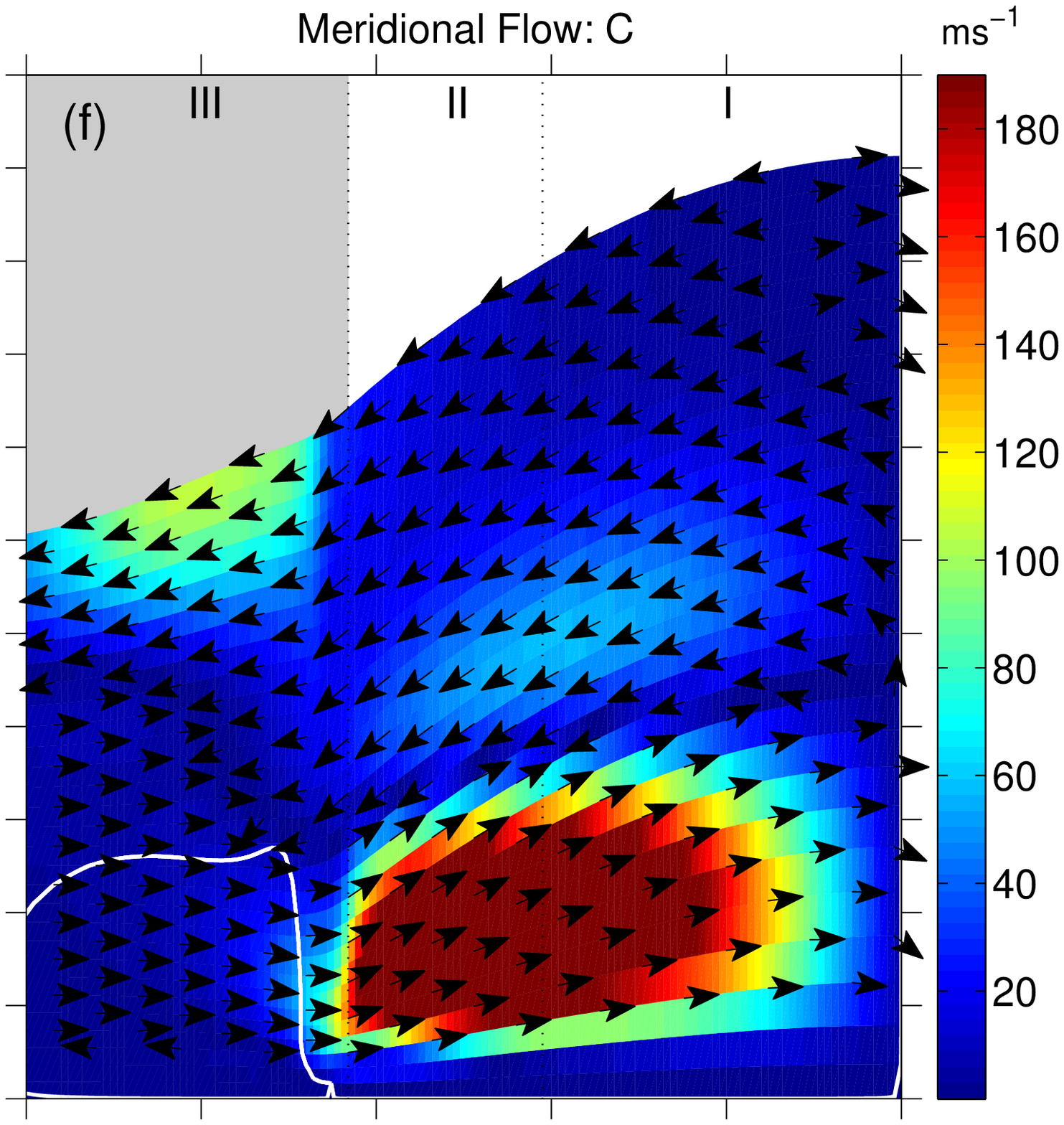}\\

    \includegraphics[width=0.66\figwidth]{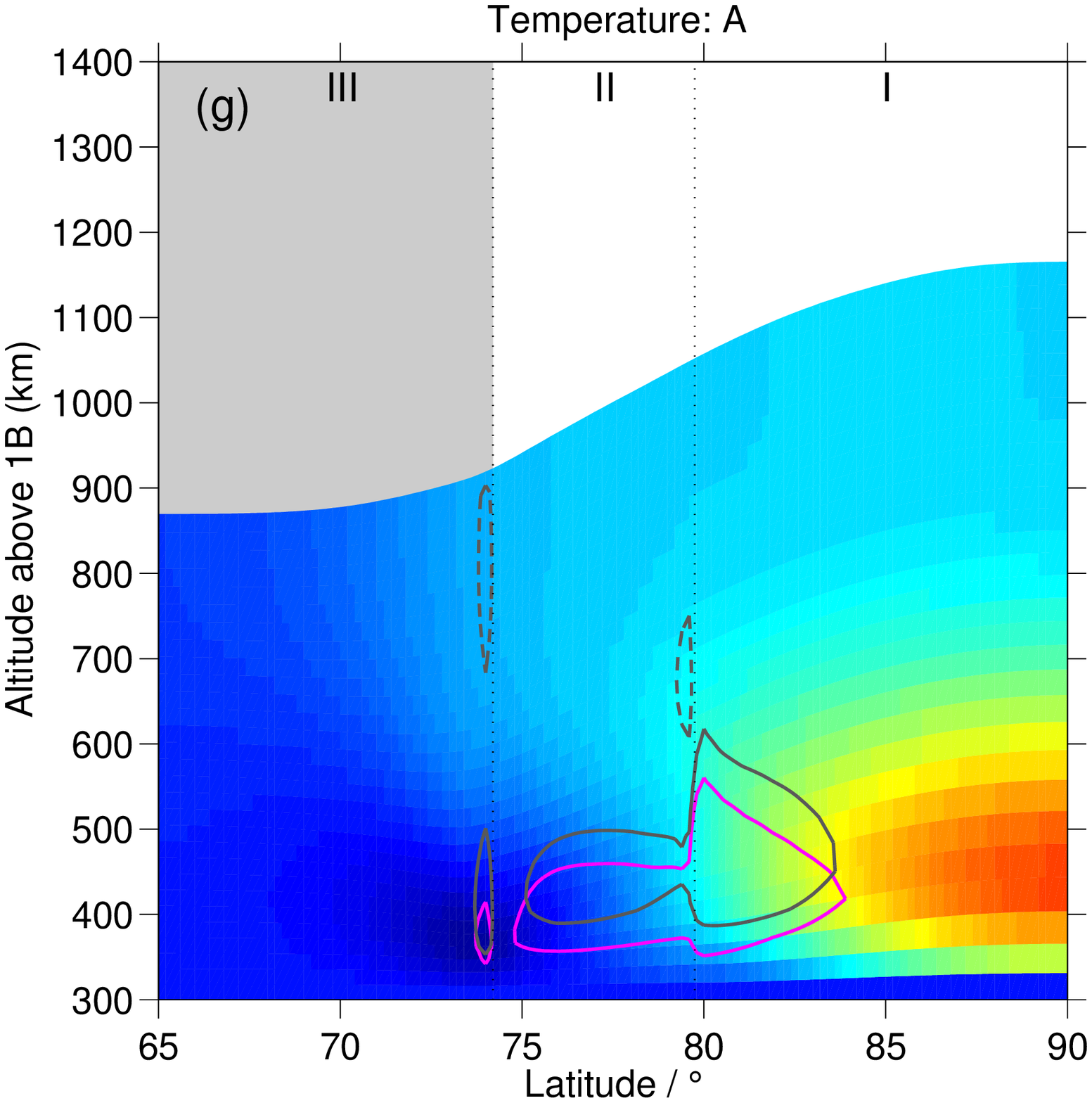}
    \includegraphics[width=0.582\figwidth]{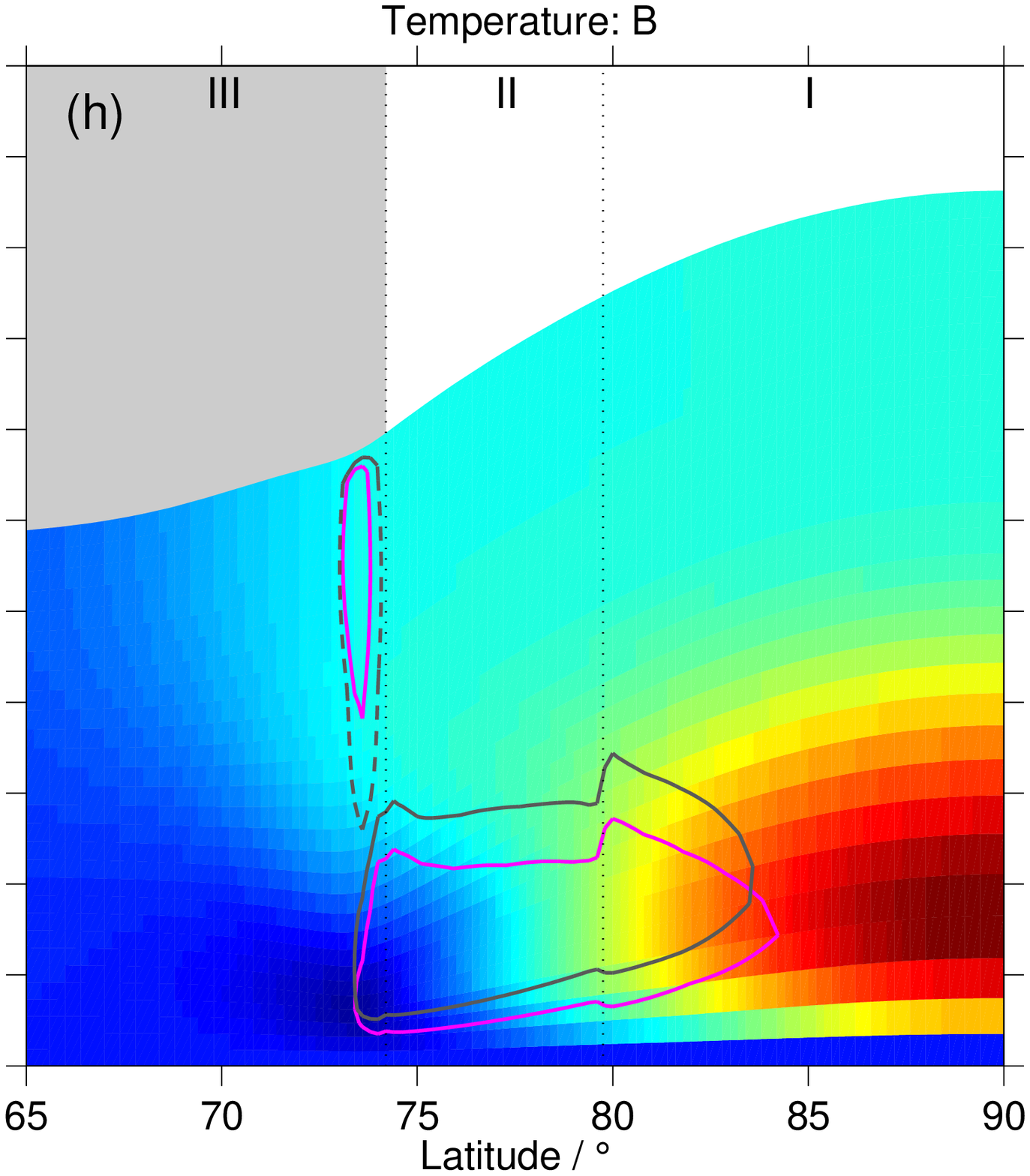}
    \includegraphics[width=0.57\figwidth]{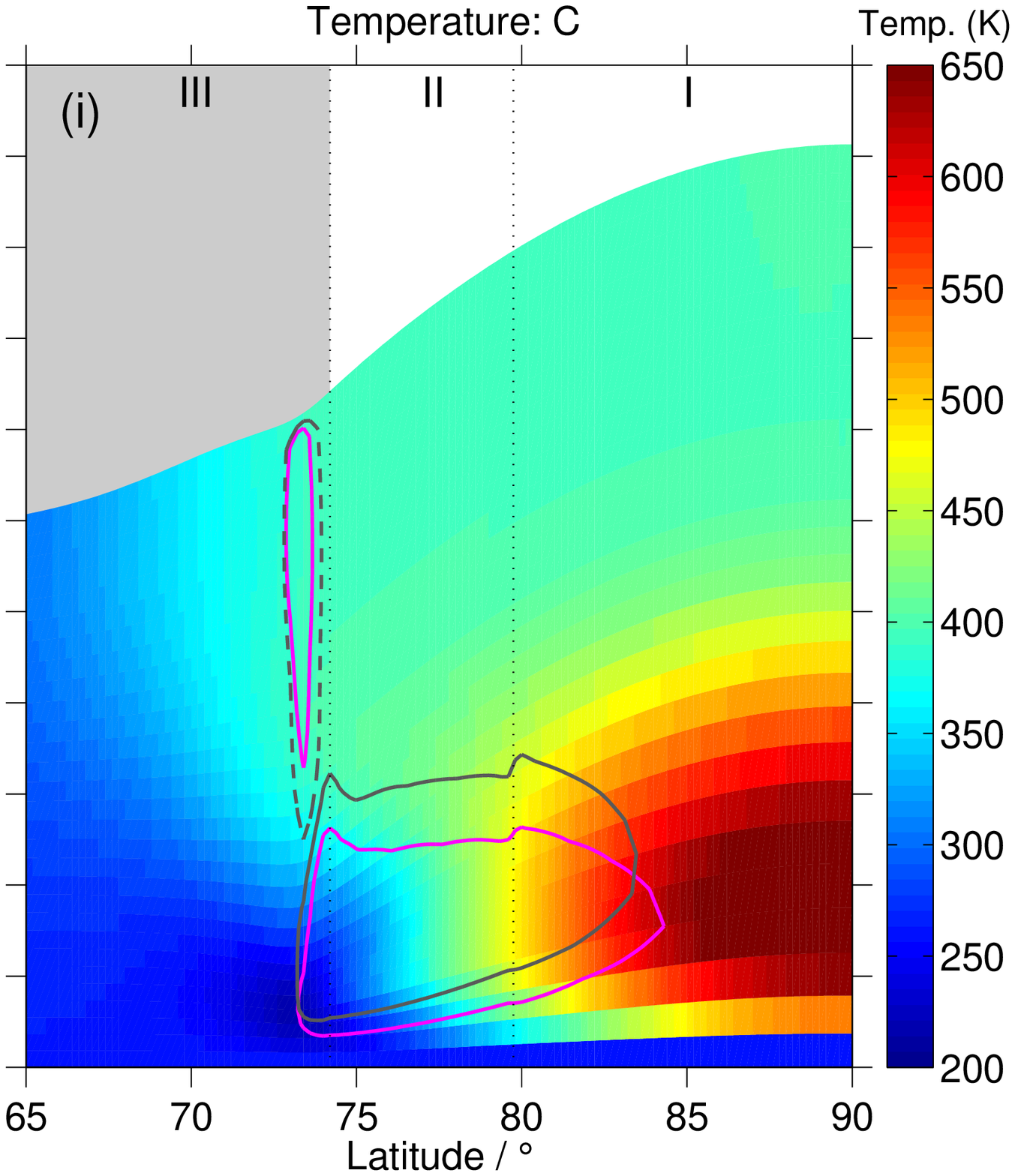}\\ 

    \includegraphics[width=0.674\figwidth]{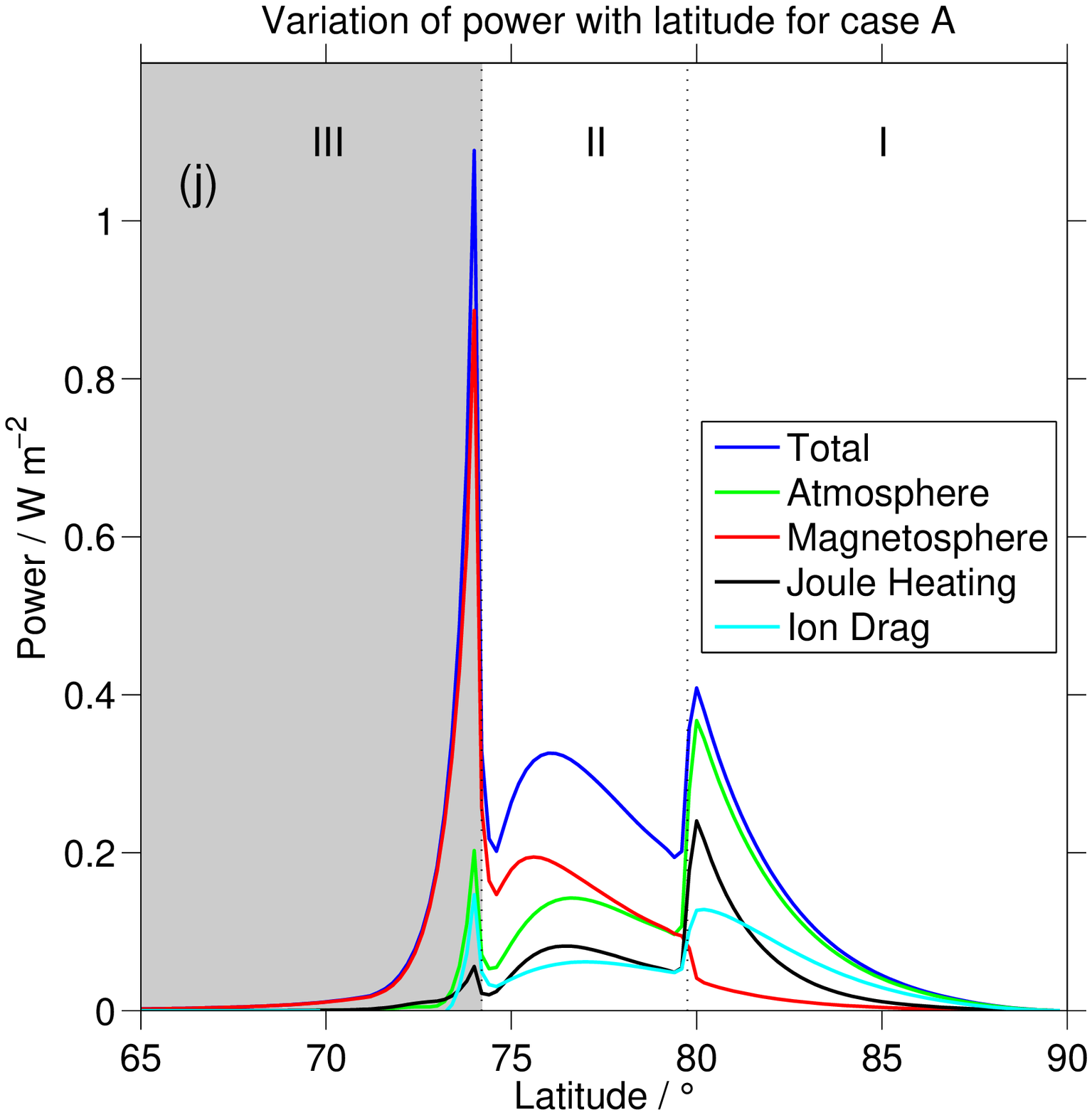}
    \includegraphics[width=0.6\figwidth]{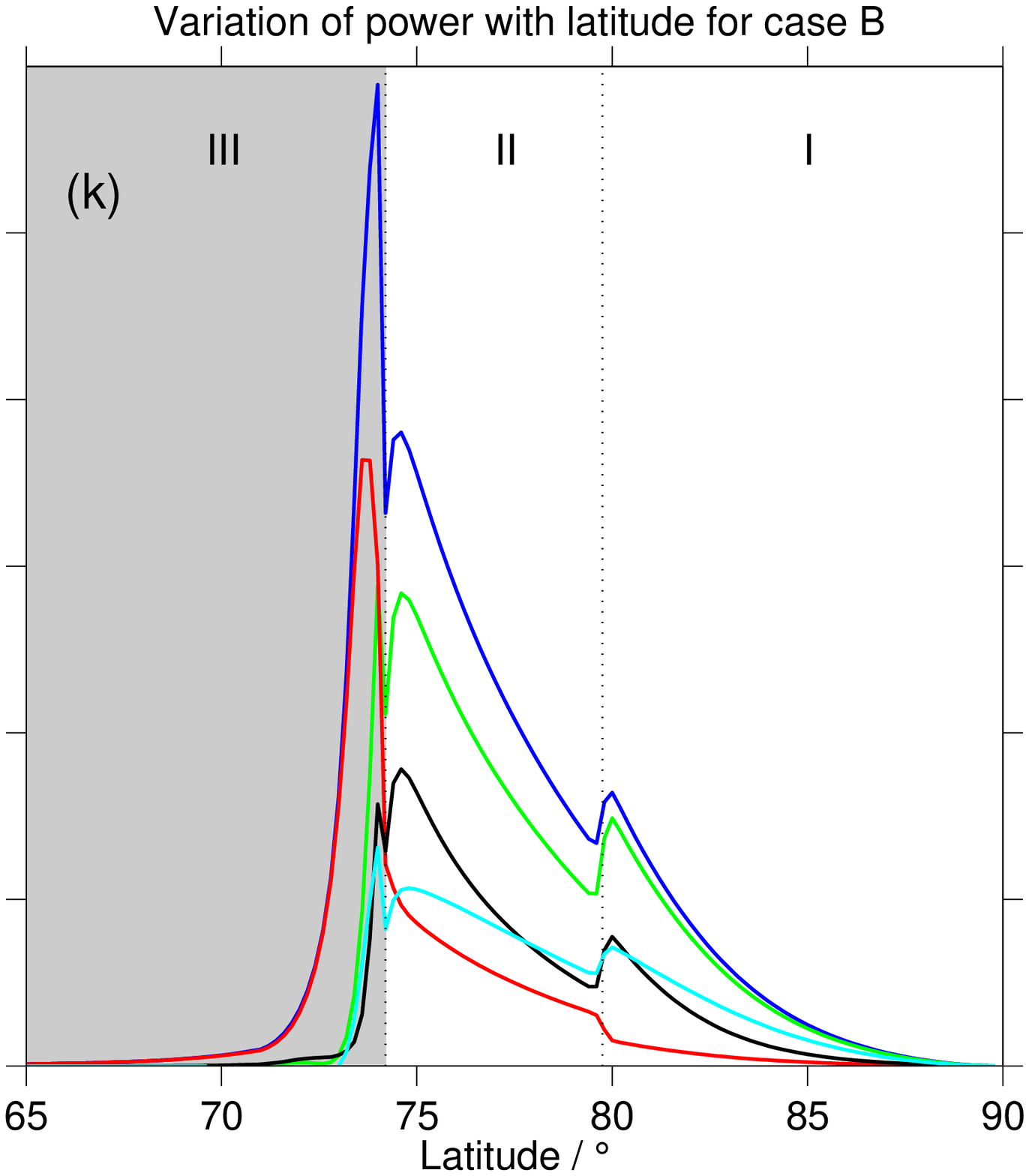}
    \includegraphics[width=0.6\figwidth]{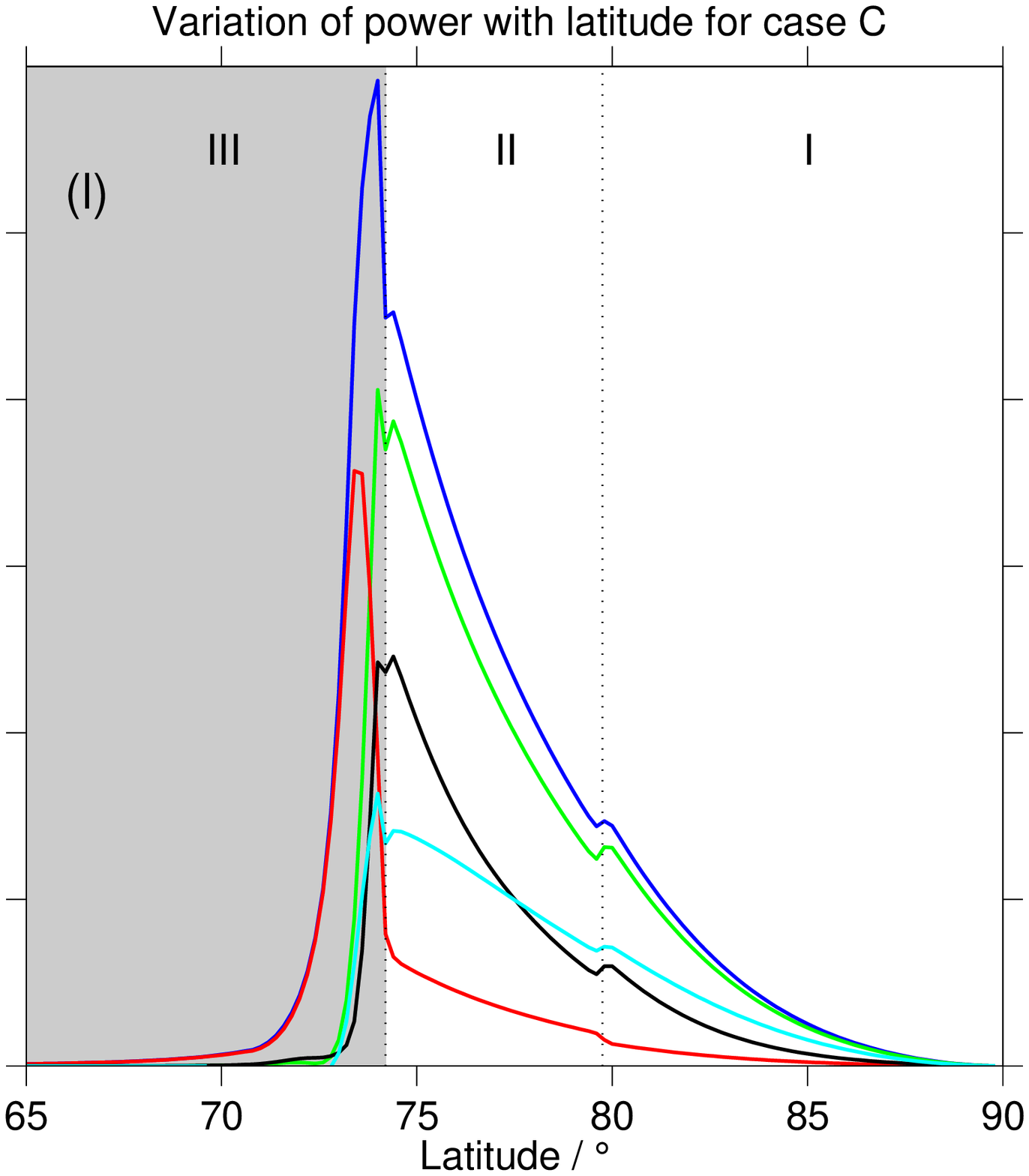}

    \caption{ \scriptsize{Figures a-c show the variation of thermospheric azimuthal velocity (colour scale) in the 
    corotating reference frame for cases A-C respectively. Positive values \Rev{dark red} represent super-corotation, whilst 
    negative values (light red to blue) represent sub-corotation. The direction of meridional flow is indicated by 
    the arrows and the white line represents the locus of rigid corotation. \Rev{Magnetospheric} regions 
    \Rev{(region III is shaded)} are labelled 
    and separated by black dotted lines. Figures d-f show \Rev{the} meridional velocity in the thermosphere for cases A-C. 
    The colour scale indicates the speed of flows. Other labels and lines are as for (a)-(c). Figures g-i show 
    thermospheric temperature distributions. Magenta contours enclose regions where Joule heating exceeds
    $\unitSI[20]{W\,kg^{-1}}$, solid grey contours enclose regions where ion drag increases the kinetic energy at 
    rates exceeding $\unitSI[20]{W\,kg^{-1}}$ and dashed grey contours enclose regions where ion drag decreases the 
    kinetic energy at rates greater than $\unitSI[20]{W\,kg^{-1}}$. Figures j-l show how the power per unit area 
    varies 
    for cases A-C. The blue line represents total power which is the sum of magnetospheric power (red line) and 
    atmospheric power (green line); atmospheric power is the sum of both Joule heating (black solid line) and ion 
    drag (cyan solid line). Other labels are as for (a)-(c). }}
		
      \label{fig:thermosphere}
  \end{figure*}

  According to SA09, meridional advection is the main process by which angular momentum is transferred 
  to the high latitude thermosphere. Advection (combination of the horizontal and vertical advection of momentum by 
  winds blowing along and across fixed pressure surfaces) is just one of the means by which momentum is changed at a 
  fixed location within the thermosphere. 
  In \Fig{\ref{fig:momentum}} we present force balance diagrams at low (a-b) and high (c-d) altitudes for cases A 
  and C. The force colour codes are in the figure caption. Considering the high altitude region first, advection and 
  other zonal force components (ion drag and Coriolis) are small. Thus, the pressure 
  gradient is balanced almost perfectly by the Coriolis force. This force balance creates a sub-corotational 
  flow with a small equatorward component. We now consider low altitudes near the Pedersen conductivity peak, where 
  the ion drag term \textbf{\emph{J\unit{\times}B}} is strong. Coriolis, pressure gradient and ion drag forces are not 
  balanced. Thus, a significant advection term arises to restore equilibrium, resulting in a region of strong 
  poleward acceleration (see \Figs{\ref{fig:thermosphere}d-f}). The resulting meridional flow at low altitudes is 
  thus polewards and transports heat to the polar region. \\
  
  \Figs{\ref{fig:thermosphere}a-c} show how \Rev{the} thermospheric azimuthal velocity in the corotating reference 
  frame \Rev{varies} within the high latitude region for the different cases. Positive (resp. negative) values of neutral 
  azimuthal velocity \Rev{indicate} super (resp. sub) -corotating regions. Arrows indicate the direction of 
  meridional flow, and the white line the locus of rigid corotation. The magnetospheric region boundaries 
  are plotted with the dotted black lines. We can see a broad azimuthal jet (blue area) in regions 
  I and II that sub-corotates to a greater degree with an increase in magnetospheric size. Also present is 
  a super-corotational jet (dark red region) just equatorward of the region III / II boundary, 
  visible in \Fig{\ref{fig:omegaplots}}. Ion drag (see \Fig{\ref{fig:momentum}}) gives rise to the sub-corotational 
  azimuthal flows seen in regions I, II and III. As the magnetosphere expands, the \textbf{\emph{J\unit{\times}B}} term 
  increases and 
  azimuthal flows sub-corotate to an even greater degree. Advection forces arise due to the lack of equilibrium at 
  low altitudes\Rev{,} causing an accelerated poleward flow \Rev{whose velocity increases by ${\sim}\unitSI[90]{\%}$ from 
  case A to C}. The effect of advection can be seen in 
  \Figs{\ref{fig:thermosphere}d-f}, which show meridional flows in the high latitude region. This accelerated flow 
  transports energy from Joule heating, depositing it at higher latitudes and forming a polar `hot spot' 
  \citep{smith2007Natur}. Super-corotation occurs at latitudes where zonal ion drag and advection forces are 
  negligible compared to 
  the Coriolis force, which can then accelerate the flow beyond corotation. At high altitudes, forces are essentially 
  balanced. Thus high altitude zonal flows now have an equatorward component. Therefore, meridional flows show a 
  poleward low-altitude flow and an equatorward high-altitude flow consistent with the previous studies of 
  \citet{smith2007Natur} \Rev{and} SA09.\\
  

  \Figs{\ref{fig:thermosphere}g-i} show thermospheric temperature distributions. The temperature scale 
  is shown on the colour bar. Magenta and solid grey contours enclose areas where Joule heating and ion drag energy 
  inputs exceed $\unitSI[20]{W\,kg^{-1}}$ and dashed grey contours highlight regions where ion drag decreases  
  kinetic energy at a rate greater than $\unitSI[20]{W\,kg^{-1}}$. A uniform rate of $\unitSI[20]{W\,kg^{-1}}$ gives 
  an integrated energy input rate of the order of \unitSI[100]{mW\,m^{-2}} 
  ($\unitSI[1]{ergs\,cm^{-2}\,s^{-1}}$ ${=}\unitSI[1]{mW\,m^{-2}}$) within the Pedersen conducting layer. This 
  integrated rate is of a similar order of magnitude to the estimated total IR auroral emission 
  (\unitSI[200]{ergs\,cm^{-2}\,s^{-1}} from \citet{drossart1993}). We see significant energy input from Joule heating 
  and ion drag at low altitudes and between \Rev{\unitSI[73]{^{\circ}}\unitSI[-85]{^{\circ}}} latitude due to their dependence 
  on the strength of the current density \Rev{\textbf{\emph{j}}. \textbf{\emph{j}}} is proportional to the difference 
  between local thermospheric and plasma angular velocity, and to the Pedersen conductivity \unit{\sigma_P} which peaks 
  at low altitudes. There lies a narrow 
  region of high altitude Joule heating just equatorward of the region III / II boundary in expanded cases due to the 
  large shear between \unit{\Omega_T} and \unit{\Omega_M}. The decrease in kinetic energy (grey dashed lines) occurs 
  as the ion drag force now acts to accelerate thermospheric flows towards corotation (see 
  \Figs{\ref{fig:momentum}c-d} where ion drag is eastwards). The remaining feature of prominence is the large `hot 
  spot' \Rev{at} low altitudes \Rev{in} region I as discussed above. The peak temperature of the `hot spots' 
  \Rev{increase from ${\sim}\unitSI[560]{K}$ in case A to ${\sim}\unitSI[695]{K}$ in C.}

\subsubsection{Atmospheric and magnetospheric energy `budget'}
\label{sec:energies}

\begin{figure}
    \centering 
    \includegraphics[width= 0.99\figwidth]{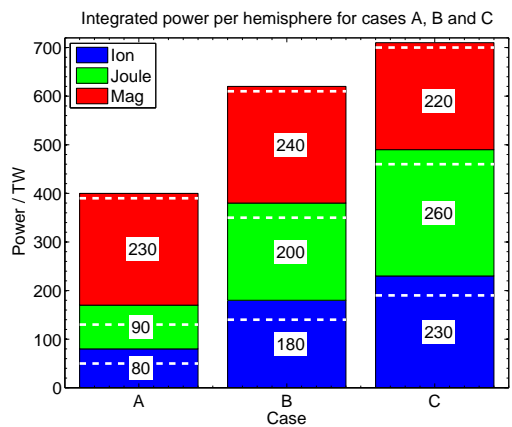}
    
    \caption{ Integrated ionospheric powers per hemisphere for cases A-C are represented in this figure. Ion drag 
    is represented by blue bars, Joule heating by green bars and magnetospheric power by red bars. The white dashed 
    line shows the division in powers between closed and open field line regions. Powers in the closed field 
    regions lie below the dashed white line whilst powers in the open field regions lie above it. Total power 
    dissipated for each mechanism (in TW) is printed on its respective colour bar. }
      \label{fig:powerall}
  \end{figure}

  \Figs{\ref{fig:thermosphere}j-l} present the total, magnetospheric, atmospheric, ion drag and Joule 
  heating power per unit area (see \Eqs{\ref{eq:totpower}{-}\ref{eq:iondrag}}) for each magnetospheric configuration. 
  The colour code indicates the form of energy dissipation (blue curve shows total power). Total power is the sum of 
  the magnetospheric and atmospheric powers and atmospheric power is the sum of Joule heating and ion drag powers. 
  Integrating the power per unit area 
  over the model hemisphere gives us the global powers for each of these mechanisms, shown in 
  \Fig{\ref{fig:powerall}}.\\

  \Figs{\ref{fig:thermosphere}j-l} exhibit peaks in power per unit area just equatorward of the interface 
  between regions III and II due to the large ionospheric current associated with the breakdown in corotation of the 
  magnetodisc (region III). Most of the energy dissipated in region III accelerates the magnetospheric plasma towards 
  corotation. This magnetospheric power dominance diminishes for the expanded magnetosphere, in which more power is 
  dissipated in the atmosphere via Joule heating and ion drag. Region II is dominated by these atmospheric power 
  terms in cases B and C whilst the same region in case A is still noticeably magnetospheric power-dominated. 
  Atmospheric power is the major form of energy dissipation in region I. The atmospheric power dominance in regions 
  II and I is mainly due to the low assumed values for magnetospheric angular velocity \unit{\Omega_M} (see 
  \Fig{\ref{fig:omegaplots}}). The difference (\unit{\Omega_T - \Omega_M}) is largest in these regions, which 
  produces relatively large ionospheric Pedersen currents and atmospheric power. The low \Rev{value of} \unit{\Omega_M} produces 
  a magnetospheric power that remains low compared to other regions. Note that relative amounts of 
  energy provided to the atmosphere and magnetosphere on any flux shell depends on the difference 
  (\unit{\Omega_T - \Omega_M}) through \Eqs{\ref{eq:torque}-\ref{eq:atmpower}}. They are equal when 
  $\unit{\Omega_{M}}{=}\unitSI[0.5]{\Omega_{J}}$. \\
  
  These figures suggest that as Jupiter's magnetosphere is compressed a higher proportion of the total power of 
  planetary rotation (in the steady state) is transferred to the magnetosphere via the magnetic field, and by 
  contrast, as the magnetosphere is expanded more heat is dissipated in the atmosphere. \\
  
  \Fig{\ref{fig:powerall}} shows the integrated ion drag (blue), Joule heating (green) and magnetospheric (red) power 
  per hemisphere for each case and how these powers are distributed in the open and closed field line regions. Powers 
  in the closed field regions lie below the dashed white line whilst powers in the open field regions lie above 
  it. Integrated powers within region I remain essentially unchanged for both 
  atmospheric (ion drag plus Joule heating) and magnetospheric mechanisms due to the assumed constant value of the 
  magnetospheric angular velocity for all cases. Atmospheric power increases significantly with magnetospheric size\Rev{,} 
  by a factor of ${\sim}\unitSI[3]{}$ from case A to C. Magnetospheric power shows a slight decrease between cases A 
  and C and is a maximum in \Rev{case} B. Magnetospheric power is proportional to the torque, which increases 
  with magnetospheric expansion, and the magnetospheric angular velocity which decreases with expansion. One would 
  thus expect that, given a large enough range of magnetospheric sizes, the magnetospheric power would display a 
  non-monotonic profile that is `modulated' by the difference between the angular velocity of the neutral 
  thermosphere and plasmadisc. \\
  
  We now compare our integrated power\Rev{s} per hemisphere with those calculated by \citet{cowley07} to take into account 
  how using a two-dimensional model of Jupiter's thermosphere changes previous theoretical power estimates. We only 
  compare cases A and B with the `intermediate' and `baseline' cases from \citet{cowley07} respectively, as they have 
  comparable magnetodisc radii. The total integrated magnetospheric powers per hemisphere are comparable in 
  magnitude; `intermediate' and `baseline' cases in \citet{cowley07} are ${\sim}\unitSI[85]{\%}$ those of our cases A 
  and B. The difference between atmospheric powers is greater, since this study uses a detailed thermospheric model. 
  In region I, cases A and B have atmospheric power\Rev{s} ${\sim}\unitSI[35]{\%}$ of \citet{cowley07}'s `intermediate' and 
  `baseline' cases. In regions III and II (closed field) the atmospheric powers in \citet{cowley07} are 
  ${\sim}\unitSI[70{-}80]{\%}$ those in cases A and B. Thus, within the closed field region of cases A and B, the 
  inclusion of a detailed thermospheric flow model has led to more energy transferred from the thermosphere to the 
  magnetosphere for accelerating the magnetospheric plasma and more heat dissipated within the thermosphere via Joule 
  heating and ion drag.

\section{Effect of outer boundary conditions}
\label{sec:pathogen}
  
  The results for case A exhibit a relatively large peak for the Pedersen conductivity and \FAC density just 
  equatorward of the boundary between regions III and II \Rev{(\Fig{\ref{fig:jpar}})}. Previous studies such as 
  \citet{southkiv2001} suggest that 
  such peaks for compressed magnetospheres should be smaller in magnitude than those for more expanded 
  magnetospheres. This is in contrast to what we observe in section~\ref{sec:results}. If the radial current at the 
  region III / II boundary, \unit{I_{\rho\infty}} is larger than realistic values for a compressed magnetosphere, 
  large FACs would develop in the poleward part of region III to satisfy the \unit{I_{\rho\infty}} boundary 
  condition. In this section we present model outputs with smaller values of \unit{I_{\rho\infty}}. We select 
  illustrative \unit{I_{\rho\infty}} values for each case which minimise the variance in the current profiles 
  throughout regions III and II. Decreasing \unit{I_{\rho\infty}} decreases current gradients in the well constrained 
  \unit{\Omega_M} model for region III and allows for a smoother transition to region II, whose \unit{\Omega_M} 
  profile is poorly constrained (due to paucity of observations).
  
\subsection{Outer boundary conditions for a compressed magnetosphere}
\label{sec:cmp_pathogen}

  To commence this part of our study, we ran case A but with $\unit{I_{\rho\infty}}{=}\unitSI[45]{MA}$, 
  to see whether any significant changes would arise in the currents at the region III / II 
  boundary. These results are shown in \Fig{\ref{fig:pathongen}}, where blue and red 
  lines represent cases \unitSI{A_{100}} and \unitSI{A_{45}} respectively (subscript denotes the value of 
  \unit{I_{\rho\infty}}).


\subsubsection{Comparison of angular velocities and currents for case A}
\label{sec:path_angcurA}
  
  In this section we compare angular velocities and current-related parameters for cases 
  \unitSI{A_{45}} and \unitSI{A_{100}} (case A in section~\ref{sec:results}). The differences between these cases 
  essentially lie within \unitSI[72{-}78]{^{\circ}} latitude and so our subsequent discussions will focus on this 
  range. \\

\begin{figure*}
    \centering 
    \includegraphics[width=0.66\figwidth]{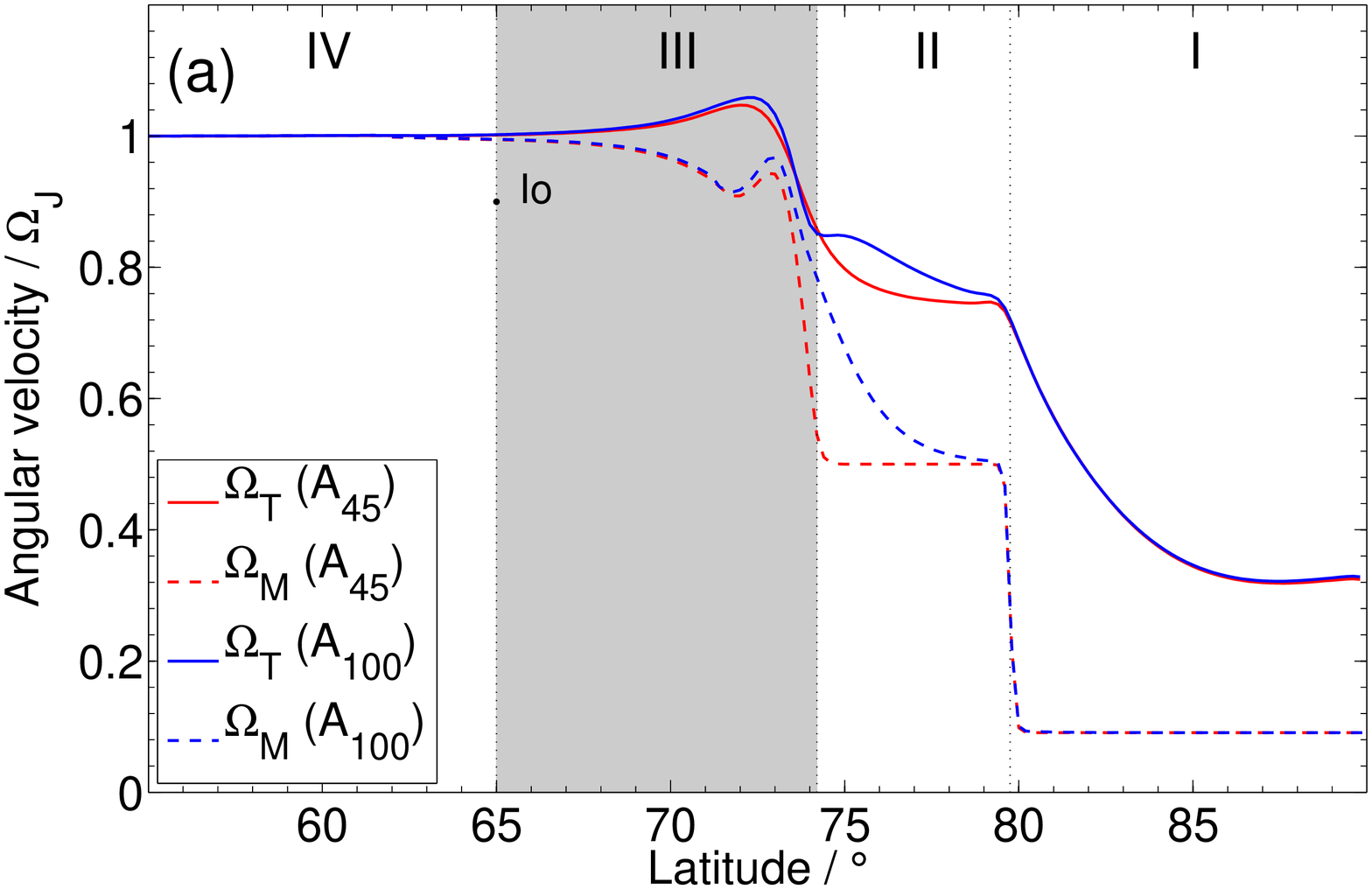}
    \includegraphics[width=0.65\figwidth]{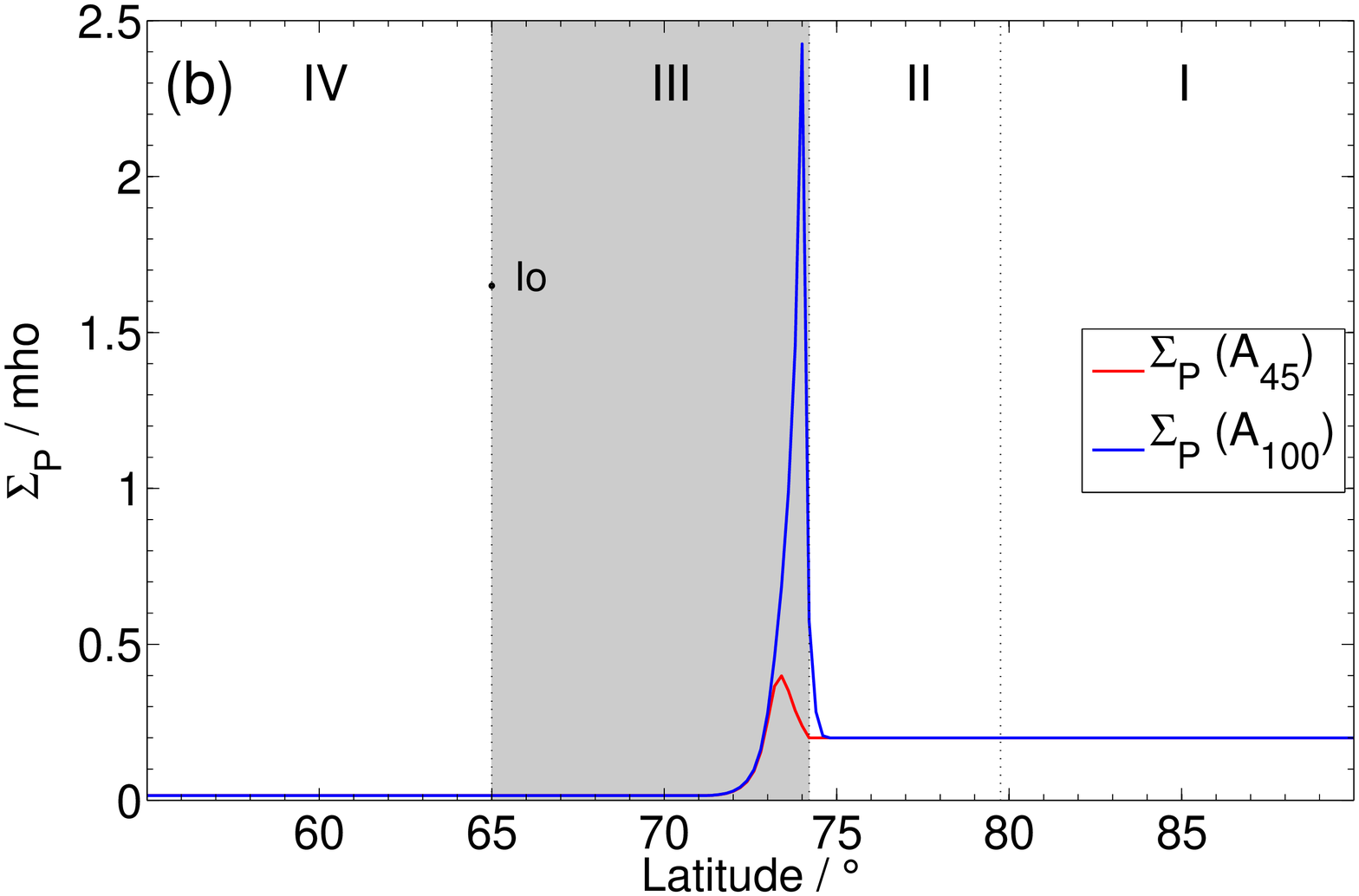}
    \includegraphics[width=0.64\figwidth]{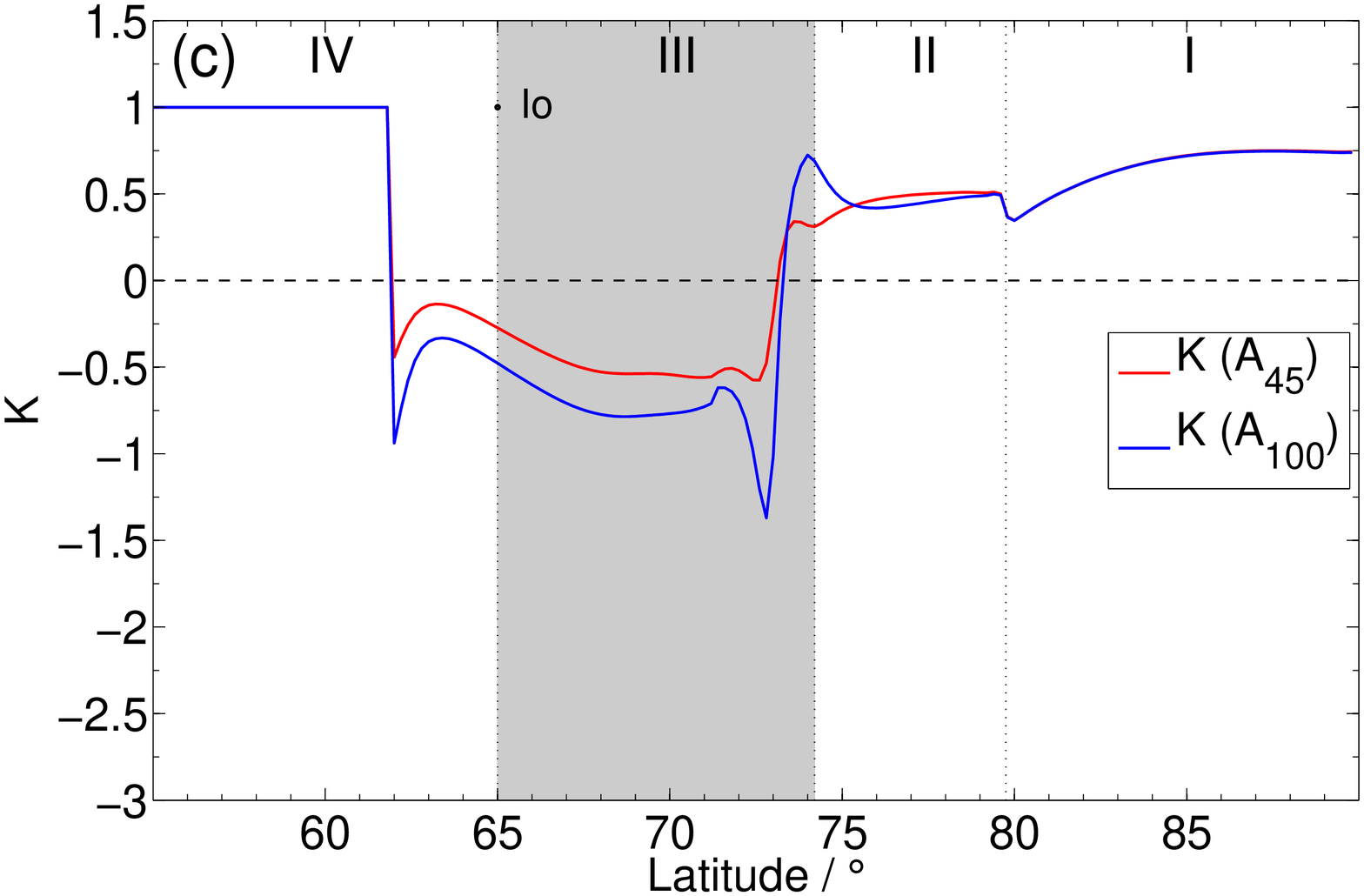}\\

    \includegraphics[width=0.66\figwidth]{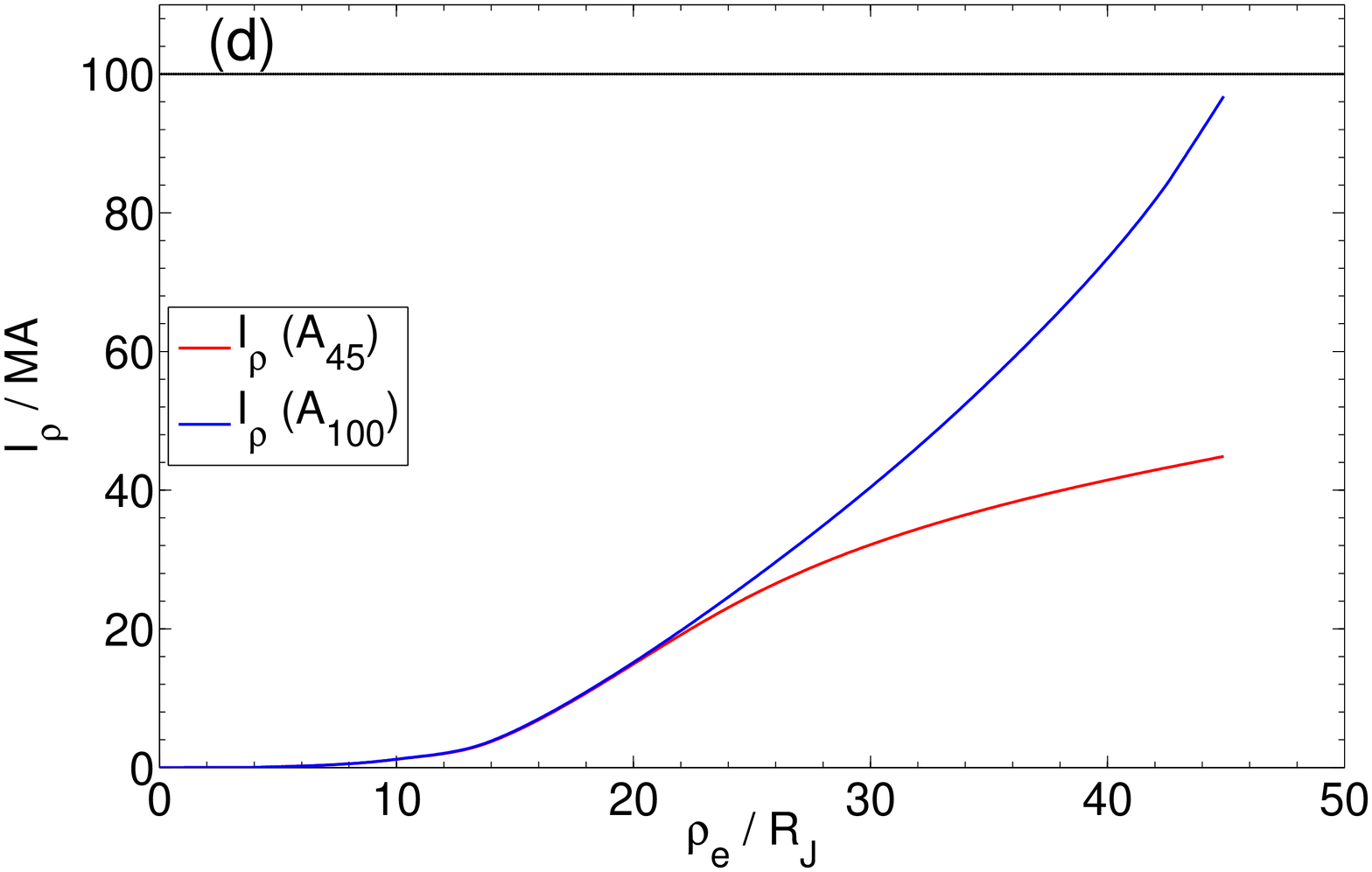}
    \includegraphics[width=0.64\figwidth]{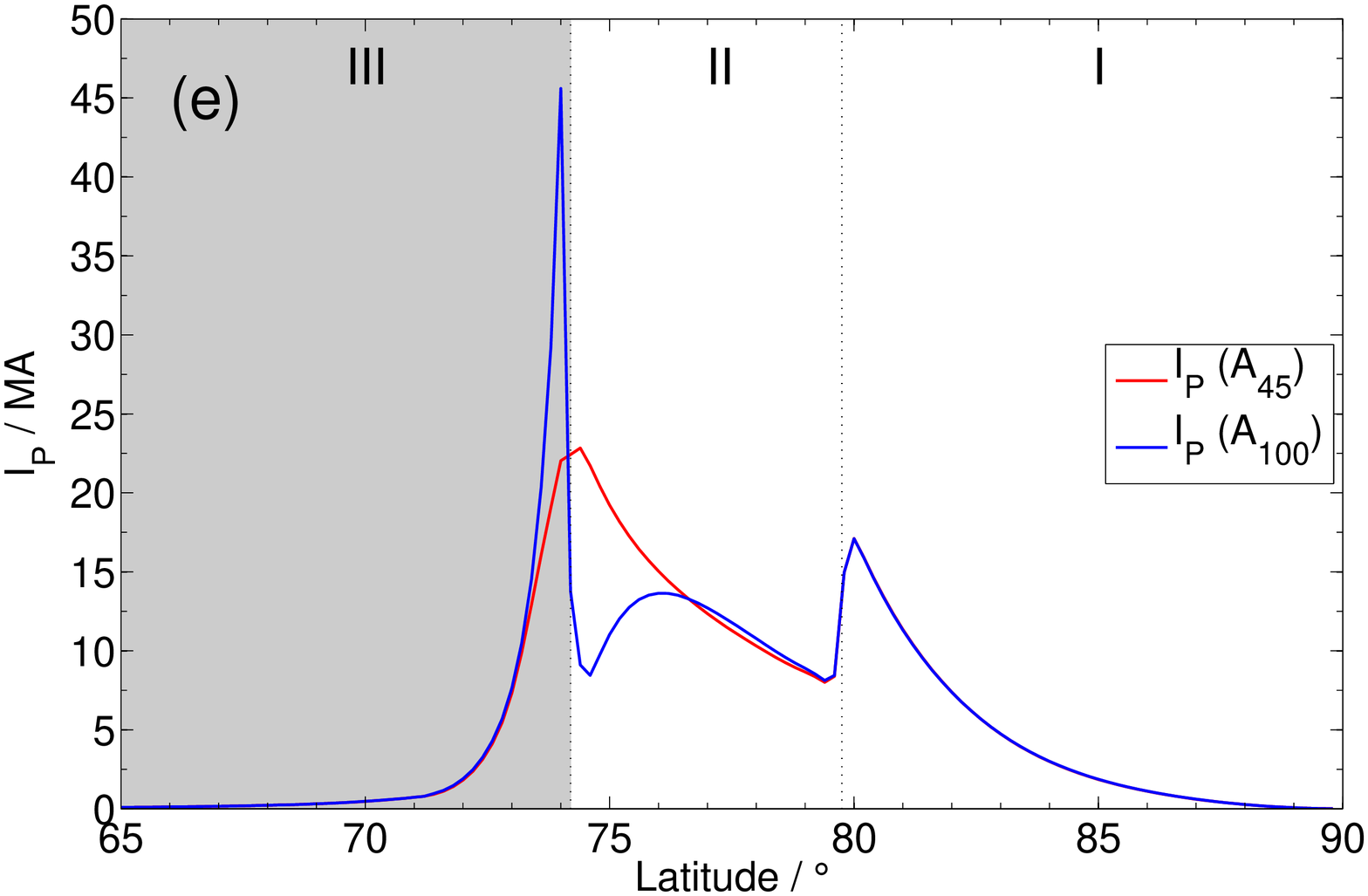}
    \includegraphics[width=0.66\figwidth]{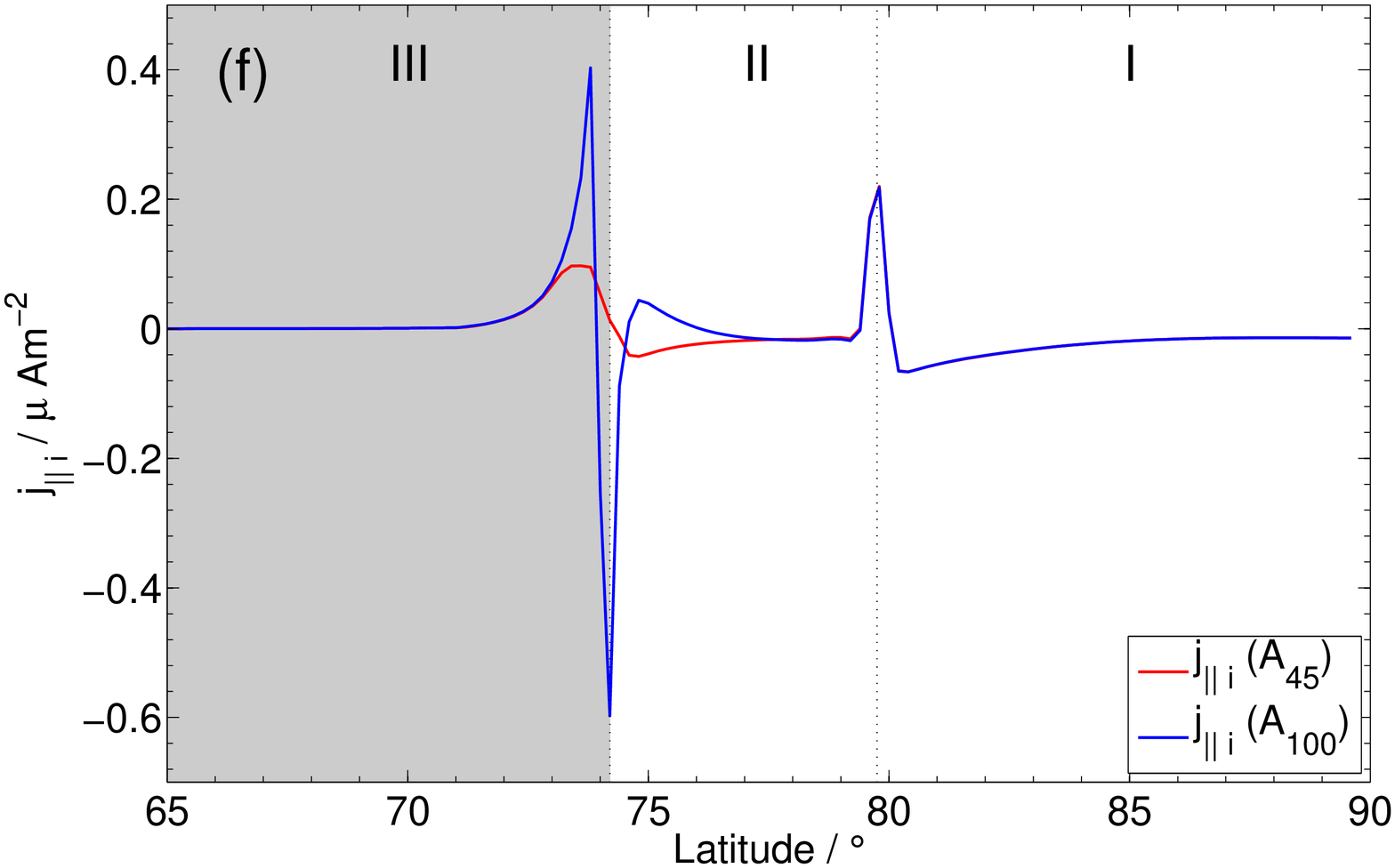}\\

    \includegraphics[width=0.70\figwidth]{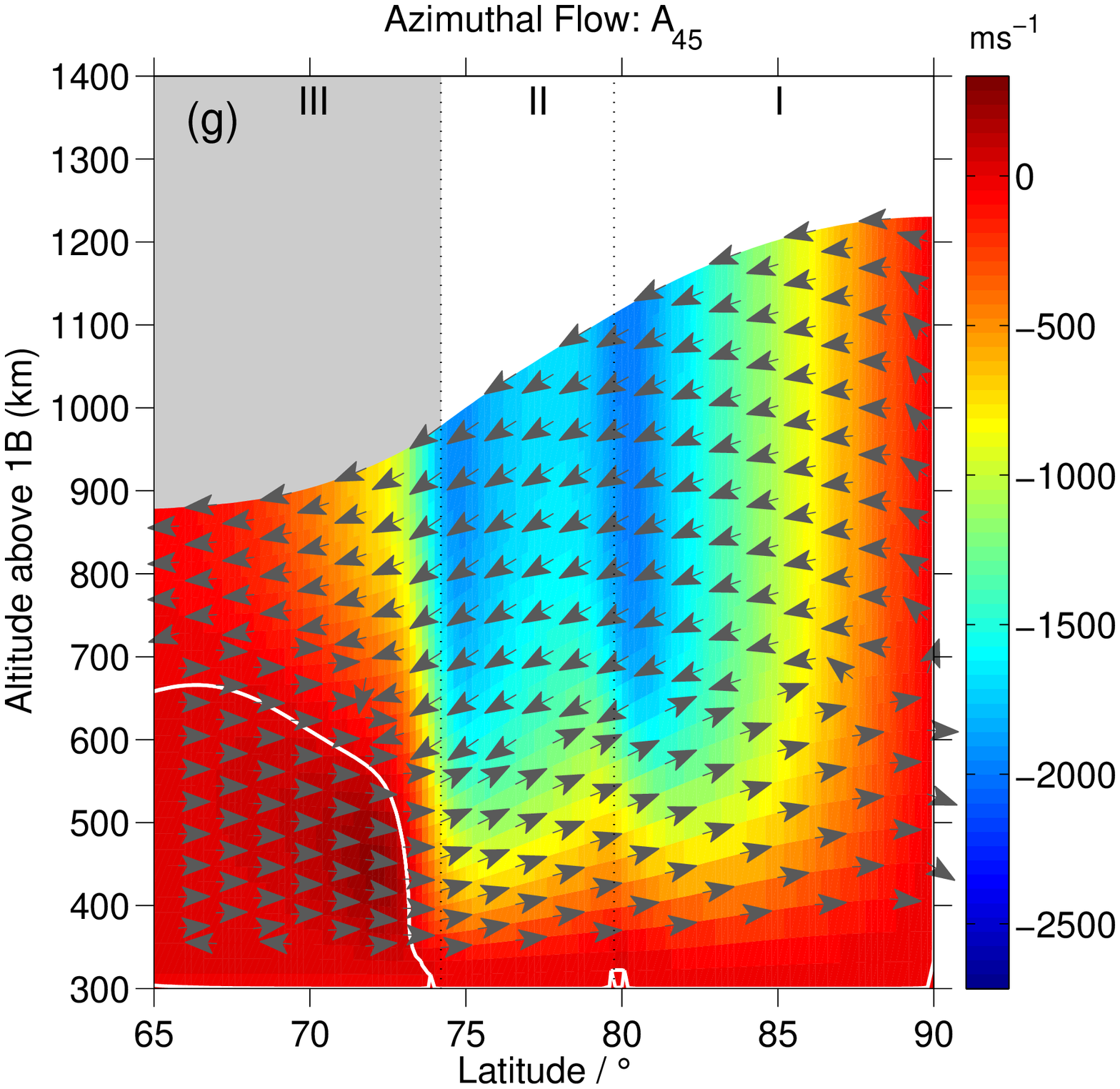}
    \includegraphics[width=0.60\figwidth]{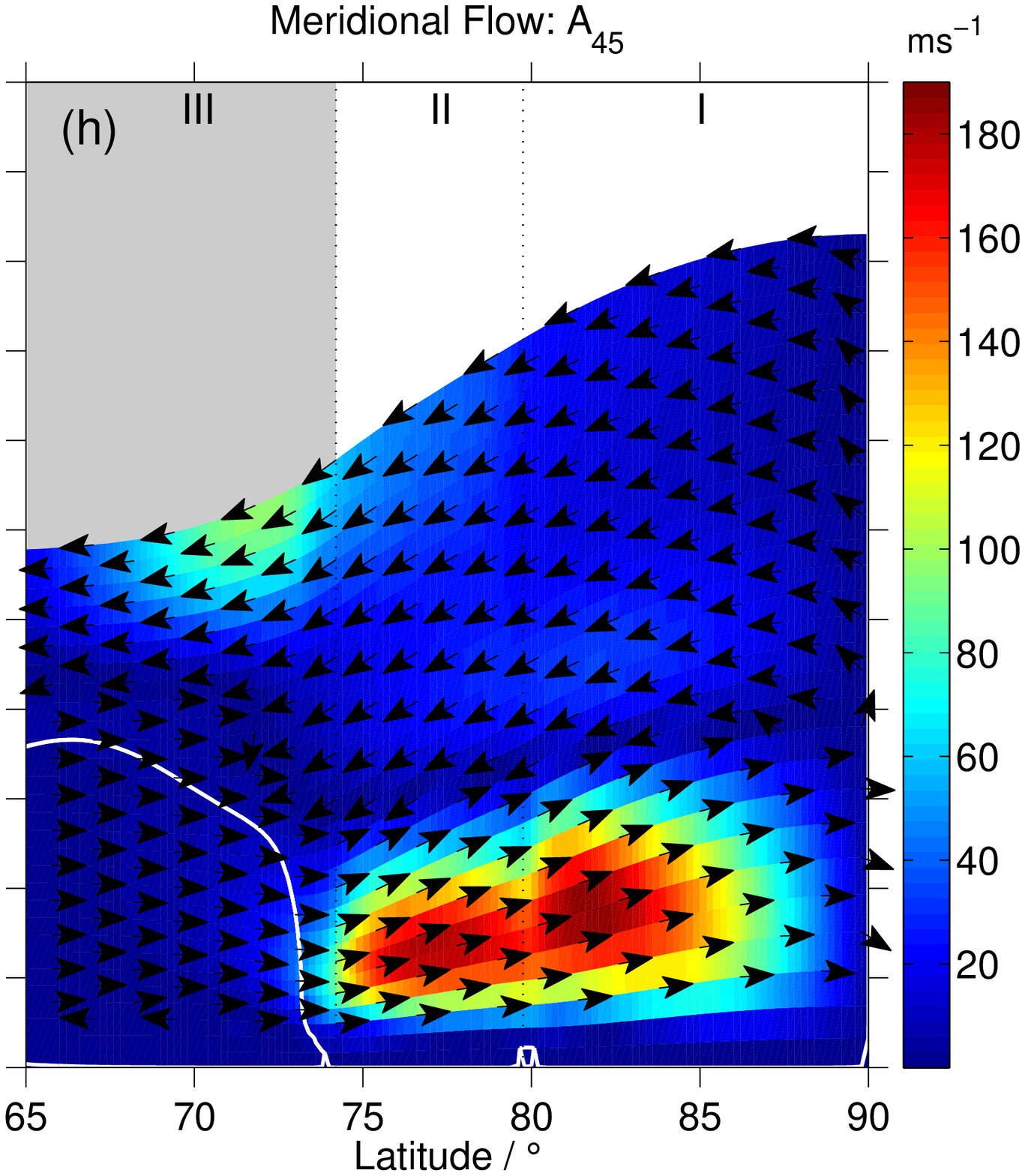}
    \includegraphics[width=0.58\figwidth]{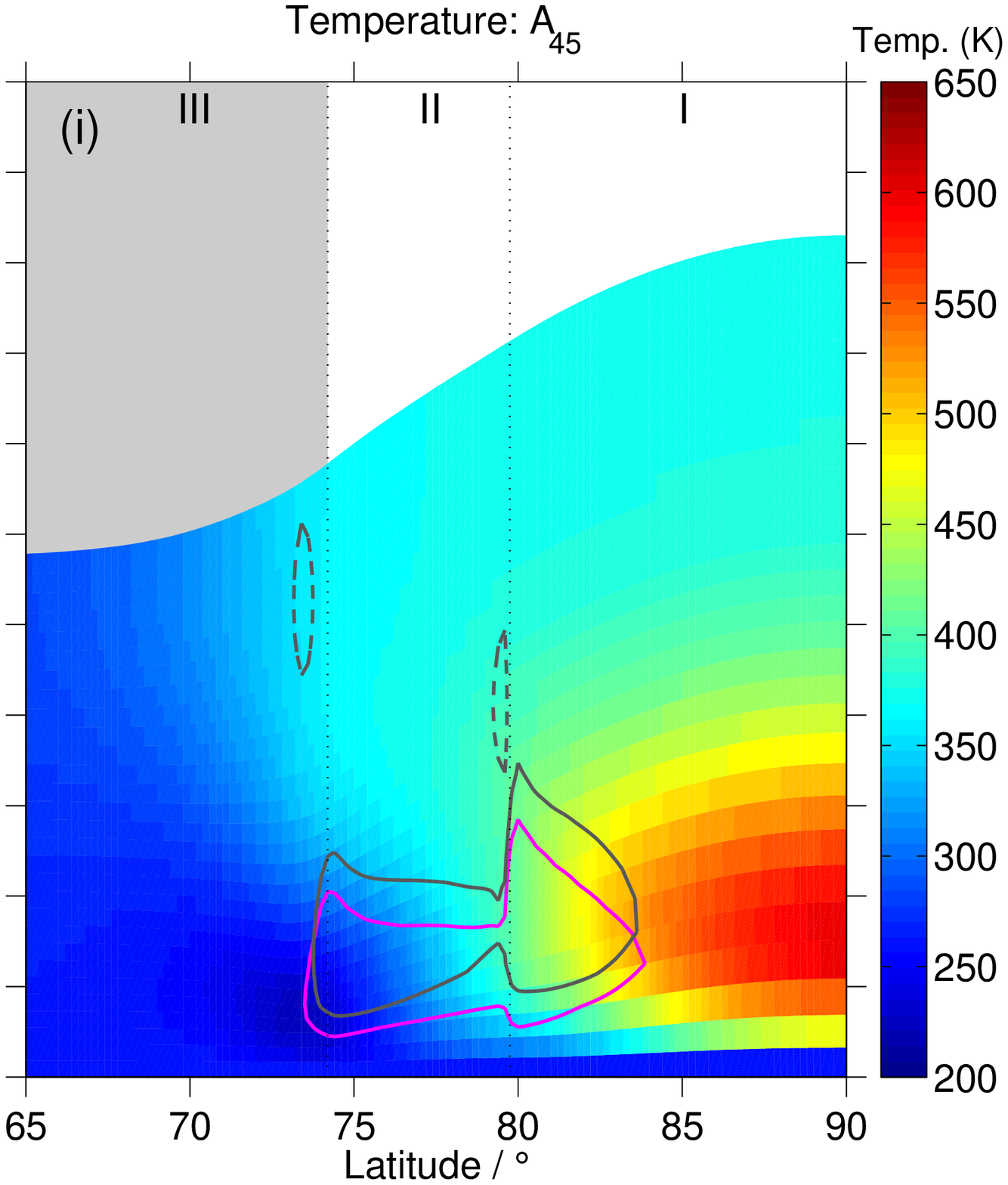}

    \caption{ Figures a-f show thermospheric and magnetospheric angular velocities, height-integrated Pedersen 
    conductivity, `slippage' parameter, azimuthally-integrated radial current, azimuthally-integrated Pedersen 
    current and \FAC density respectively for case A with $\unit{I_{\rho\infty}}{=}\unitSI[45]{MA}$ (\unitSI{A_{45}}) 
    represented by red lines and case A with $\unit{I_{\rho\infty}}{=}\unitSI[100]{MA}$ (\unitSI{A_{100}}) 
    represented 
    by blue lines. Note that case \unitSI{A_{100}} is the same as case A in section~\ref{sec:results}. For (a) the 
    solid lines represent \Rev{the} thermospheric angular velocity and \Rev{the} dashed lines represent \Rev{the} 
    magnetospheric angular velocity. Magnetospheric regions \Rev{(region III is shaded)} 
    are labelled and separated by dotted black lines. Figures g-i show 
    \Rev{this} thermospheric azimuthal velocity, meridional velocity and temperature distributions in the high latitude region 
    for case \unitSI{A_{45}}. Arrows, contours and colour bars are the same as in \Fig{\ref{fig:thermosphere}}.}
    
    \label{fig:pathongen}
  \end{figure*}

  \Fig{\ref{fig:pathongen}a} shows the influence of \unit{I_{\rho\infty}} on the \unit{\Omega_M} and \unit{\Omega_T} 
  values. Case \unitSI{A_{100}} with the higher \unit{I_{\rho\infty}} value also has a higher torque 
  on the disc plasma, which results in a smaller difference \unit{\Omega_T - \Omega_M} near the disc boundary 
  (\unitSI[74{-}75]{^{\circ}}). Equatorward of \unitSI[74]{^{\circ}}, the \unit{\Omega_T} and \unit{\Omega_M} 
  profiles are very similar for both cases. The outer disc region \unitSI[74{-}75]{^{\circ}} thus develops a strong 
  \FAC in case \unitSI{A_{100}} in order to satisfy the higher \unit{I_{\rho\infty}} imposed.\\
  
  Due to the smaller values of \unit{\Omega_{M}} and radial current for case \unitSI{A_{45}}, Pedersen 
  conductivity values (\Fig{\ref{fig:pathongen}b}) are significantly smaller near 
  the region III / II boundary compared to case \unitSI{A_{100}}. The different gradients in \unit{\Omega_M} and 
  \unit{\Omega_T} for case \unitSI{A_{45}} cause a slight equatorward shift in the conductivity peak 
  compared to \unitSI{A_{100}}. The `slippage' parameter for \unitSI{A_{45}} in \Fig{\ref{fig:pathongen}c} differs 
  only slightly from that of case \unitSI{A_{100}} due to its smaller \unit{\Omega_T - \Omega_M} differences. \\
  
  The azimuthally-integrated radial, Pedersen and FACs for cases \unitSI{A_{45}} (red 
  line) and \unitSI{A_{100}} (blue line) are shown in \Figs{\ref{fig:pathongen}d-f} respectively. The other labels 
  are the same as in \Fig{\ref{fig:currents}}. For \unitSI{A_{100}} the magnetosphere near-rigidly corotates with 
  \unit{\Omega_{J}} throughout regions IV and most of III \citep{nichols04}. The \unitSI{A_{45}} radial current 
  profile resembles those for expanded cases, due to the magnetosphere 
  sub-corotating to a greater degree (see \Fig{\ref{fig:pathongen}a}). The Pedersen current for case \unitSI{A_{45}} 
  has a smooth, almost linear transition across and through regions III and II as opposed to the abrupt 
  cutoff at the region III / II boundary in \unitSI{A_{100}}. The \unitSI{A_{45}} profile quantitatively resembles 
  Pedersen currents for expanded cases and those in \citet{cowley07}. For \unitSI{A_{45}}, \FAC profiles are similar 
  to those for \unitSI{A_{100}} with the exception that the magnitude of the peaks just 
  equatorward of the region III / II boundary and the trough are significantly smaller. For lower 
  \unit{I_{\rho\infty}} values, then, the main auroral oval would be significantly dimmer, and possibly more 
  similar to the putatively weak auroral signature at the region I / II boundary.

\subsubsection{Thermospheric distributions for case A}
\label{sec:path_distA}

  Here we discuss the changes in azimuthal and meridional velocity as well as the temperature 
  distribution, which arise from setting $\unit{I_{\rho\infty}}{=}\unitSI[45]{MA}$ for case A. 
  All conventions and colours in \Figs{\ref{fig:pathongen}g-h} are the same as those in 
  \Fig{\ref{fig:thermosphere}}.\\

  \Figs{\ref{fig:pathongen}g-h} show the distribution of azimuthal and meridional velocities across the 
  high latitude region for case \unitSI{A_{45}}. In \unitSI{A_{100}} (\Fig{\ref{fig:thermosphere}a}) there is a large 
  sub-corotational jet 
  in regions II and I with the strongest degree of sub-corotation just poleward of the region II / I boundary. In 
  \unitSI{A_{45}}, the large sub-corotating jet now has two regions of strong sub-corotation, the new one being just 
  poleward of the region III / II boundary. These two strong sub-corotational jets within the larger jet are 
  evident in \Fig{\ref{fig:pathongen}a} at the region boundaries (dotted black lines), where there are large 
  changes in magnetospheric angular velocity. Meridional velocities for case \unitSI{A_{45}} follow the same trend as 
  in \unitSI{A_{100}} (\Fig{\ref{fig:thermosphere}d}) where there is a poleward flow at low altitudes and an 
  opposite flow at high altitudes. The main difference between \Rev{the} meridional flows is that localised accelerated flows 
  (high altitude in region III and low altitude in regions II and I) have larger velocities in case 
  \unitSI{A_{45}} due to larger advection terms.\\
   
  \Fig{\ref{fig:pathongen}i} shows the temperature distribution for case \unitSI{A_{45}}. Comparing this 
  with case \unitSI{A_{100}} (\Fig{\ref{fig:thermosphere}g}) indicates that energy input via Joule heating (magenta 
  contours) and ion drag (solid grey contours) is greater in \unitSI{A_{45}}. The larger energy input, predominantly 
  in region II, is caused by larger shear between thermospheric and magnetospheric angular velocities. This leads 
  to a slight increase in thermospheric temperature \Rev{(${\sim}\unitSI[6]{\%}$)} most evident in region I, the polar `hotspot' 
  into which auroral heat energy is transported by meridional winds.

\subsection{Outer boundary conditions for the baseline magnetosphere}
\label{sec:bsn_pathogen}

  For our baseline \Rev{case, case B}, the smallest variance in current profiles occurred with 
  $\unit{I_{\rho\infty}}{=}\unitSI[68]{MA}$. We compare this case \unitSI{B_{68}} with the original \unitSI{B_{100}} 
  case in \Fig{\ref{fig:pathB}}.

\subsubsection{Comparison of angular velocities and currents for case B}
\label{sec:path_angcurB}

   As for the compressed magnetosphere, we compare angular velocities and current-related parameters of the 
   \unitSI{B_{68}} and \unitSI{B_{100}} models in the \unitSI[72{-}75]{^{\circ}} latitude range where significant 
   differences arise. \Fig{\ref{fig:pathB}a} compares the variation of magnetospheric and thermospheric angular velocities 
   for cases \unitSI{B_{68}} and \unitSI{B_{100}}. In region III, both the magnetosphere and 
   thermosphere for \unitSI{B_{68}} are slightly sub-corotating compared to the \unitSI{B_{100}}. The Pedersen conductivity 
   and `slippage' parameter for \unitSI{B_{68}} \Rev{also} have similar profiles, but with smaller 
   magnitudes in region III, to those for \unitSI{B_{100}}. These minor differences are caused by smaller 
   thermospheric and magnetospheric angular velocities in region III.\\

   \Fig{\ref{fig:pathB}d-e} show azimuthally-integrated radial and Pedersen currents. \Rev{The corresponding \FAC density 
   as a function of latitude is shown in \Fig{\ref{fig:pathB}f}.} The radial current profile is smaller in magnitude than that 
   of \unitSI{B_{100}}. The \unitSI{B_{68}} Pedersen current has a single small peak at the region III / II boundary 
   in contrast with the sharp peak in \unitSI{B_{100}}. The \FAC density has a smaller, slightly broader peak in 
   region III, 
   suggesting a low intensity auroral oval compared to the \unitSI{B_{100}} case. The absence of strong downward 
   \FAC for \unitSI{B_{68}} also suggests that the method used to join the region 
   III currents with the region II currents could produce artefacts for relatively large values of 
   \unit{I_{\rho\infty}}.\\

   As with case A, the fine structure around the boundary between the middle and outer magnetosphere in 
   \Figs{\ref{fig:pathB}a-f} has been removed by decreasing the value of \unit{I_{\rho\infty}} from \unitSI[100]{MA} 
   to \unitSI[68]{MA} for our baseline case.

\subsubsection{Thermospheric distributions for case B}
\label{sec:path_distB}

    Here we discuss the minor changes in Jupiter's thermospheric flows and temperature distribution made by 
    changing the radial current boundary condition value for our baseline case. All conventions and colours are the 
    same as those in \Fig{\ref{fig:thermosphere}}. \\ 
     
\begin{figure*}
    \centering 
    \includegraphics[width=0.66\figwidth]{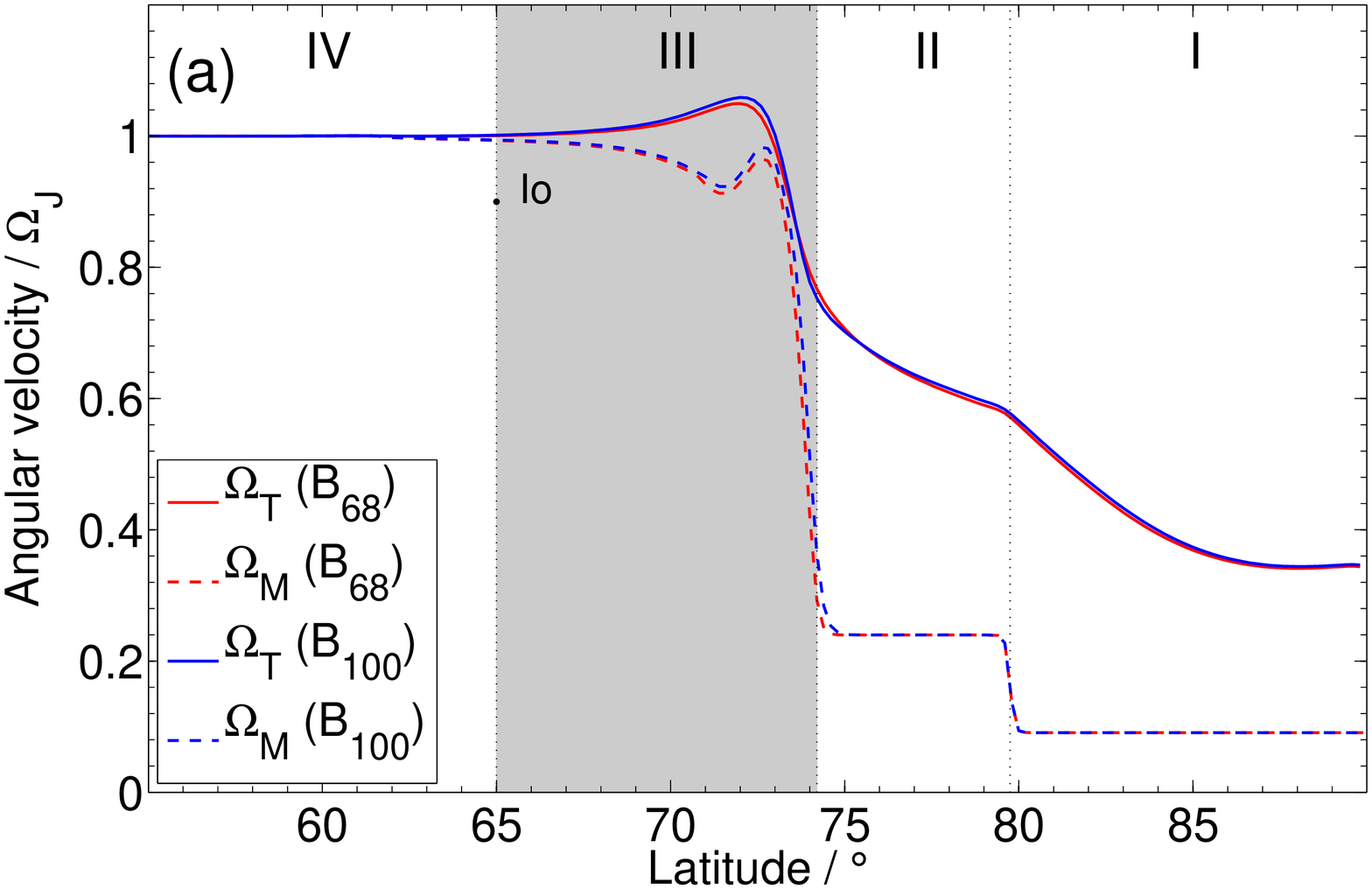}
    \includegraphics[width=0.65\figwidth]{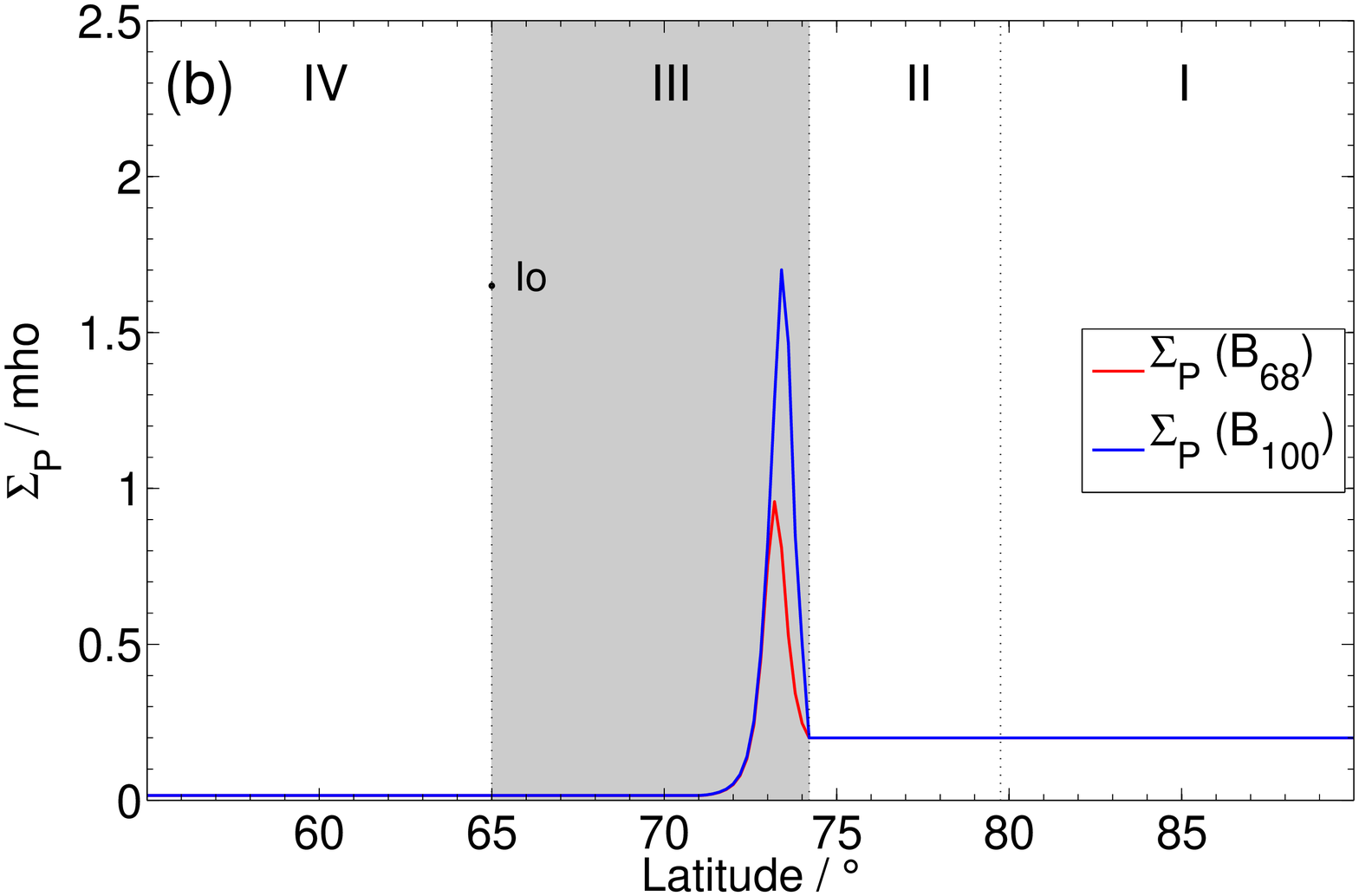}
    \includegraphics[width=0.64\figwidth]{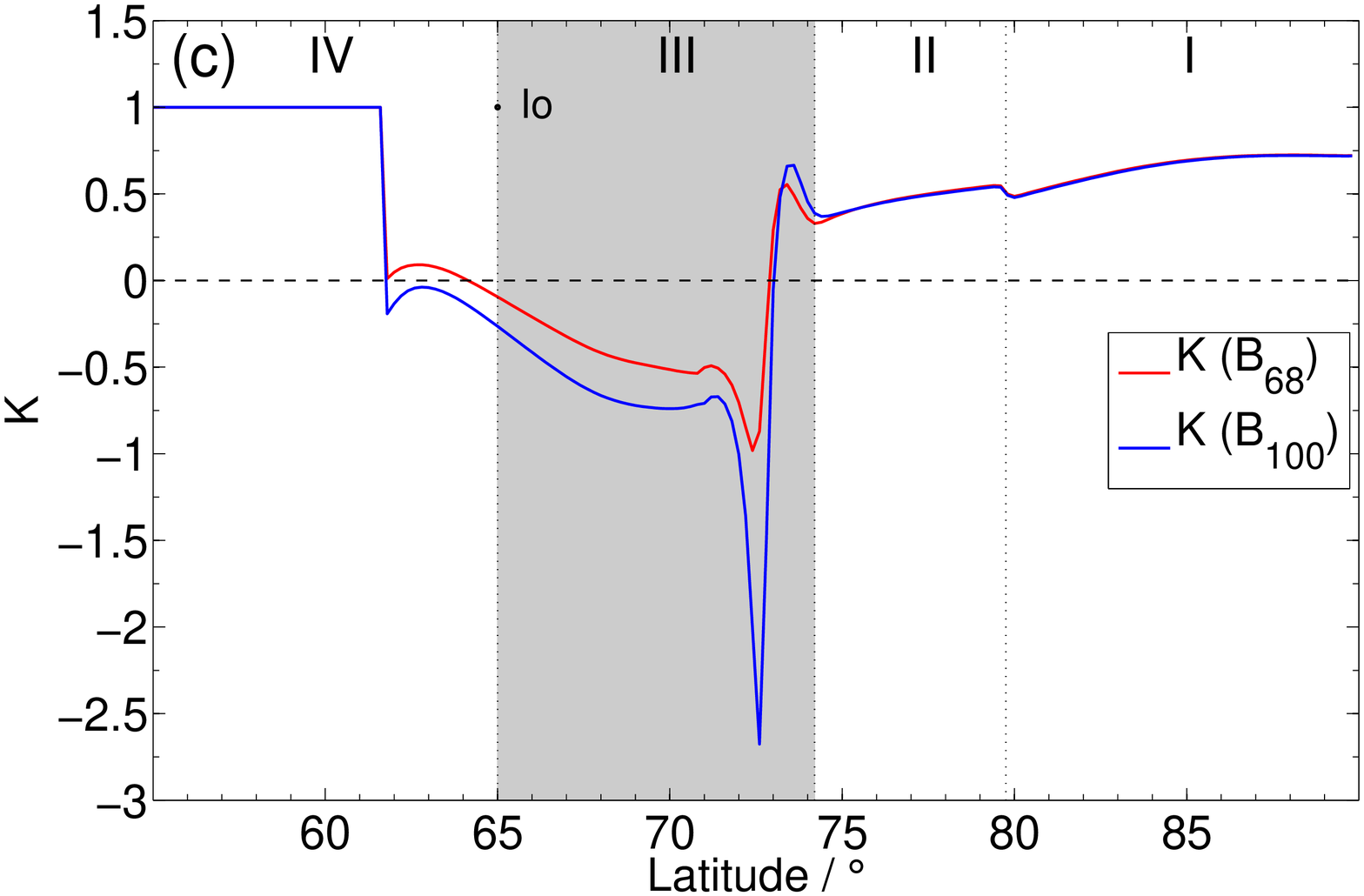}\\

    \includegraphics[width=0.66\figwidth]{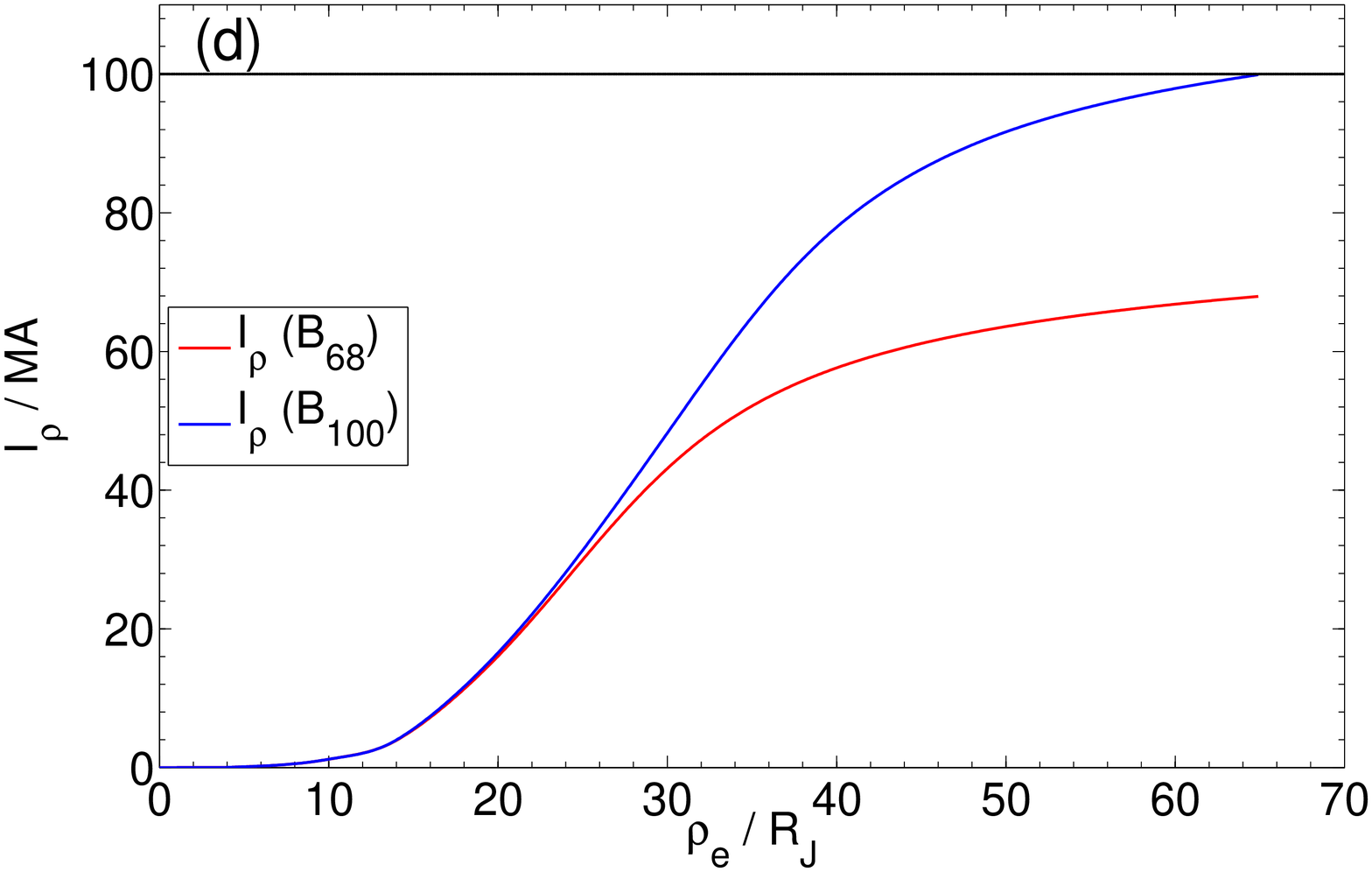}
    \includegraphics[width=0.64\figwidth]{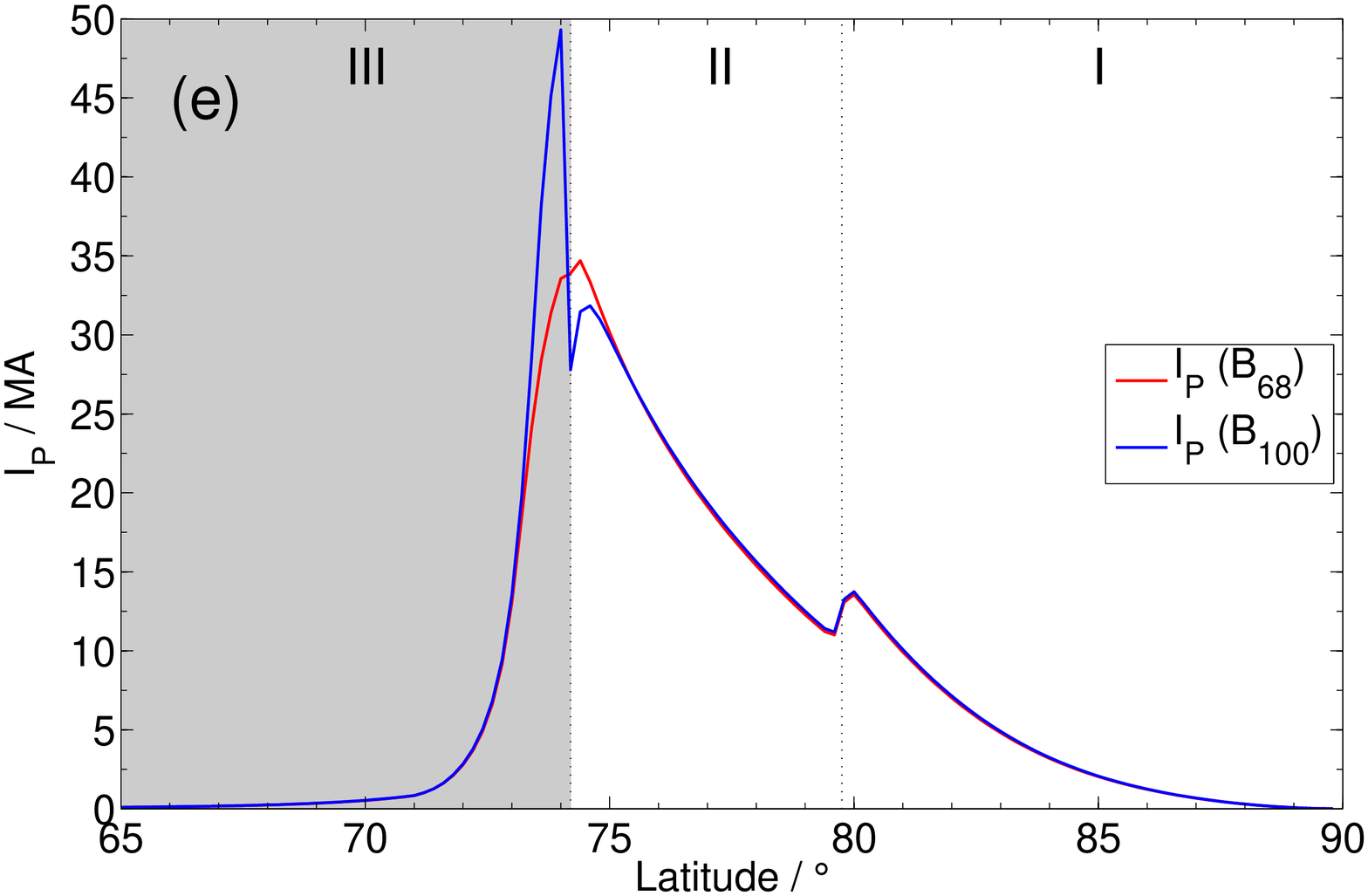}
    \includegraphics[width=0.66\figwidth]{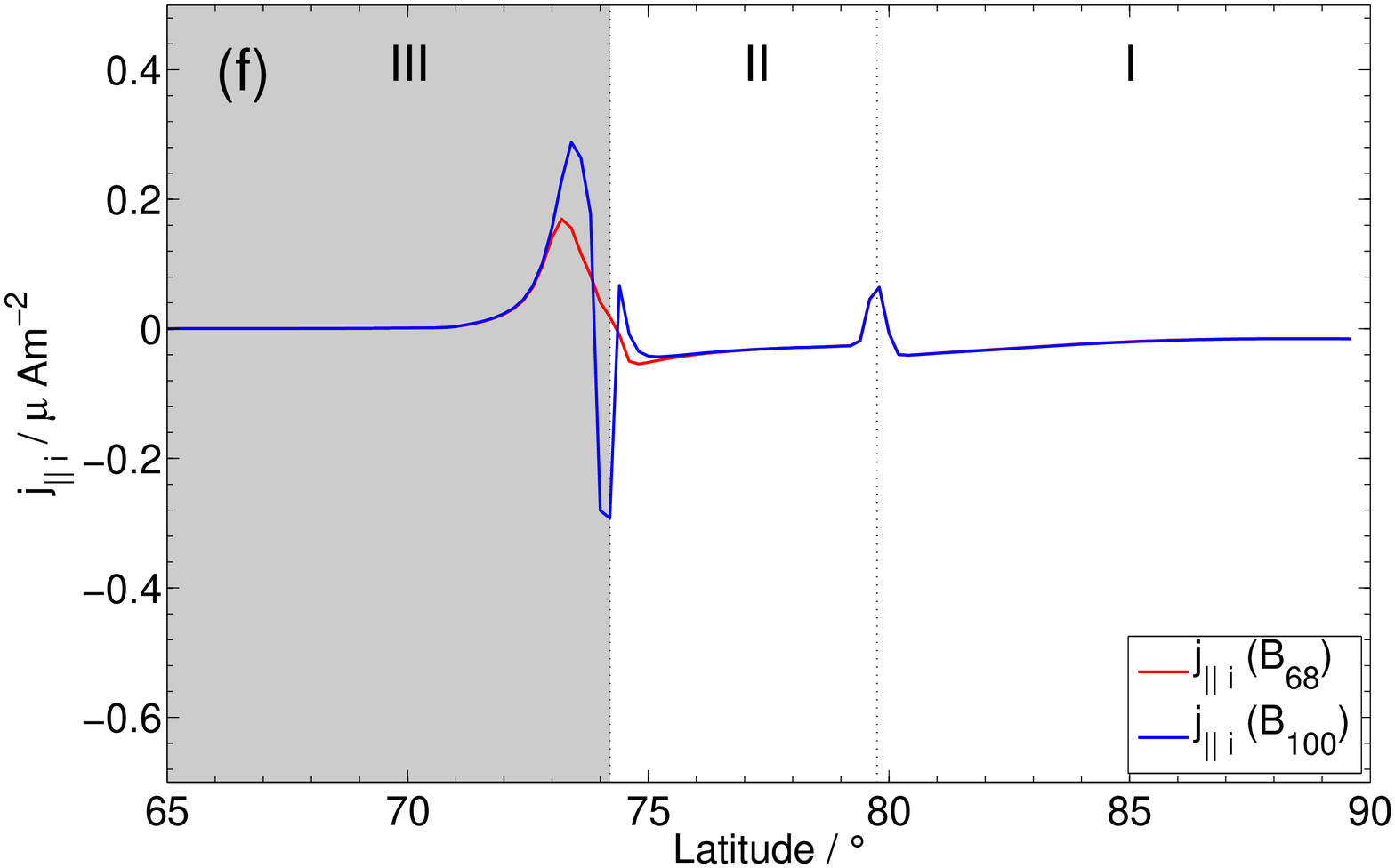}\\

    \includegraphics[width=0.7\figwidth]{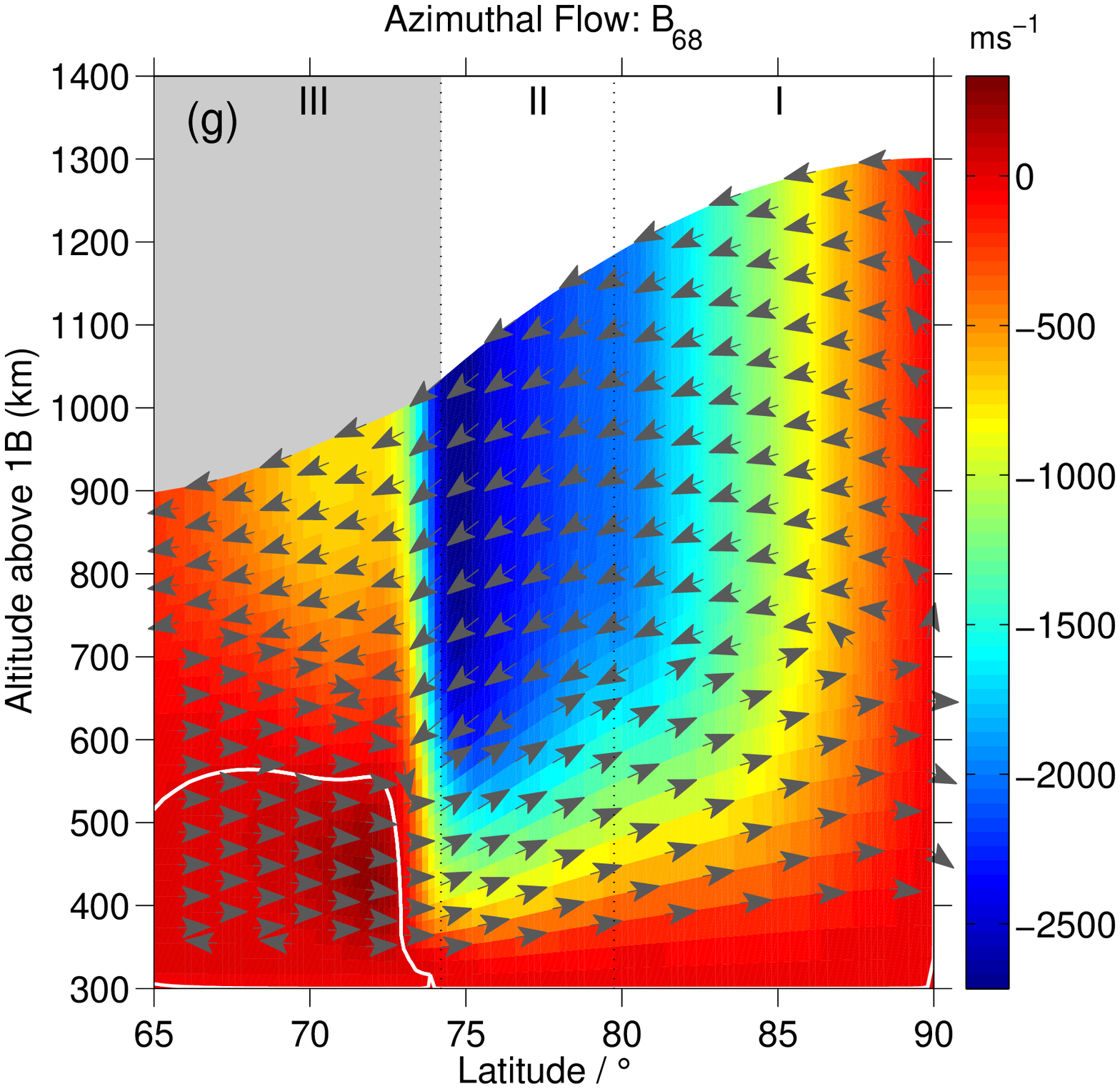}
    \includegraphics[width=0.6\figwidth]{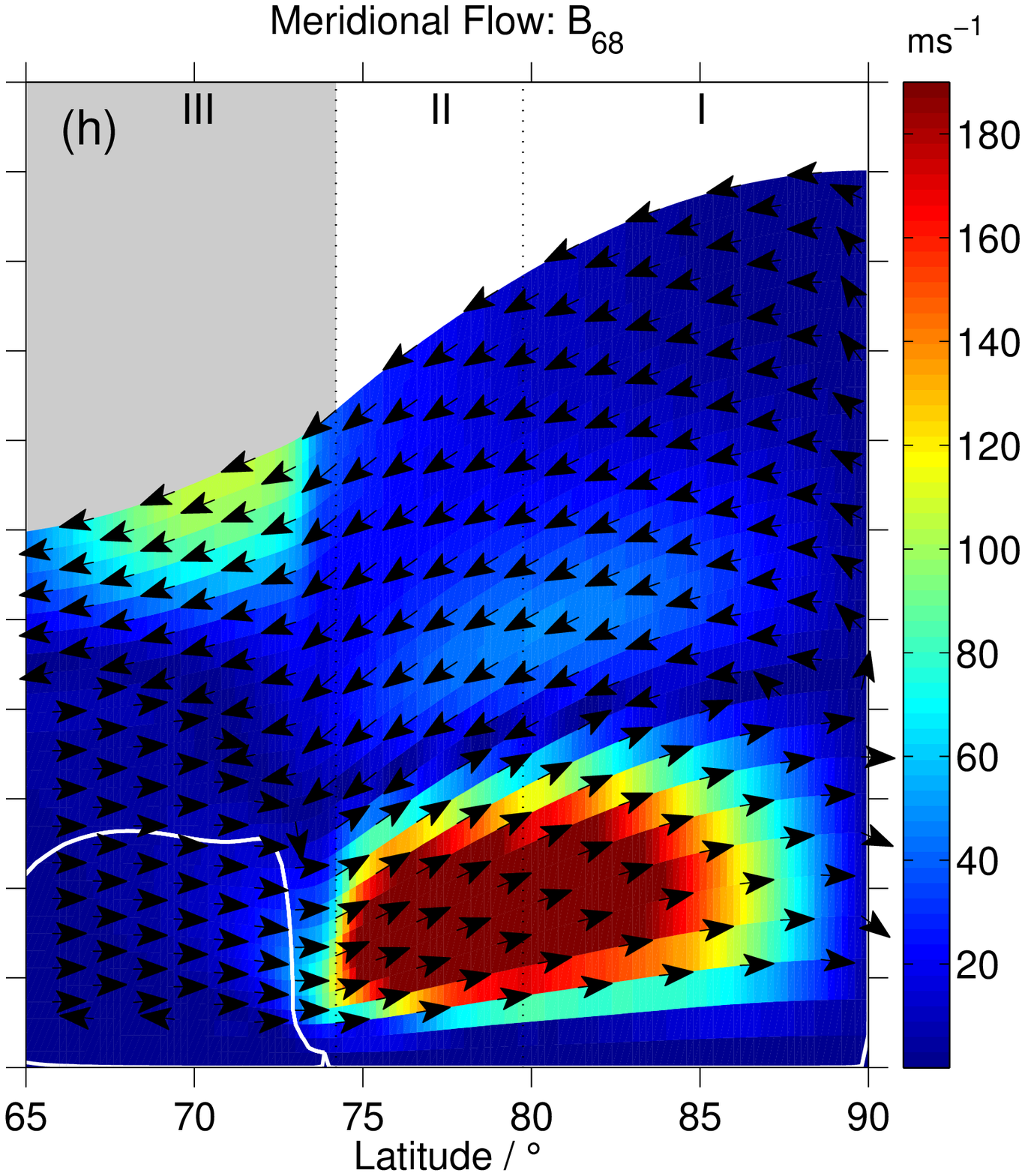}
    \includegraphics[width=0.58\figwidth]{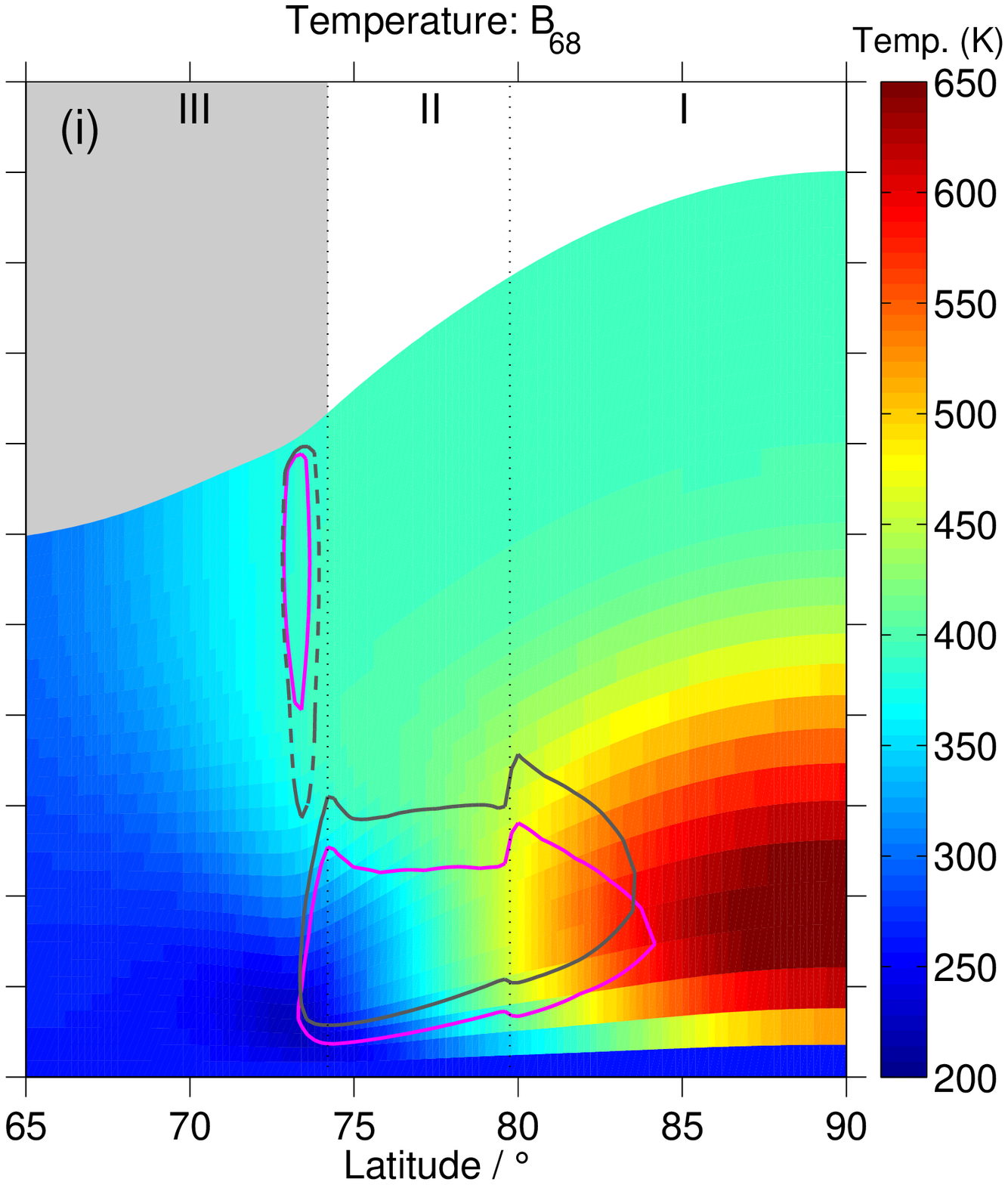}

    \caption{Figures a-f show thermospheric and magnetospheric angular velocities, height-integrated Pedersen 
    conductivity, `slippage' parameter, azimuthally-integrated radial current, azimuthally-integrated Pedersen 
    current and \FAC density respectively for case B with $\unit{I_{\rho\infty}}{=}\unitSI[68]{MA}$ 
    (\unitSI{B_{68}}) represented by red lines and case B with $\unit{I_{\rho\infty}}{=}\unitSI[100]{MA}$ 
    (\unitSI{B_{100}}) represented by blue lines. Note that case \unitSI{B_{100}} is the same as case B in 
    section~\ref{sec:results}. For (a) the solid lines represent \Rev{the} thermospheric angular velocity and \Rev{the} 
    dashed lines 
    represent \Rev{the} magnetospheric angular velocity. Magnetospheric regions \Rev{(region III is shaded)} 
    are labelled and separated by dotted black 
    lines. Figures g-i show \Rev{this} thermospheric azimuthal velocity, meridional velocity and temperature distribution in 
    the high latitude region for case \unitSI{B_{68}}. Arrows, contours and colour bars are the same as in 
    \Fig{\ref{fig:thermosphere}}.}
      \label{fig:pathB}
  \end{figure*}

    The azimuthal velocity in the high latitude region is shown in 
    \Fig{\ref{fig:pathB}g}. We expect a slight increase in sub-corotation throughout 
    region III (see \Fig{\ref{fig:pathB}a}) compared to \unitSI{B_{100}}. This is evident by comparing 
    \Fig{\ref{fig:pathB}g} with 
    \Fig{\ref{fig:thermosphere}a} where we can see that the region of super-corotation (dark red) has diminished for 
    \unitSI{B_{68}}. The meridional velocity distribution is shown in \Fig{\ref{fig:pathB}h}. The high 
    altitude localised accelerated flow in region III is slightly faster than in case \unitSI{B_{100}} because the 
    pressure gradient and advection terms are \unitSI[27{-}40]{\%} larger in this region of \unitSI{B_{68}}. This 
    would lead to a minimal temperature increase \Rev{${\sim}\unitSI[3]{\%}$}, most notably in regions II and I 
    (see \Fig{\ref{fig:pathB}i}).

\subsection{Outer boundary conditions for an expanded magnetosphere}
\label{sec:exp_pathogen}

    For our expanded configuration, we found that $\unit{I_{\rho\infty}}{=}\unitSI[80]{MA}$ gave a smooth 
    profile (least variance in \FAC density). This change in \unit{I_{\rho\infty}} produced corresponding changes in 
    model outputs, which are far less significant than those from our compressed case. Both magnetospheric and 
    thermospheric angular velocities in case \unitSI{C_{80}} have slightly smaller 
    magnitudes in the magnetodisc region when compared to \unitSI{C_{100}}. The current profiles calculated by 
    setting $\unit{I_{\rho\infty}}{=}\unitSI[80]{MA}$ produced currents strongly resembling those for case 
    \unitSI{B_{68}}. Peak values for currents at the region III / II boundary are ${\sim}\unitSI[70\mbox{--}80]{\%}$ 
    of those in case \unitSI{C_{100}}. Thermospheric flows for \unitSI{C_{80}} differ slightly from \unitSI{C_{100}}\Rev{,} 
    most significantly \Rev{in} the larger degree of sub-corotation in \unitSI{C_{80}}. The `hotspot' in the polar region is 
    ${\sim}\unitSI[2]{\%}$ hotter than in case \unitSI{C_{100}} because of faster poleward flows transporting heat 
    more efficiently. These faster flows are due to stronger advection in case \unitSI{C_{80}} producing stronger 
    acceleration compared to \unitSI{C_{100}}.

\subsection{Effect of outer boundary conditions on ionospheric powers}
\label{sec:path_powers}

    In this section, we examine figures for ionospheric power per unit area and their respective integrated 
    power per hemisphere for cases \unitSI{A_{45}}, \unitSI{B_{68}} and \unitSI{C_{80}}.\\
    
\begin{figure*}
    \centering 

    \includegraphics[width=0.695\figwidth]{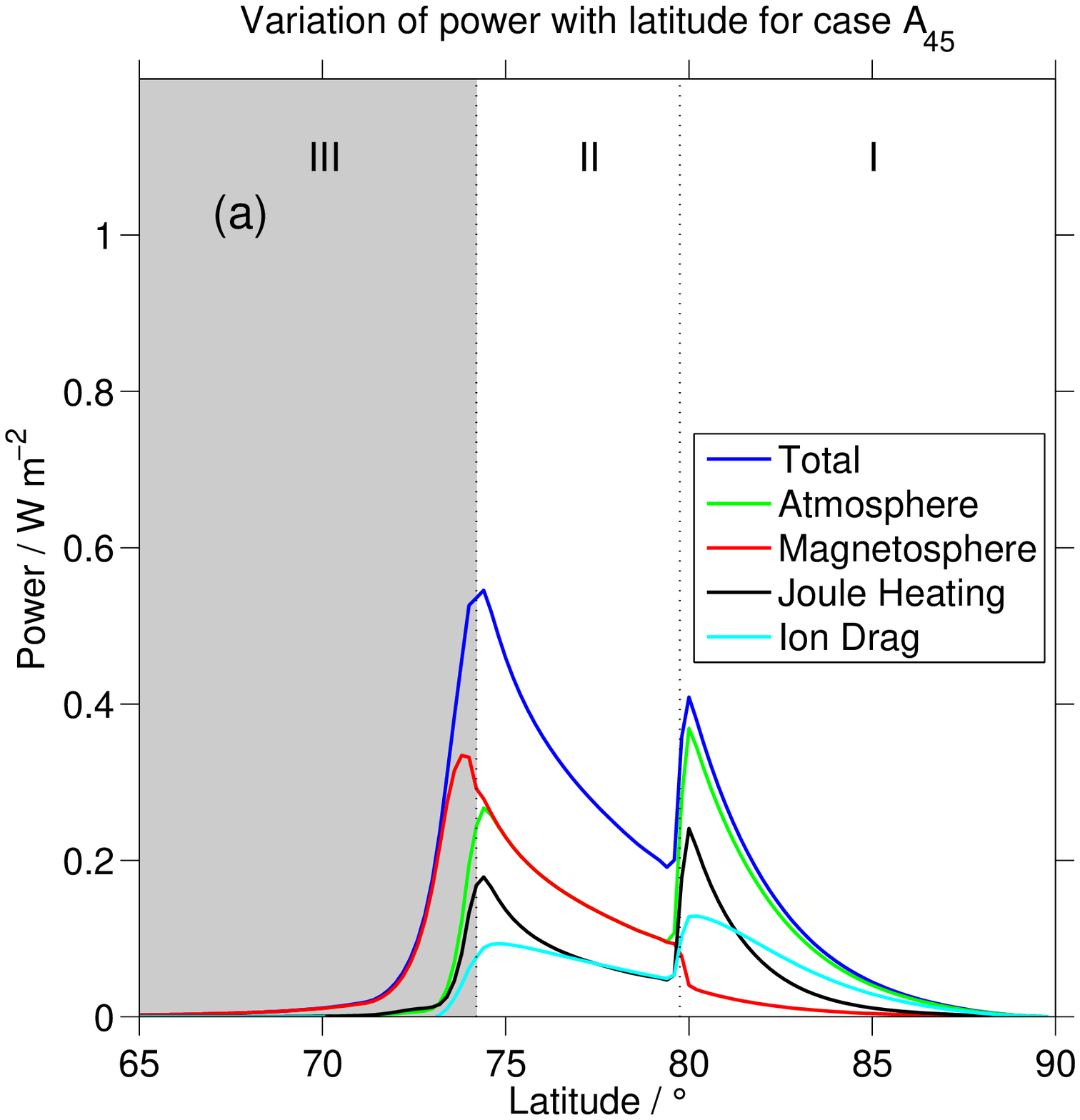}
    \includegraphics[width=0.620\figwidth]{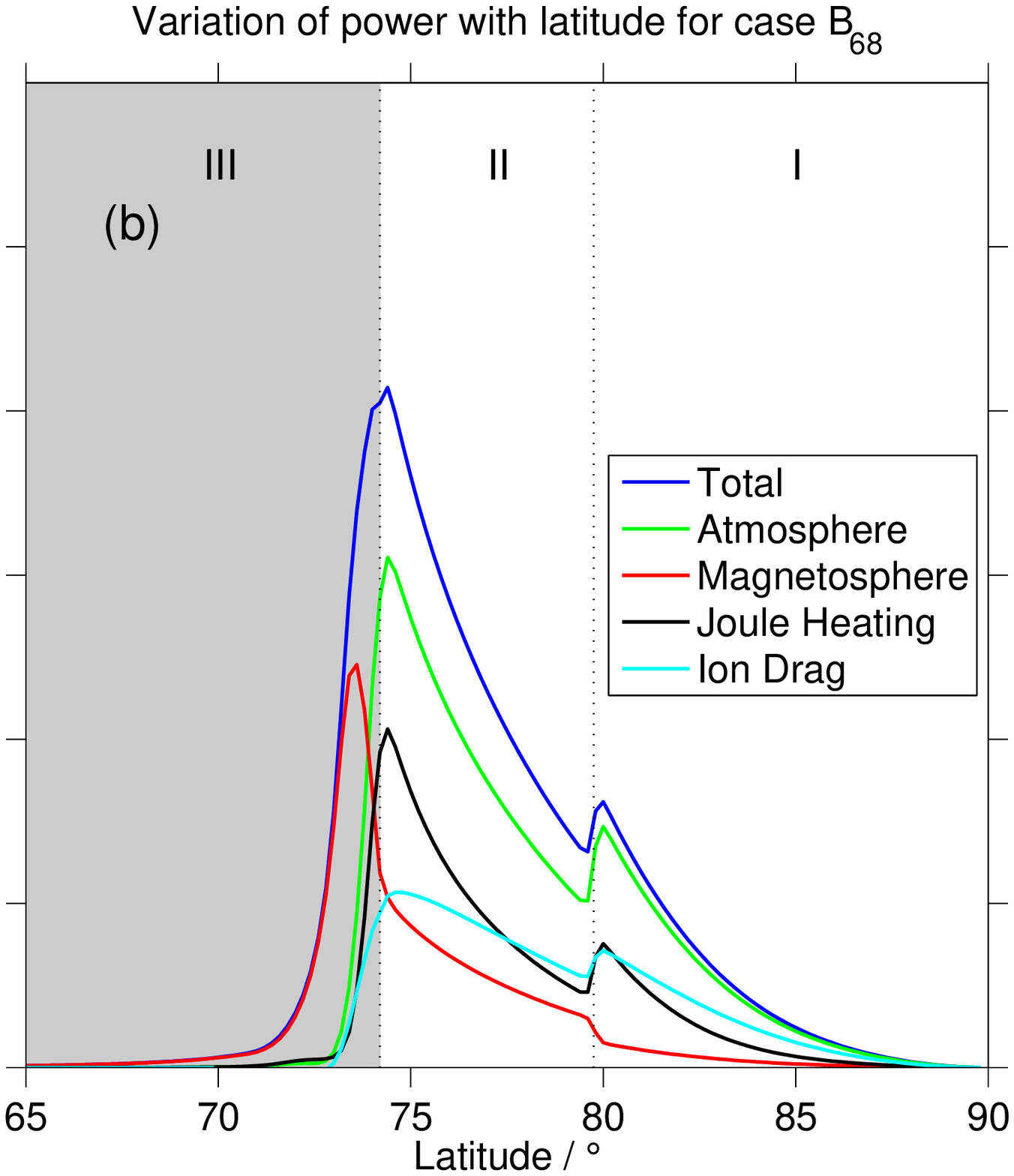}
    \includegraphics[width=0.620\figwidth]{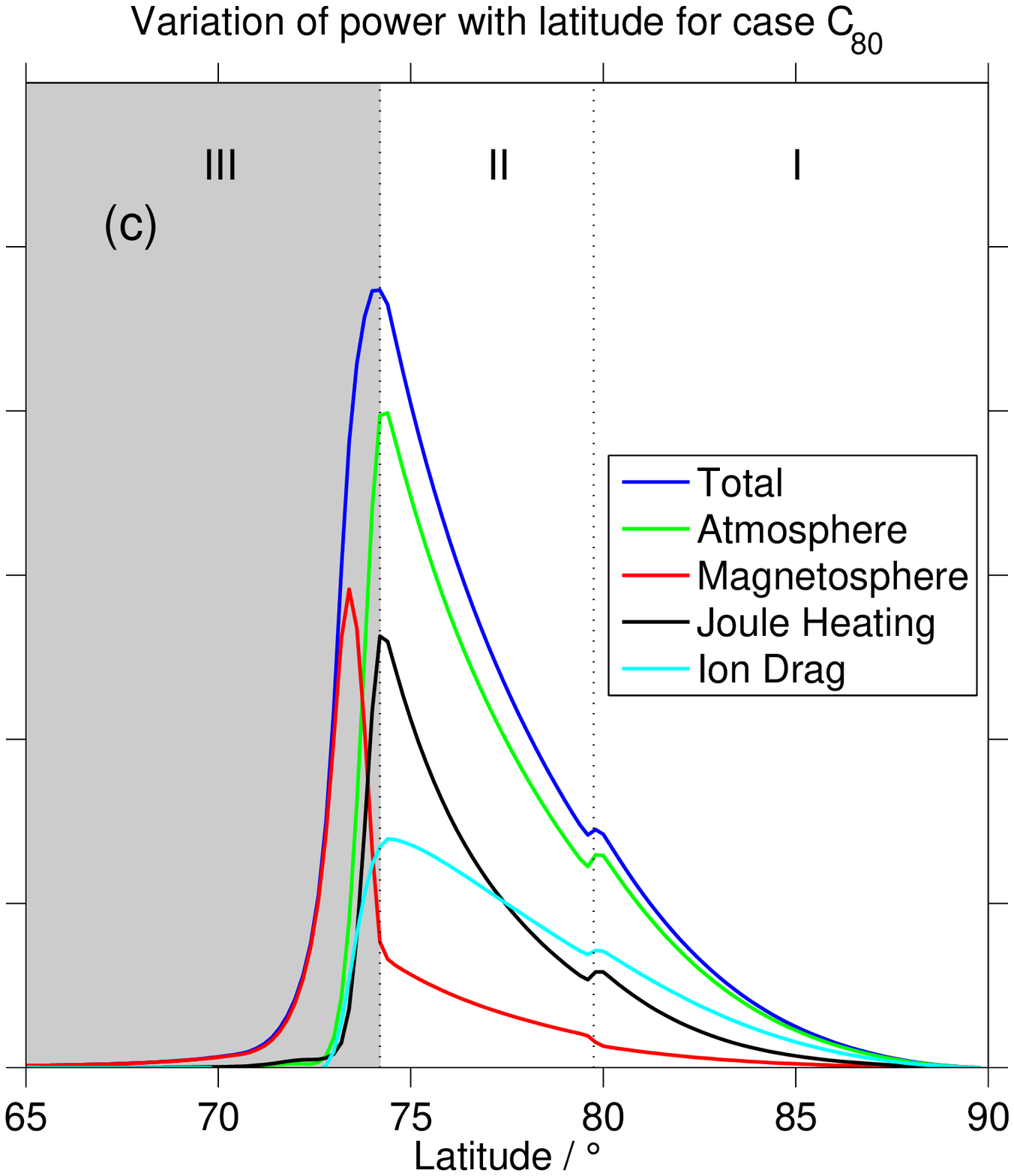}\\

    \includegraphics[width=0.705\figwidth]{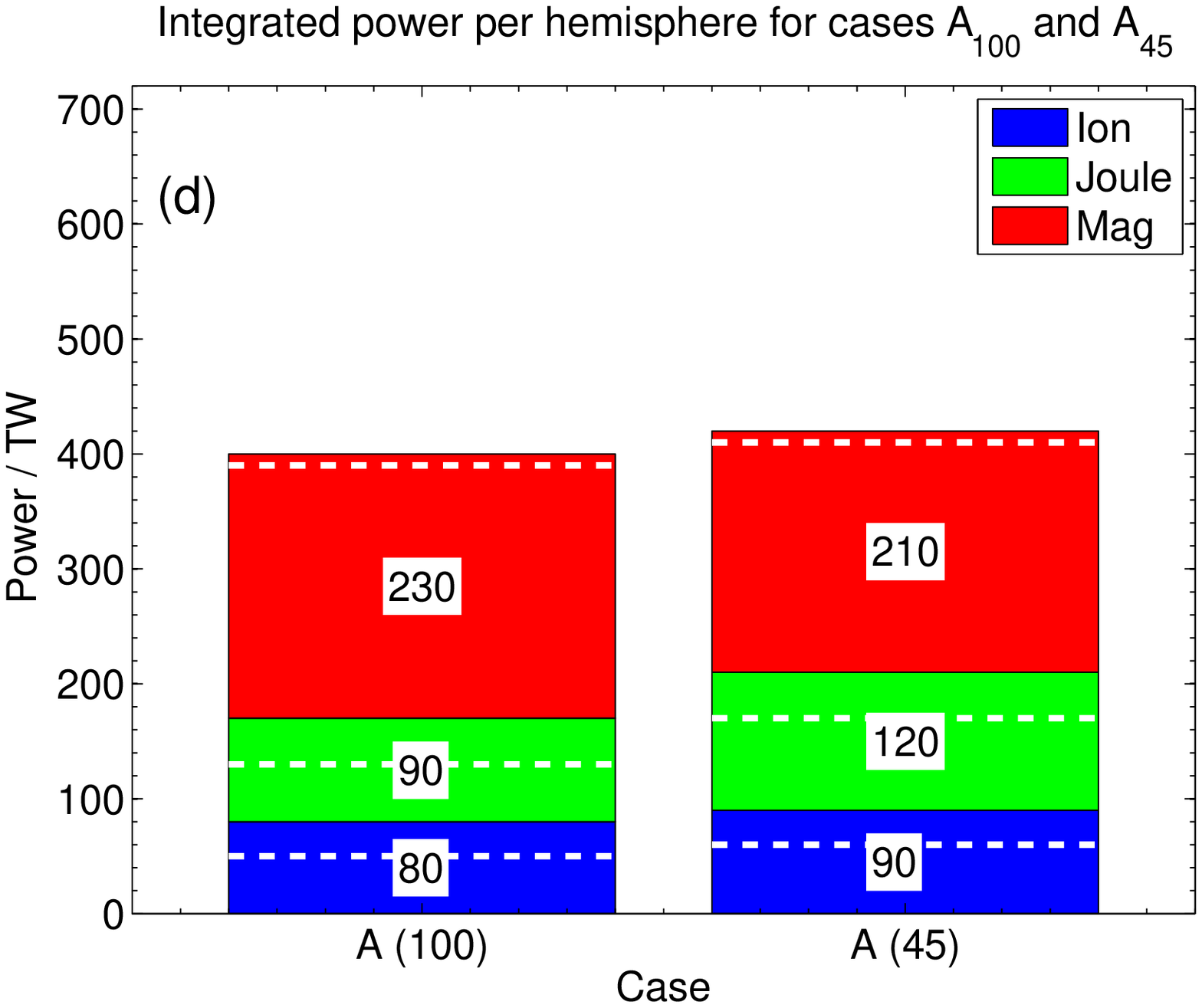}
    \includegraphics[width=0.62\figwidth]{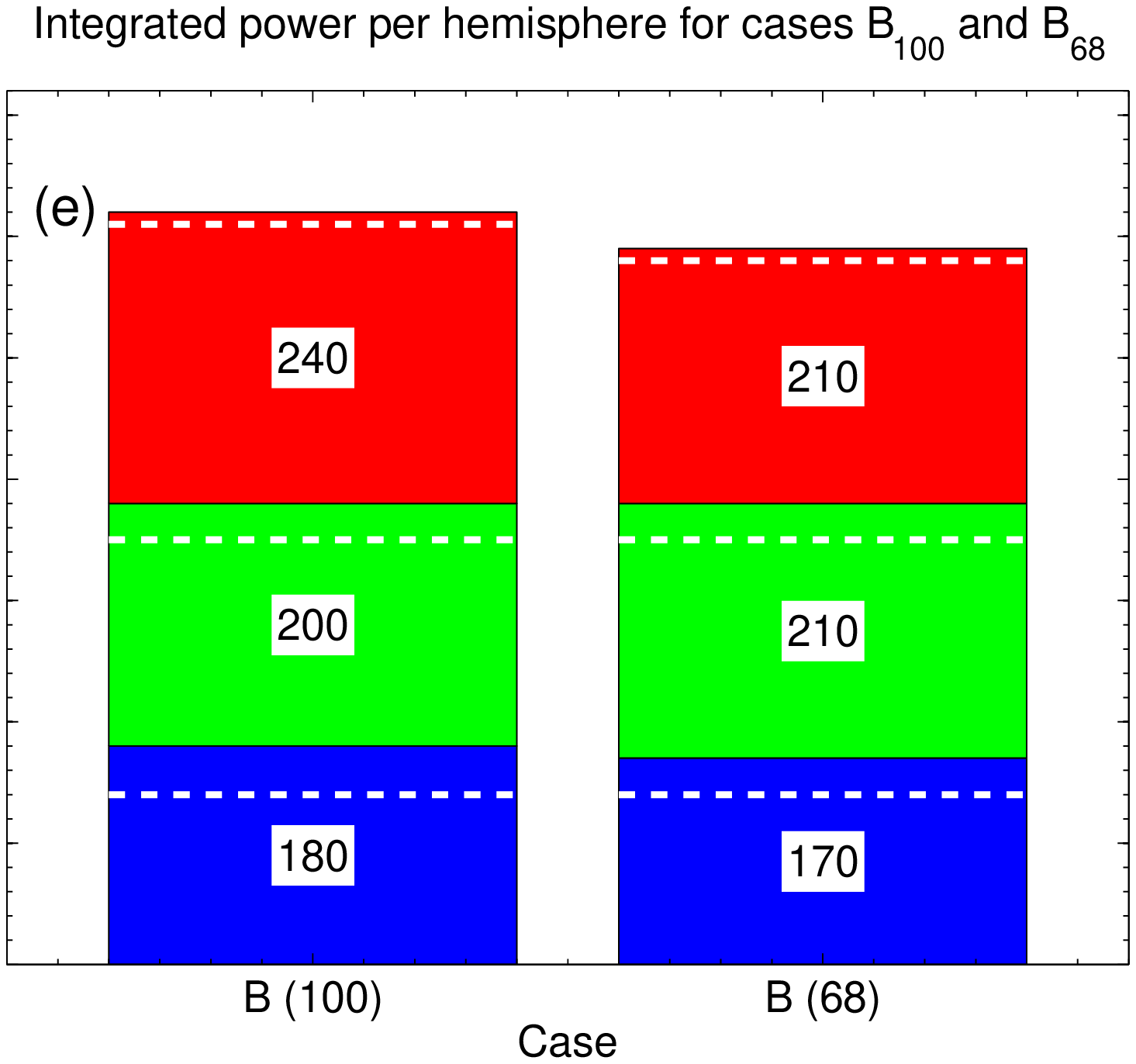}
    \includegraphics[width=0.62\figwidth]{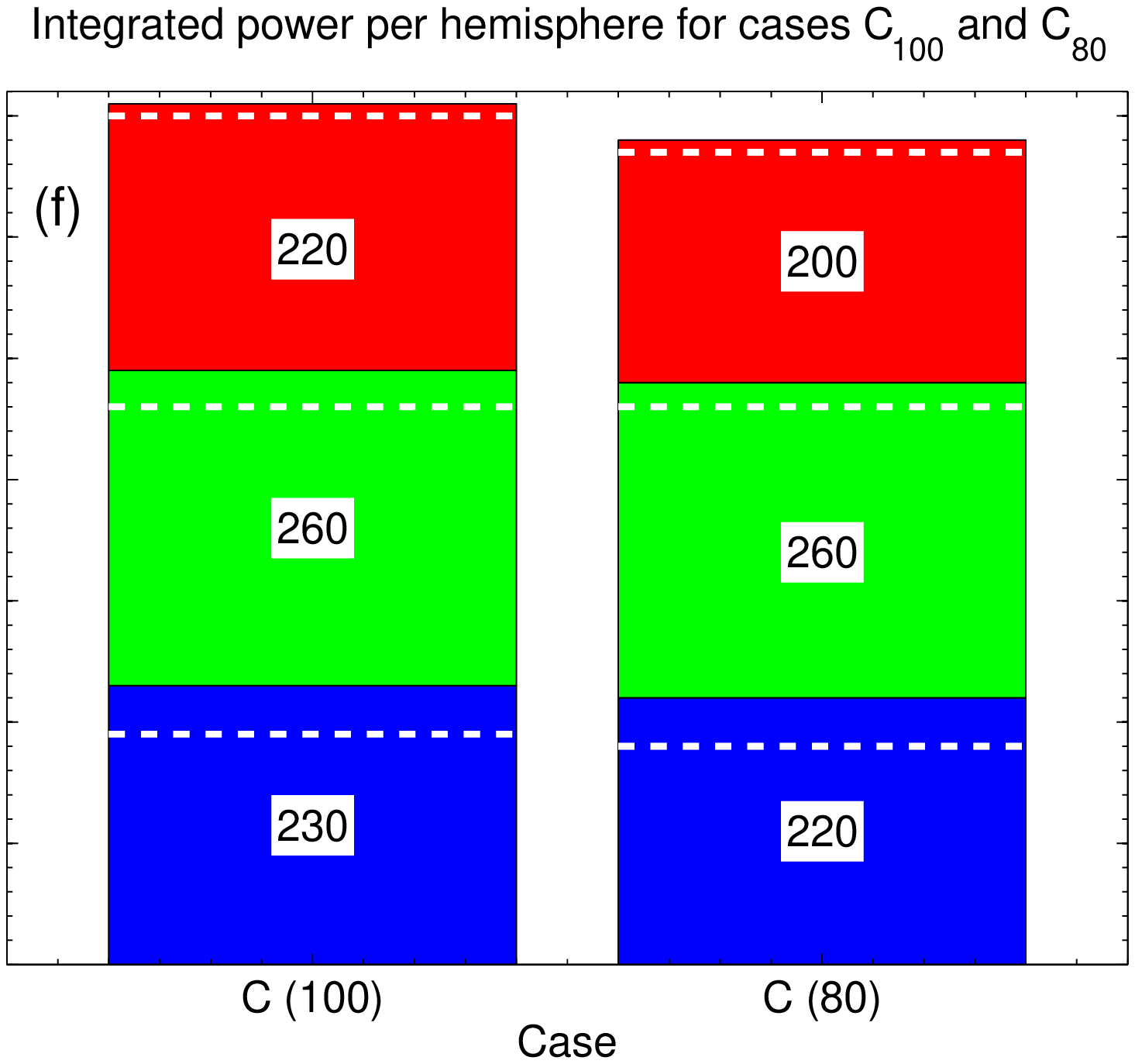}

    \caption{Figures a-c show ionospheric powers per unit area in the high latitude region for cases \unitSI{A_{45}}, 
    \unitSI{B_{68}} and \unitSI{C_{80}} respectively. Total power per unit area is represented by the blue line, 
    magnetospheric power by the red line, atmospheric power by the green line, Joule heating by the black line and 
    ion drag power by the cyan line. Magnetospheric regions are labelled and separated by dotted black lines. 
    Figures d-f show integrated powers per hemisphere for cases 
    \unitSI{A_{45}}, \unitSI{B_{68}} and \unitSI{C_{80}} respectively. Ion drag, Joule heating and magnetospheric 
    powers 
    are indicated by blue, green and red bars. Powers in the closed field line regions lie below the white dashed 
    line whilst powers in the open field regions lie above it. Total power dissipated for each mechanism (in TW) 
    is printed on its respective colour bar. }
      \label{fig:pathP}
  \end{figure*}

  The power per unit area for case \unitSI{A_{45}} in the high latitude region is shown in 
  \Fig{\ref{fig:pathP}a}. Colour conventions and labels are the same as those in 
  \Figs{\ref{fig:thermosphere}j-l}. Powers per unit area (in \Fig{\ref{fig:pathP}a}) are integrated over each 
  hemisphere and are shown in \Fig{\ref{fig:pathP}d}. The transition between region III and II is the most 
  interesting for comparison; 
  \unitSI{A_{100}} has a large prominent peak in magnetospheric power whereas \unitSI{A_{45}} has a significantly 
  reduced peak due to smaller values of \unit{\Omega_M}. In region II $\unit{\Omega_M}{\sim}\unitSI[0.5]{\Omega_J}$ 
  (see \Fig{\ref{fig:pathongen}a}) implying that magnetospheric and atmospheric power in this region are equal. 
  Joule heating and ion drag are thus increased in \unitSI{A_{45}} compared to \unitSI{A_{100}} to meet the above 
  requirement. These results suggest that smaller \unit{I_{\rho\infty}} values will generally dissipate more heat 
  in the atmosphere but less efficiently maintain corotation in the magnetosphere. \\
  
  For cases \unitSI{B_{68}} and \unitSI{C_{80}} the power per unit area and integrated powers per hemisphere 
  are shown in \Figs{\ref{fig:pathP}b-c and e-f} respectively. In comparing these two 
  cases with \unitSI{B_{100}} and \unitSI{C_{100}} we find only small differences in atmospheric powers (Joule 
  heating 
  and ion drag), predominantly at the region III / II boundary where there are two peaks with a 
  small trough in between. The integrated atmospheric powers per hemisphere thus remain relatively uniform with the 
  changes in \unit{I_{\rho\infty}} specified for all baseline (\unitSI{B_{68}} and \unitSI{B_{100}}) and expanded 
  (\unitSI{C_{80}} and \unitSI{C_{100}}) cases. The magnetospheric power per unit area for \unitSI{B_{68}} and 
  \unitSI{C_{80}} has significantly smaller magnitudes in region III compared to their 
  $\unit{I_{\rho\infty}}{=}\unitSI[100]{MA}$ counterparts. Therefore, for these configurations of the magnetosphere, 
  decreasing the value of \unit{I_{\rho\infty}} decreases the efficiency with which the atmosphere can accelerate the 
  magnetosphere towards corotation, but has no significant effect on atmospheric powers.

\section{Conclusion}
\label{sec:conclusion}

  In this study, we \Rev{have} expanded on the model of SA09 and described the effects of different 
  solar wind dynamic pressures on the coupled ionosphere-magnetosphere system at Jupiter. We constructed three 
  typical magnetospheric profiles 
  (see \Table{\ref{tb:cases}})\Rev{,} compressed, baseline (average) and expanded. These were then coupled to our global 
  two-dimensional thermospheric model \citep{smith08saturn} and a global conductivity model of the ionosphere 
  (GG). This allowed for a comparison with results from SA09, but also \Rev{provided} a first quantitative investigation 
  of how ionospheric, thermospheric and magnetospheric parameters were affected by differing solar wind conditions.\\

  Our results confirm many results from previous studies such as \Rev{those of} \Rev{\citet{southkiv2001,cowbun2003b} 
  and \citet{cowley07}}. We see an increase (resp. decrease) in thermospheric and magnetospheric angular velocities for 
  compressed (resp. expanded) magnetospheres relative to our baseline. The thermosphere 
  super-corotates just equatorward of the middle / outer magnetosphere boundary similarly to SA09. We solve for 
  \unit{\Omega_M} 
  self-consistently in the magnetodisc in all cases using the equations of disc dynamics. The \unit{\Omega_M} value 
  in the outer magnetosphere is a constant, dependent on disc radius i.e solar wind pressure \citep{cowley05}. 
  Magnetospheric angular velocities in the polar cap, are also fixed at a set fraction (${\sim}\unitSI[10]{\%}$) of 
  rigid corotation (\unit{\Omega_J}) \citep{isbell1984}. We also found that the coupling currents showed an 
  increase (${\sim}\unitSI[20]{\%}$) in intensity when going from an average to a more expanded magnetosphere and a 
  decrease (${\sim}\unitSI[40]{\%}$) when going from average to compressed. \\

  Our thermospheric model was used to simulate azimuthal and meridional neutral velocities. We see super-corotation 
  in the azimuthal flows equatorward of the edge of the magnetodisc flux shells. There lies a strong 
  sub-corotational jet at mid to upper altitudes in the mapped ionospheric locations of the outer magnetosphere and 
  polar cap. The spatial size of the 
  strong sub-corotation region increases with increased magnetospheric size due to the weaker magnetic field strength 
  in expanded magnetospheres; thus the transfer of angular momentum is less effective at maintaining corotation. 
  \Rev{ Angular momentum is transferred from the thermosphere to the magnetosphere, in order to accelerate the latter 
  towards corotation. If the thermosphere itself is significantly sub-corotating, then there is a lower `reservoir' of 
  available angular momentum that can be transferred. This results in a decreased plasma angular velocity in these 
  outer regions of the magnetosphere.} We see a meridional flow directed polewards at low altitudes and equatorwards at 
  high altitudes. From the 
  poleward edge of the magnetodisc to the centre of the polar cap, \Rev{a region of accelerated poleward flow} exists whose 
  velocity magnitude increases from a compressed to an expanded 
  magnetosphere. This occurs because there is a force imbalance in this region that increases advection of momentum 
  in expanded 
  magnetospheres. Advection restores balance which results in the acceleration discussed above. This accelerated flow 
  produces a `hotspot' in the polar cap, with a maximum temperature \Rev{increase of ${\sim}\unitSI[130]{K}$} from compressed 
  to expanded magnetosphere. The size of the `hotspot' also increases with an expanding magnetosphere. 
  We find that the outer magnetosphere and polar cap are most strongly heated by Joule heating and ion drag. This 
  heat is then distributed across the polar region via advection rather than viscous transport, whilst more 
  equatorial regions are significantly cooled. This aspect of thermospheric flow is consistent with those 
  presented in SA09. These results also suggest that accurate measurements of ionospheric temperature in the polar 
  region could potentially be used to probe magnetospheric conditions.\\

  We also showed that the power dissipated in the upper atmosphere (consisting of both Joule heating and ion 
  drag) increases with \Rev{an} expanded magnetospheric configuration. The power used to accelerate the magnetospheric 
  plasma initially increases as we expand the magnetosphere from compressed to average configurations, but then 
  decreases with an expansion from average to expanded. This suggests that power used to accelerate the magnetosphere 
  has a `local' maximum for a magnetosphere size somewhere between a compressed and expanded configuration. The total 
  power extracted from planetary rotation is the net sum of the atmospheric and magnetospheric powers and this is 
  positively correlated with magnetosphere size. Comparing our compressed and average magnetospheres with the 
  `intermediate' and `baseline' cases in \citet{cowley07}, we showed that the use of a 
  two-dimensional thermosphere model results in the transfer of ${\sim}\unitSI[20]{\%}$ more energy from the 
  thermosphere to the magnetosphere in order to accelerate the plasma in the magnetodisc. Using our more realistic 
  model of thermospheric flow also produced increased dissipation of energy in the thermosphere via Joule heating and 
  ion drag than the cases presented in \citet{cowley07}. \\

  We have shown that our original compressed case has some unusual current density features due to a 
  relatively high value for \Rev{the} radial current at the outer disc boundary. In order to confirm this we decreased the 
  boundary value \Rev{of} \unit{I_{\rho\infty}} for each case in order to produce alternative models with minimum variance in 
  their \FAC profiles. This led to the selection of \unit{I_{\rho\infty}} of \unitSI[45]{MA}, \unitSI[68]{MA} and 
  \unitSI[80]{MA} for the compressed, average and expanded cases respectively.\\

  Decreasing the radial current \unit{I_{\rho\infty}} at the boundary between the middle and outer 
  magnetospheres resulted in all 
  magnetosphere-ionosphere coupling currents being reduced in accordance with the new value of \unit{I_{\rho\infty}}. 
  This is expected under the assumption of current continuity. The main differences between 
  cases with large \Rev{and reduced} radial current\Rev{s} lies mainly within the magnetodisc. For \Rev{the}
  \FAC density, changes related to \unit{I_{\rho\infty}} were also significant throughout the outer magnetosphere. 
  Thermospheric and magnetospheric angular velocities changed only slightly
  for the baseline and expanded case but much more substantially for our compressed case. For azimuthal flows we 
  found that decreasing \unit{I_{\rho\infty}} also generally increased 
  the level of sub-corotation throughout high latitudes. For meridional flows we found slight increases in 
  localised regions of accelerated flow, most evident in the alternate compressed case. We also found that the polar 
  region 
  becomes slightly warmer with a decrease in \unit{I_{\rho\infty}}; peak temperatures for the alternative 
  configurations increasing relative to their 
  $\unit{I_{\rho\infty}}{=}\unitSI[100]{MA}$ counterparts. The total integrated powers increased 
  with decreasing \unit{I_{\rho\infty}} for our compressed case, but decreased for our baseline and expanded 
  cases. The integrated magnetospheric power for all cases decreased along with \unit{I_{\rho\infty}}, 
  whilst atmospheric power increased by ${\sim}\unitSI[20]{\%}$ for the alternate compressed case but remained almost 
  equal for 
  our baseline and expanded cases. Thus, it seems that decreasing the boundary radial current \unit{I_{\rho\infty}} 
  effectively decreases the `ability' of the thermosphere to transfer angular momentum to the magnetosphere. This 
  \Rev{behaviour, as expected} decreases the intensity of auroral emissions and produces a slightly warmer polar region.\\

  Our calculations suggest that main oval auroral emissions and brightness for an expanded magnetosphere 
  would generally be greater than that of a compressed one. The detailed 
  structure of the \FAC density profile in the magnetodisc is most sensitive to the value of \unit{I_{\rho\infty}} 
  for the compressed case. Compressed magnetospheres in the steady state have 
  larger field strength than expanded ones and are more efficient at maintaining the co-rotating magnetodisc plasma 
  at larger distances. This leads to a smaller shear in angular velocity between the magnetosphere 
  and thermosphere and, consequently, smaller thermospheric temperatures. As a result, auroral emission is brightest 
  for the most expanded magnetospheric systems. We also saw that auroral emissions would increase at the boundary 
  between the outer magnetosphere and the polar region with magnetospheric compression due 
  to the large change in plasma angular velocity at this boundary. Better observational constraints of 
  \unit{\Omega_M} are required to confirm this prediction. \\
  
  This aspect warrants further investigation since we have not attempted to model the change in polar cap angular 
  velocity with solar wind dynamic pressure. Furthermore, the 
  caveat with these predictions is that the system is in a steady-state (where there is no explicit time 
  dependence of the model outputs). We thus view this study as an initial step \Rev{towards} developing a model to study the 
  transient effects of rapid changes in the solar wind dynamic pressure. Results of such studies 
  could provide further insights to the `energy crisis' at Jupiter (SA09), and the physical origin of transient 
  auroral features. \\
  
  Finally, the results presented in this study contribute to a larger set of theoretical investigations which have 
  provided useful quantitative predictions of how the Jovian aurorae would respond to changes in solar wind dynamic 
  pressure. Such results are useful for interpreting auroral observations, and \Rev{for} making more extensive use of such 
  data as remote diagnostics of the physical state of the Jovian magnetosphere.

\section*{Acknowledgement} 
  JNY was supported by an STFC studentship award.

\bibliographystyle{elsarticle-harv}
\bibliography{bibliography} \pagebreak

    \begin{table}[H!]
    
    \caption{Table showing the three different magnetospheric configurations used in this study. The radii  
	      of the magnetodisc \unit{R_{MM}} and magnetopause \unit{R_{MP}} are shown along with the values of the 
	      perturbation field. Note \unit{R_{MP}} is calculated as in \citet{cowley07}. Solar wind dynamic pressure 
	      (\unit{P_{SW}}) is also shown for both \citet{joy2002} and \citet{huddleston1998} magnetopause models (J 
	      or H respectively).}
    \begin{tabular}{ l  r  r  r }
      \hline
      Case &  A &  B & C \\ \hline
      \hline
      \unit{R_{MM}}/ \unitSI{R_J} & \unitSI[45]{} & \unitSI[65]{} & \unitSI[85]{}\\ \hline
      \unit{R_{MP}}/ \unitSI{R_J} & \unitSI[75]{} & \unitSI[86]{} & \unitSI[101]{}\\ \hline
      \unit{\Delta B_z}/ \unitSI{nT} & \unitSI[-1.16]{} & \unitSI[0.0]{} & \unitSI[0.19]{}\\ \hline
      \unit{P_{SW J}}/ \unitSI{nPa} & \unitSI[0.121]{} & \unitSI[0.060]{} & \unitSI[0.020]{}\\ \hline
      \unit{P_{SW H}}/ \unitSI{nPa} & \unitSI[0.034]{} & \unitSI[0.018]{} & \unitSI[0.008]{}\\ \hline
    \end{tabular}
    \label{tb:cases} 
  \end{table}

\end{document}